\documentstyle[psfig,aps,prl,amsfonts,amssymb]{revtex}

\begin{document}
\title{Geometric Phases of the Uhlmann and Sj\"oqvist {\it et al}
Types for $O(3)$-Orbits of $n$-Level Gibbsian Density Matrices}
\author{Paul B. Slater}
\address{ISBER, University of
California, Santa Barbara, CA 93106-2150\\
e-mail: slater@itp.ucsb.edu,
FAX: (805) 893-7995}

\date{\today}

\draft
\maketitle
\vskip -0.1cm

\begin{abstract}
We accept the implicit challenge of A. Uhlmann in his 1994 paper,
``Parallel Lifts and Holonomy along Density Operators: Computable
Examples Using $O(3)$-Orbits,'' by, in fact, computing the holonomy
invariants for rotations of certain $n$-level Gibbsian density matrices
($n=2,\ldots,11$). From these we derive, by the tracing operation,
the associated geometric phases and visibilities, which we analyze
and display.
We then proceed analogously, implementing the alternative methodology
presented by E. Sj\"oqvist {\it et al} in their letter, ``Geometric Phases
for Mixed States in Interferometry'' (Phys. Rev. Lett. 85, 2845 [2000]).
For the Uhlmann case, we are able to 
also compute several {\it higher-order} holonomy
invariants. We compare the various geometric phases and visibilities
for different values of $n$, and also directly compare the two 
forms of analysis. By setting one parameter ($a$)
in the Uhlmann analysis to zero, we find that the so-reduced form 
of the first-order holonomy
invariant is simply equal to $(-1)^{n+1}$ times the holonomy invariant
in the Sj\"oqvist {\it et al} method. Additional phenomena 
of interest are
reported too.

\end{abstract}

\pacs{PACS Numbers 03.65.Vf, 05.30.-d, 05.70.-a}

\tableofcontents

\vspace{.1cm}
\section{Introduction} \label{r1}
In a recent paper, Sj\"oqvist {\it et al} \cite{sjo1} provided
``a new formalism of the geometric phase for mixed states in the 
experimental context of quantum interferometry''.
Only in passing, did these authors 
 note that ``Uhlmann was probably the first to
address the issue of mixed state holonomy, but as a purely
mathematical problem''.

Interested in possible 
(previously uninvestigated) relationships between the work of Sj\"oqvist
{\it et al} \cite{sjo1} 
and that of Uhlmann \cite{uhl1,uhl2}, the present author compared the two
approaches in terms of two spin-${1 \over 2}$ scenarios 
\cite{slaternewest}. In the first of these,
the spin-${1 \over 2}$ systems undergo unitary evolution along geodesic
triangles \cite{uhl1}, while in the second the unitary evolution takes place along
circular paths \cite{uhl2}. 
In\cite{uhl2}, Uhlmann had also proposed an additional scenario, which was
(initially) left unanalyzed in \cite{slaternewest}.
It involves ``the Gibbsian states of the form'',
\begin{equation}
\rho ={ e^{\alpha \overrightarrow{n} 
\overrightarrow{J}} \over \mbox{trace} \quad e^{\alpha \overrightarrow{n}
 \overrightarrow{J}}},
\end{equation}
which fill for a given value of $\alpha$ a 2-sphere called
$\mathbf{S}^{\alpha}_{j}$ if 
$\overrightarrow{n}$ runs through all directions in 3-space.
Now, for the starting point of the unitary evolution, one
sets $\overrightarrow{n} =(0,0,1)$, while 
$\overrightarrow{n} = (0,\sin{\theta},\cos{\theta})$ is chosen
as rotational axis.
The curve of state evolution is given by
\begin{equation} \label{fq1}
\phi \rightarrow U(\phi) \rho_{0} U(-\phi), \qquad
U(\phi) = e^{-i \phi (\sin{\theta} J_{y} + \cos{\theta} J_{z})},
\end{equation}
and the associated parallel lift of this curve with initial value
$\rho^{1/2}$ is
\begin{equation}
\phi \rightarrow  U(\phi) \rho_{0}^{1/2} V(\phi), 
\quad V(\phi) = e^{i \phi \tilde{H}},
\end{equation}
where 
\begin{equation} \label{aequation}
\tilde{H} = \cos{\theta} J_{z} +  a \sin{\theta} J_{y},
\qquad \mbox{and}\quad  a ={1 \over  \cosh{\alpha \over 2}} .
\end{equation}
It is possible to regard $V(\phi)$ as a rotation with angle
\begin{equation} \label{KAP}
\tilde{\phi} = \kappa \phi , \qquad \kappa = \sqrt{\cos^{2}{\theta} 
+a^2 \sin^{2}{\theta}} \leq 1
\end{equation}
and rotation axis
\begin{equation}
\overrightarrow{\xi} = (0, {\sin{\theta} \over \kappa},
{a \cos{\theta} \over \kappa}).
\end{equation}
The holonomy invariant can then  be written as
\begin{equation} \label{hi}
(-1^{2 j}) \rho_{0}^{1/2} e^{2 \pi i \tilde{H}} \rho_{0}^{1/2} =
(-1^{2 j}) \rho_{0}^{1/2} e^{2 \pi i \kappa \overrightarrow{\xi}
\overrightarrow{J}} \rho_{0}^{1/2}.
\end{equation}
(All the preceding equations are directly adopted from \cite{uhl1}.)

\begin{figure}
\centerline{\psfig{figure=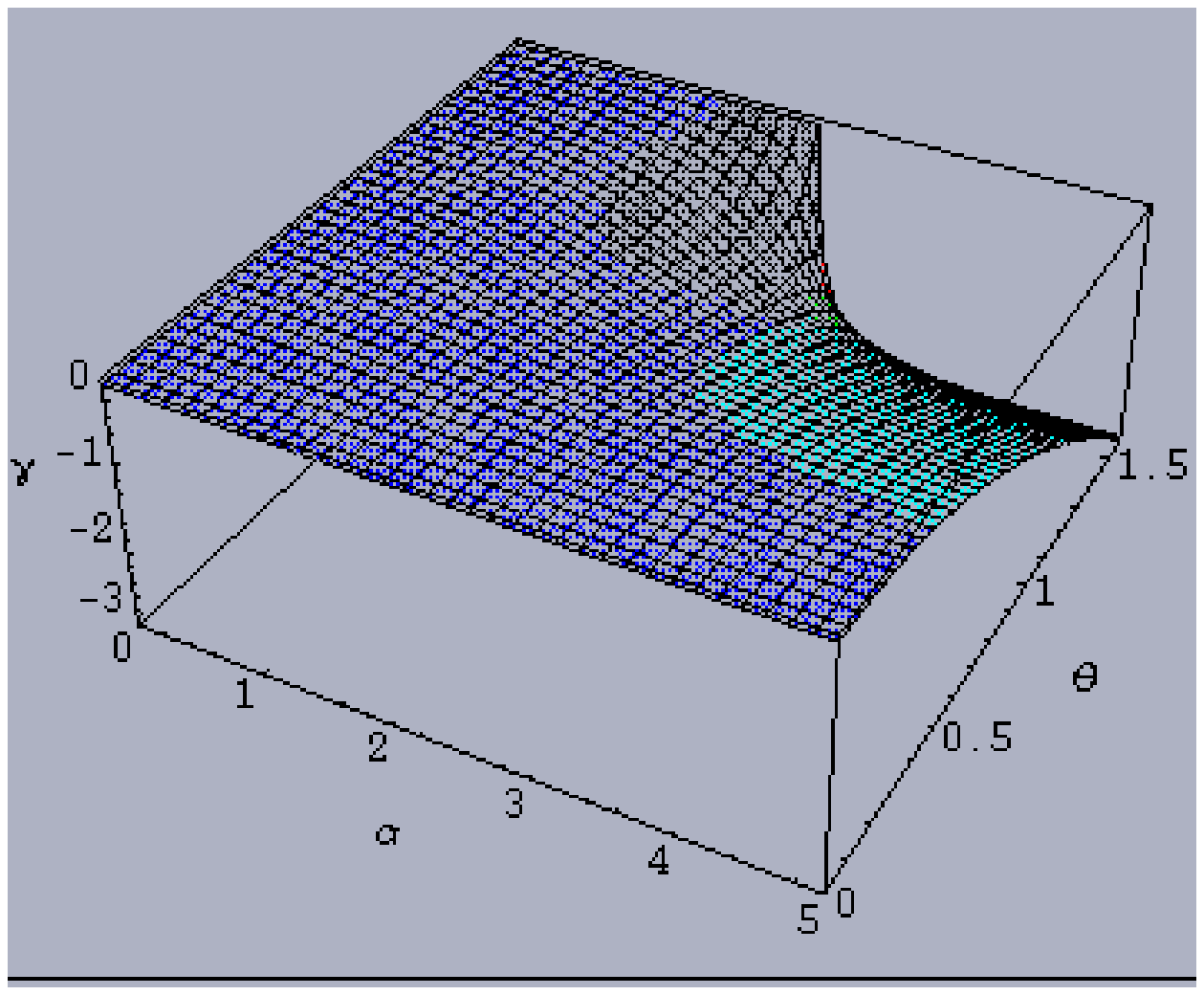}}
\caption{Uhlmann geometric phase for Gibbsian spin-${1 \over 2}$ systems}
\label{g2}
\end{figure}

\begin{figure}
\centerline{\psfig{figure=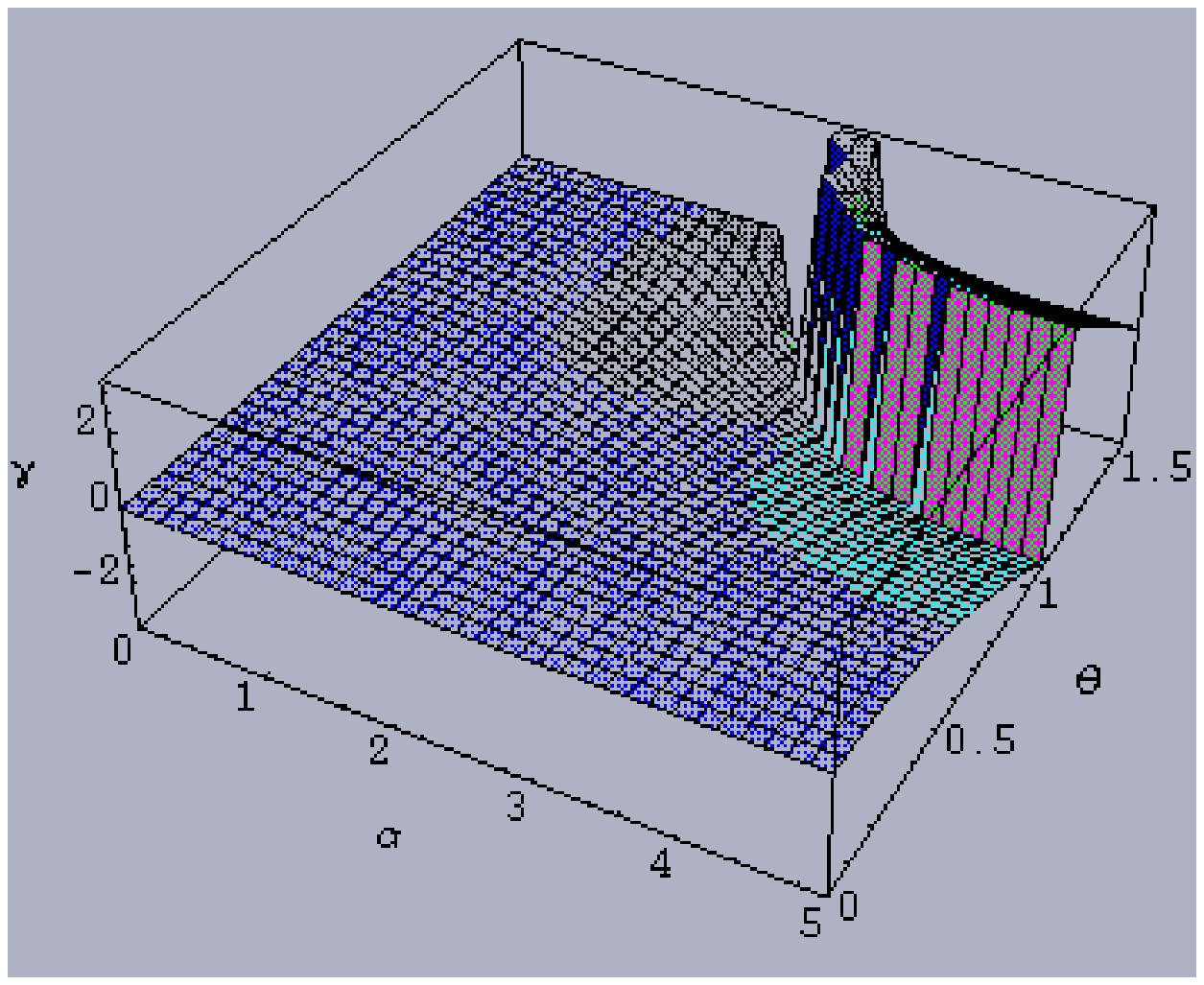}}
\caption{Uhlmann geometric phase for Gibbsian spin-1 systems}
\label{g3}
\end{figure}

\begin{figure}
\centerline{\psfig{figure=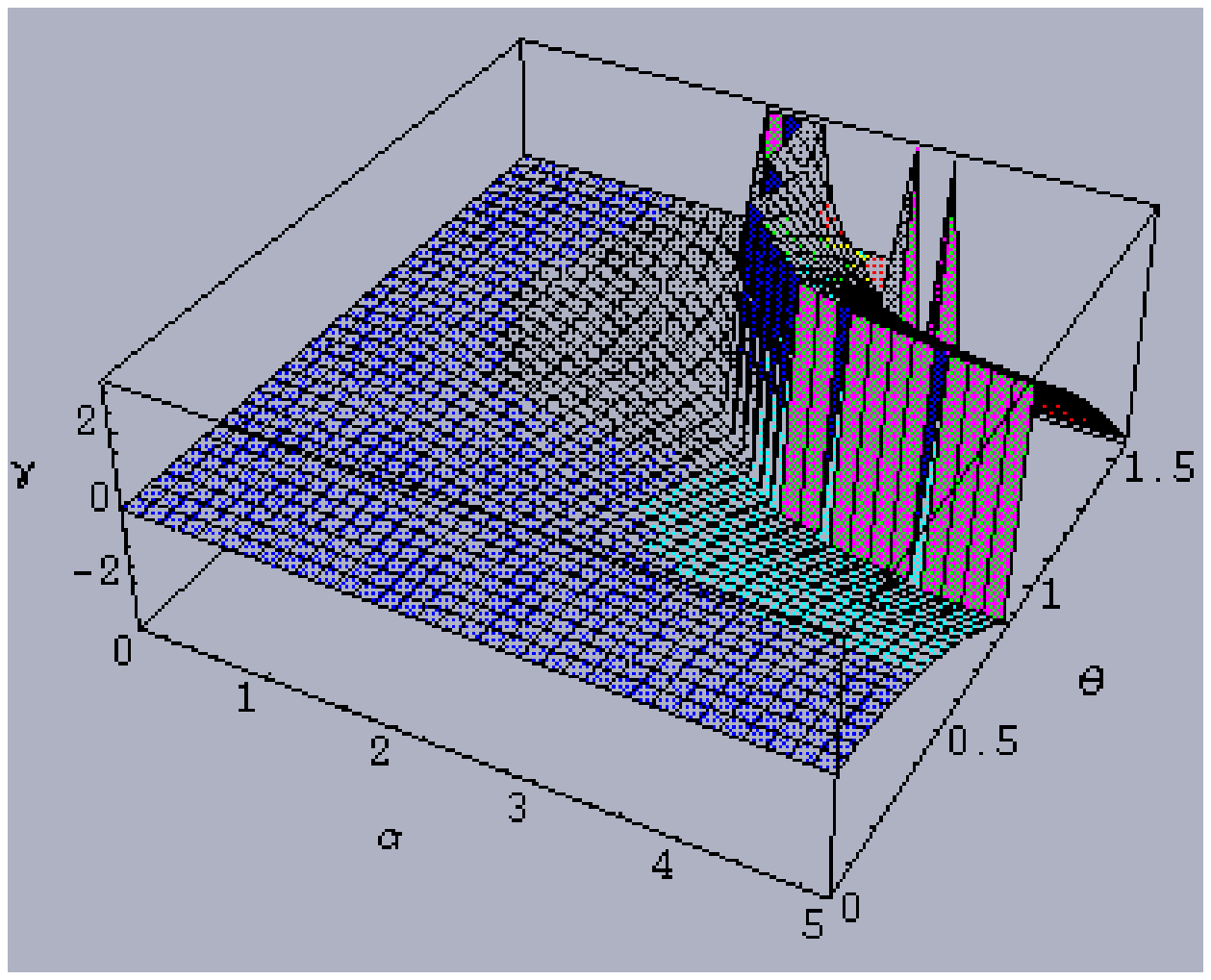}}
\caption{Uhlmann geometric phase for Gibbsian spin-${3 \over 2}$ systems}
\label{g4}
\end{figure}

\begin{figure}
\centerline{\psfig{figure=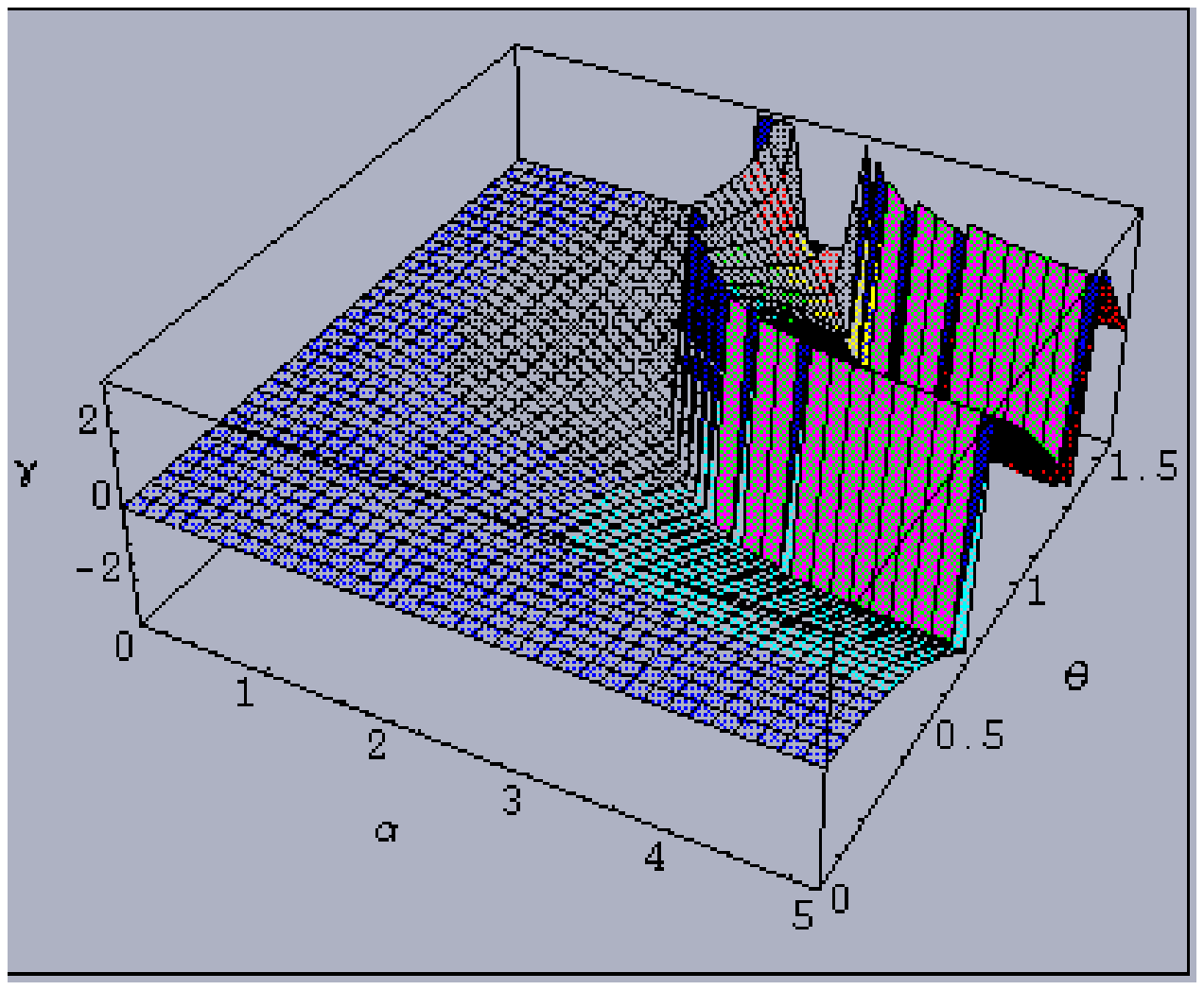}}
\caption{Uhlmann geometric phase for Gibbsian spin-2 systems}
\label{g5}
\end{figure}

\begin{figure}
\centerline{\psfig{figure=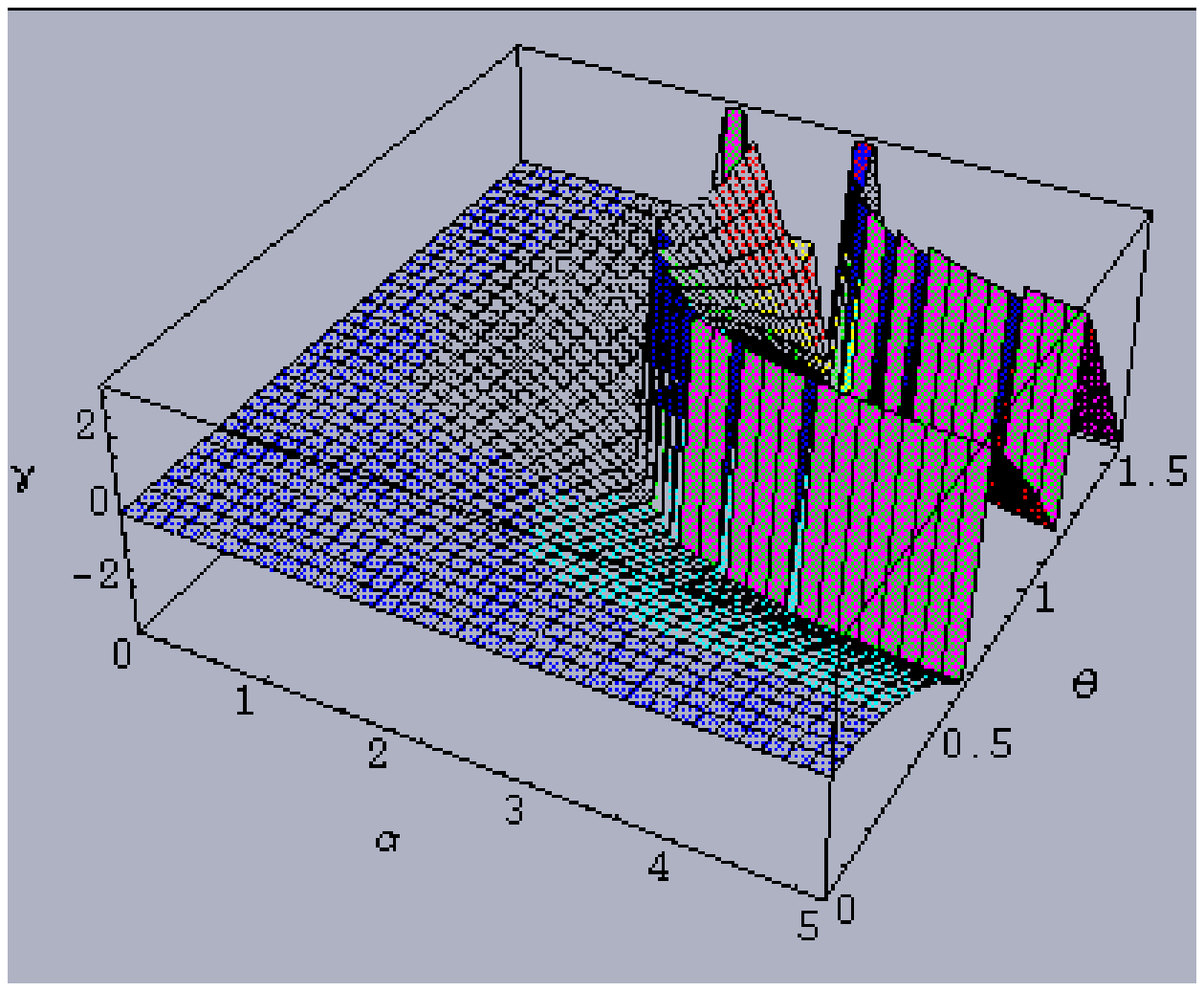}}
\caption{Uhlmann geometric phase for Gibbsian spin-${5 \over 2}$ systems}
\label{g6}
\end{figure}

\begin{figure}
\centerline{\psfig{figure=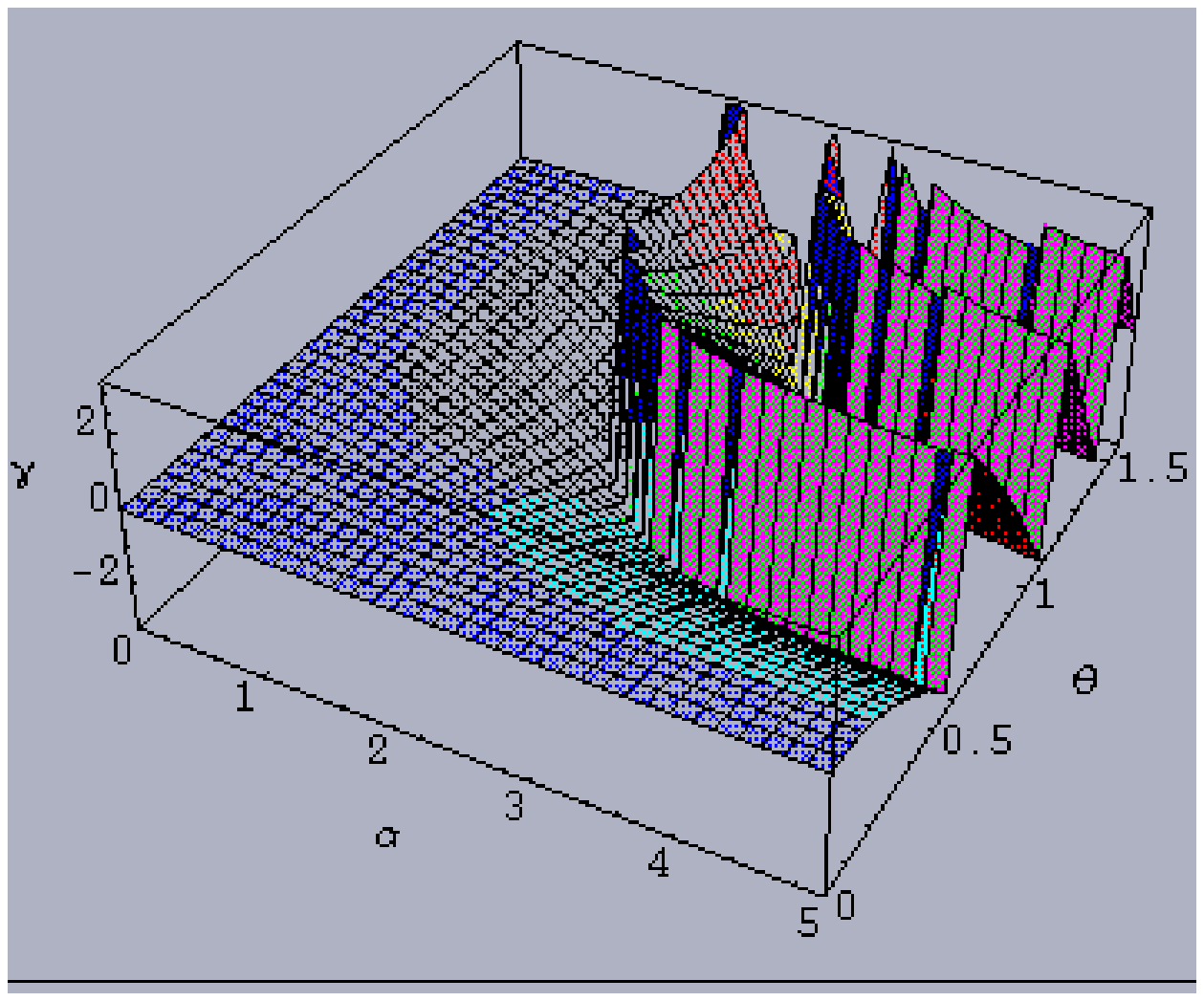}}
\caption{Uhlmann geometric phase for Gibbsian spin-3 systems}
\label{g7}
\end{figure}

\begin{figure}
\centerline{\psfig{figure=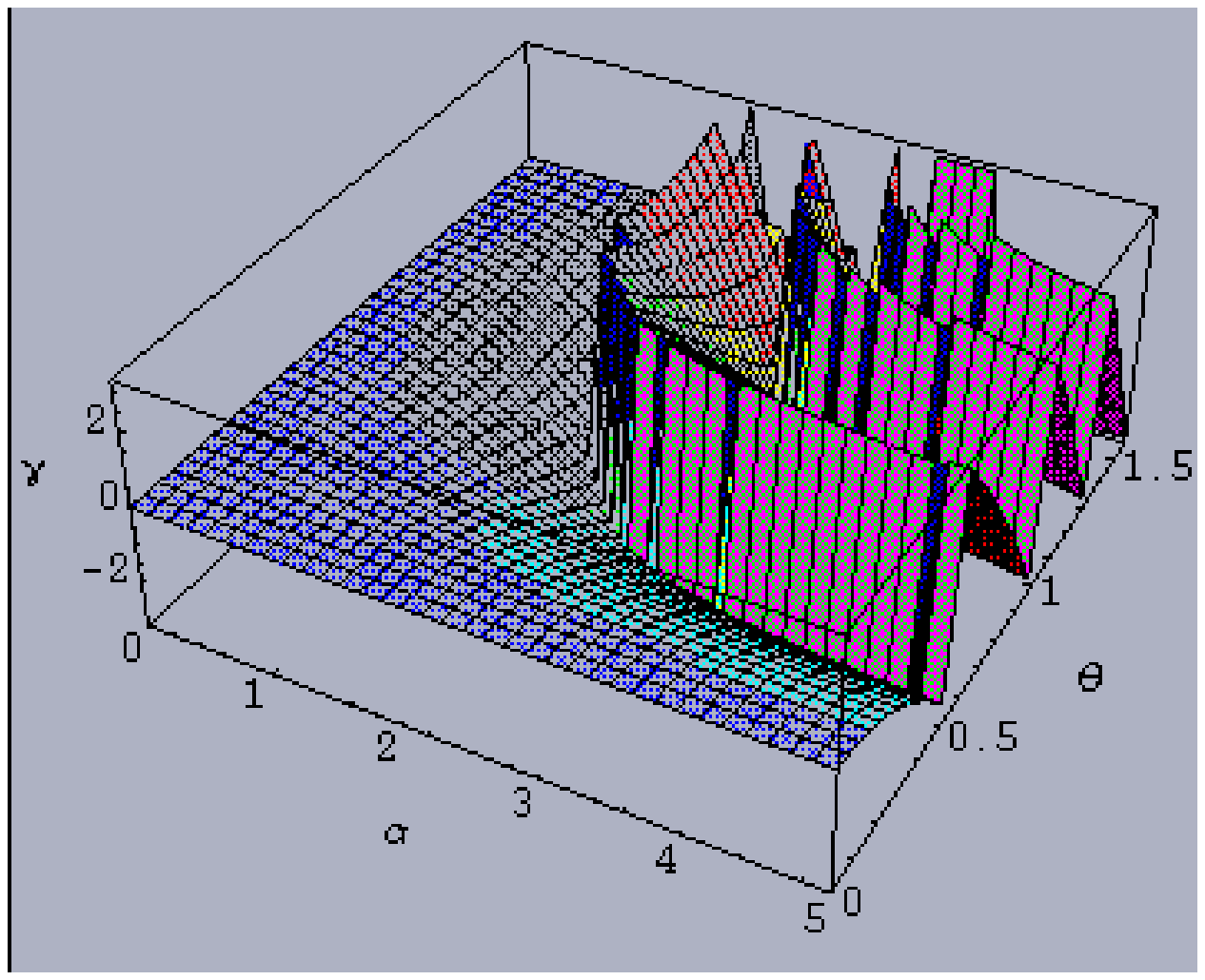}}
\caption{Uhlmann geometric phase for Gibbsian spin-${7 \over 2}$ systems}
\label{g8}
\end{figure}

\begin{figure}
\centerline{\psfig{figure=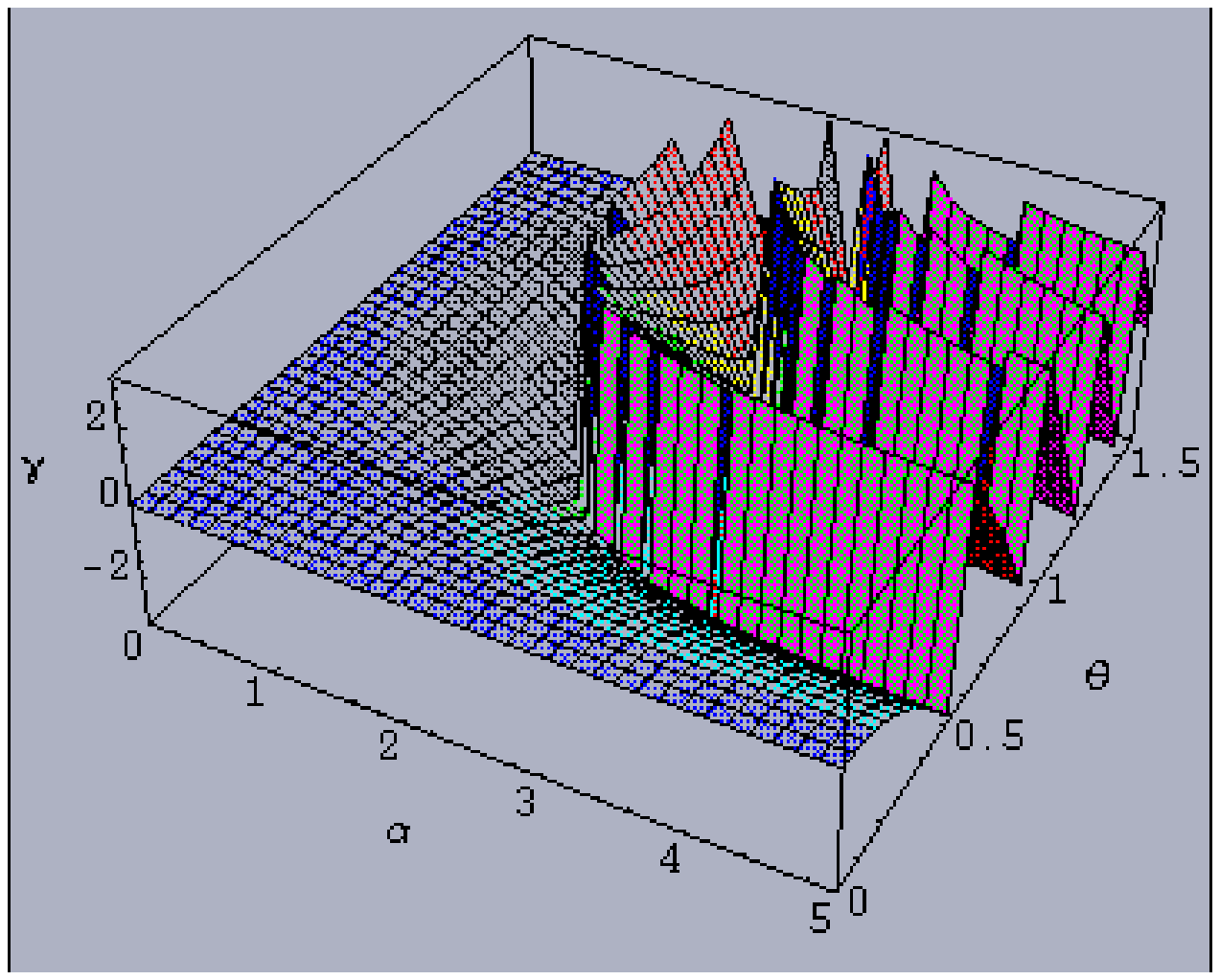}}
\caption{Uhlmann geometric phase for Gibbsian spin-4 systems}
\label{g9}
\end{figure}

\begin{figure}
\centerline{\psfig{figure=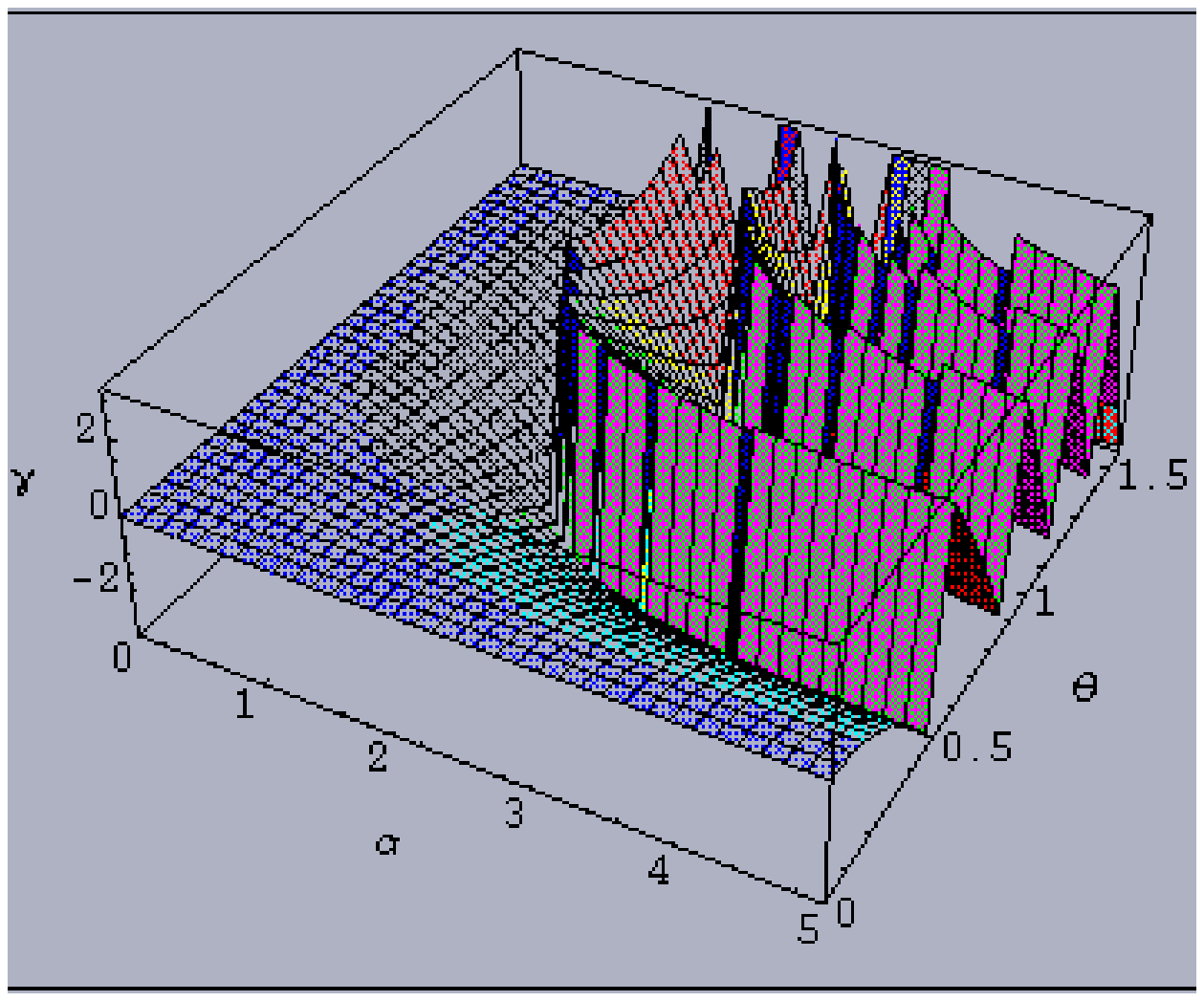}}
\caption{Uhlmann geometric phase for Gibbsian spin-${9 \over 2}$ systems}
\label{g10}
\end{figure}

\begin{figure}
\centerline{\psfig{figure=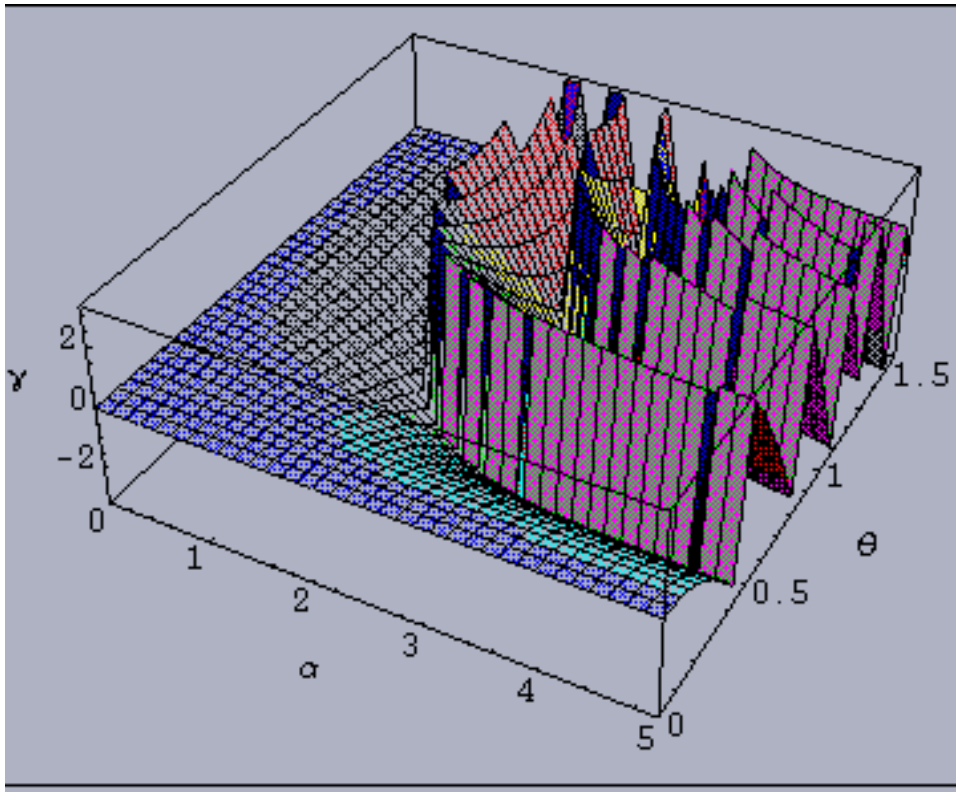}}
\caption{Uhlmann geometric phase for Gibbsian spin-5 systems}
\label{g11}
\end{figure}

In this paper, to begin,  we compute the trace of this invariant (\ref{hi})
for all $j= {1 \over 2}, 1, {3 \over 2},\ldots, {9 \over 2}$.
We plot the arguments of these traces, that is the corresponding
geometric phases ($\gamma_{j}$), 
in Figs.~\ref{g2} - \ref{g11} (sec.~\ref{r2}), and their absolute
values, that is the visibilities ($\nu_{j}$), in Figs.~\ref{v2} - \ref{v11}
(sec.~\ref{r3}).
(For $j = {1 \over 2}$, the results are equivalent to the first 
(non-Gibbsian) set of models in \cite{uhl1}, 
which was compared with the analyses
of Sj\"oqvist {\it et al} \cite{sjo1} in
\cite{slaternewest}.) Let us note that
the value $\alpha=0$ corresponds to the fully
mixed (classical) state, while $\alpha = \pm \infty$ correspond to pure
states.

In sec.~\ref{r4}, we compare certain aspects of these results
{\it across} the ten distinct values of $n$. All the various results 
up to this point can
be considered as {\it first-order} in nature.
In sec.~\ref{r5}, on the other hand, 
we  compute certain {\it higher-order} invariants.
In secs.~\ref{r6}-\ref{rr0}, we perform analyses precisely analogous to
those in secs.~\ref{r2}-\ref{r4}, but now in terms not of the approach of
Uhlmann to mixed state holonomy, but that of Sj\"oqvist {\it et al}
\cite{sjo1}.
In our final analytical section (sec.~\ref{r8}), before the
summary (sec.~\ref{r9}), we directly compare
results obtained by the two different procedures.

One of our main findings is that if one sets the implicit parameter $a$ 
in the Uhlmann holonomy invariant (\ref{hi}) to zero, then its resultant
trace is simply equal to $(-1)^{n+1}$ times the holonomy invariant
\cite[eq. (15)]{sjo1}
yielded by the methodology of Sj\"oqvist {\it et al}.
\section{Uhlmann geometric phases for Gibbsian $n$-level systems
($n=2,\ldots,11$)} \label{r2}
\section{Uhlmann visibilities for Gibbsian $n$-level systems
($n=2,\ldots,11$)} \label{r3}
\begin{figure}
\centerline{\psfig{figure=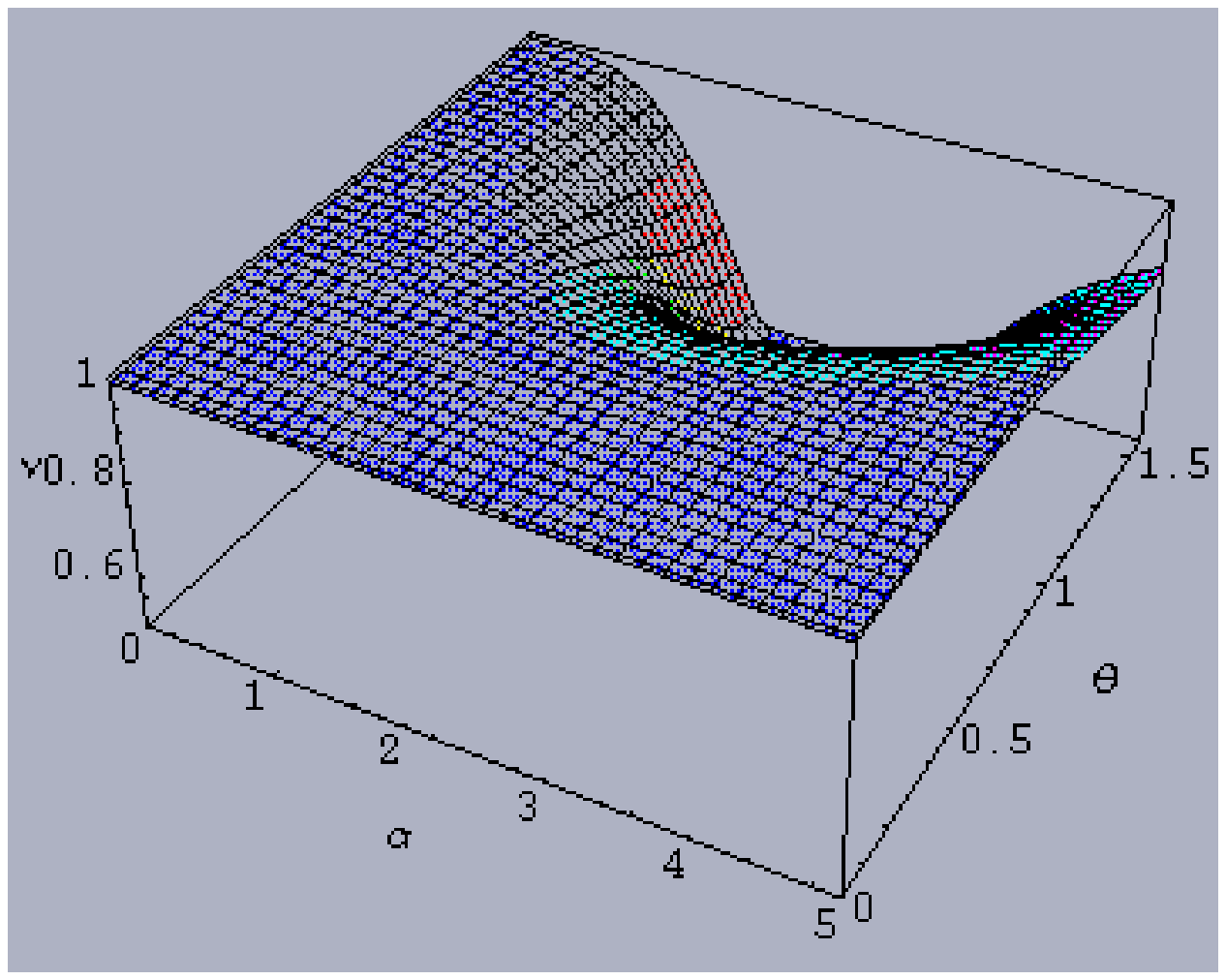}}
\caption{Uhlmann visibility for Gibbsian spin-${1 \over 2}$ systems}
\label{v2}
\end{figure}
\begin{figure}
\centerline{\psfig{figure=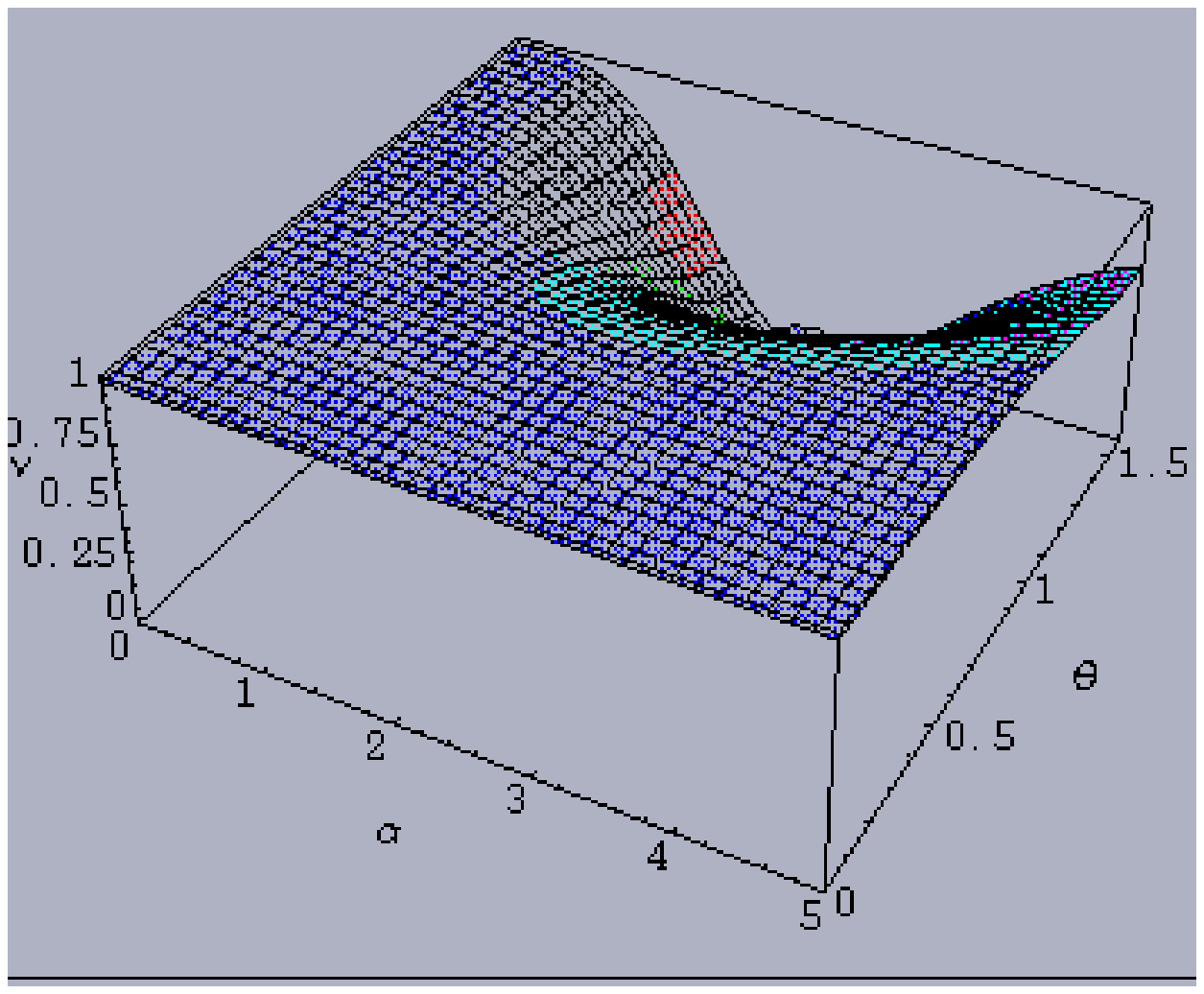}}
\caption{Uhlmann visibility for Gibbsian spin-1 systems}
\label{v3}
\end{figure}
\begin{figure}
\centerline{\psfig{figure=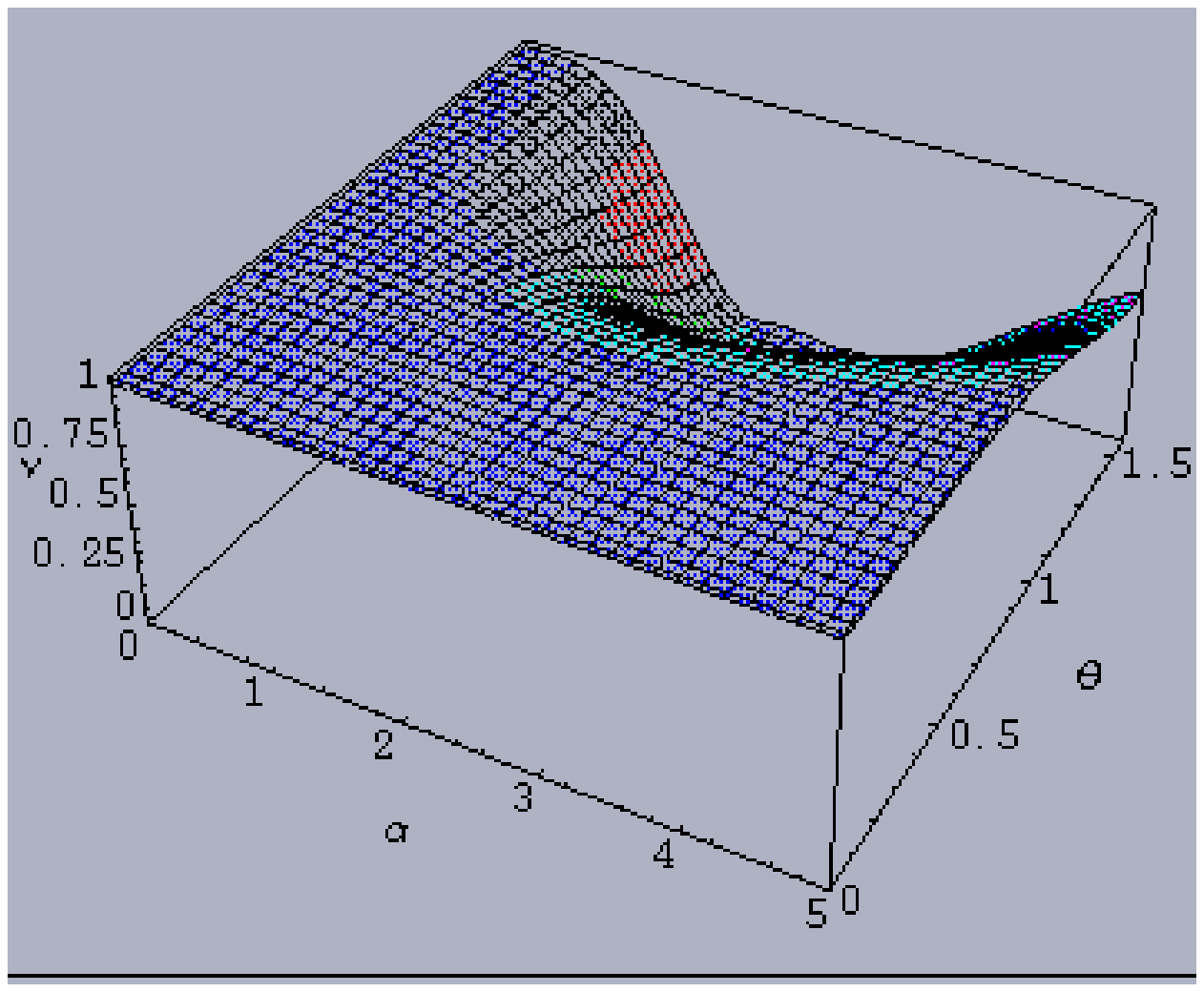}}
\caption{Uhlmann visibility for Gibbsian spin-${3 \over 2}$ systems}
\label{v4}
\end{figure}
\begin{figure}
\centerline{\psfig{figure=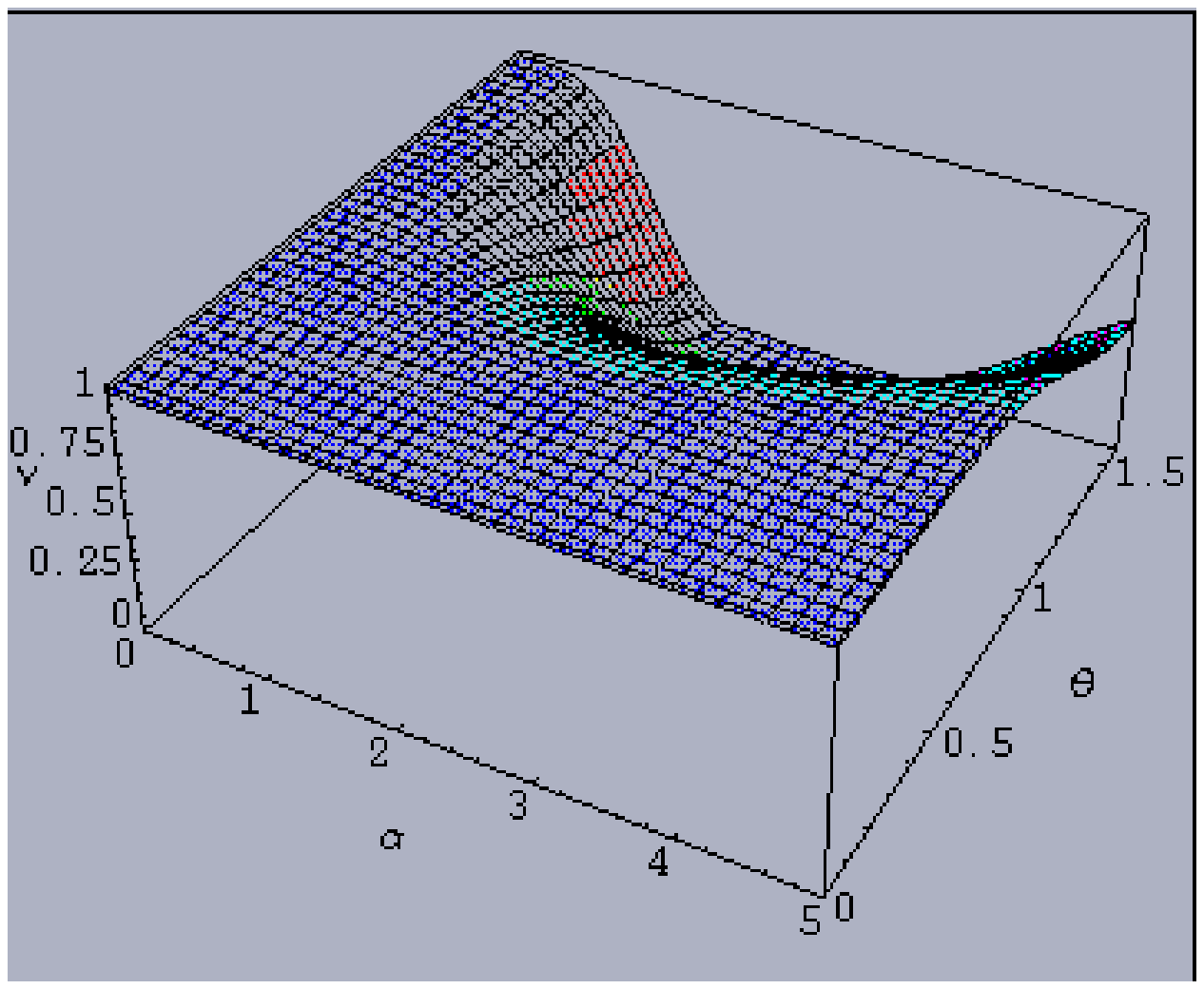}}
\caption{Uhlmann visibility for Gibbsian spin-2 systems}
\label{v5}
\end{figure}
\begin{figure}
\centerline{\psfig{figure=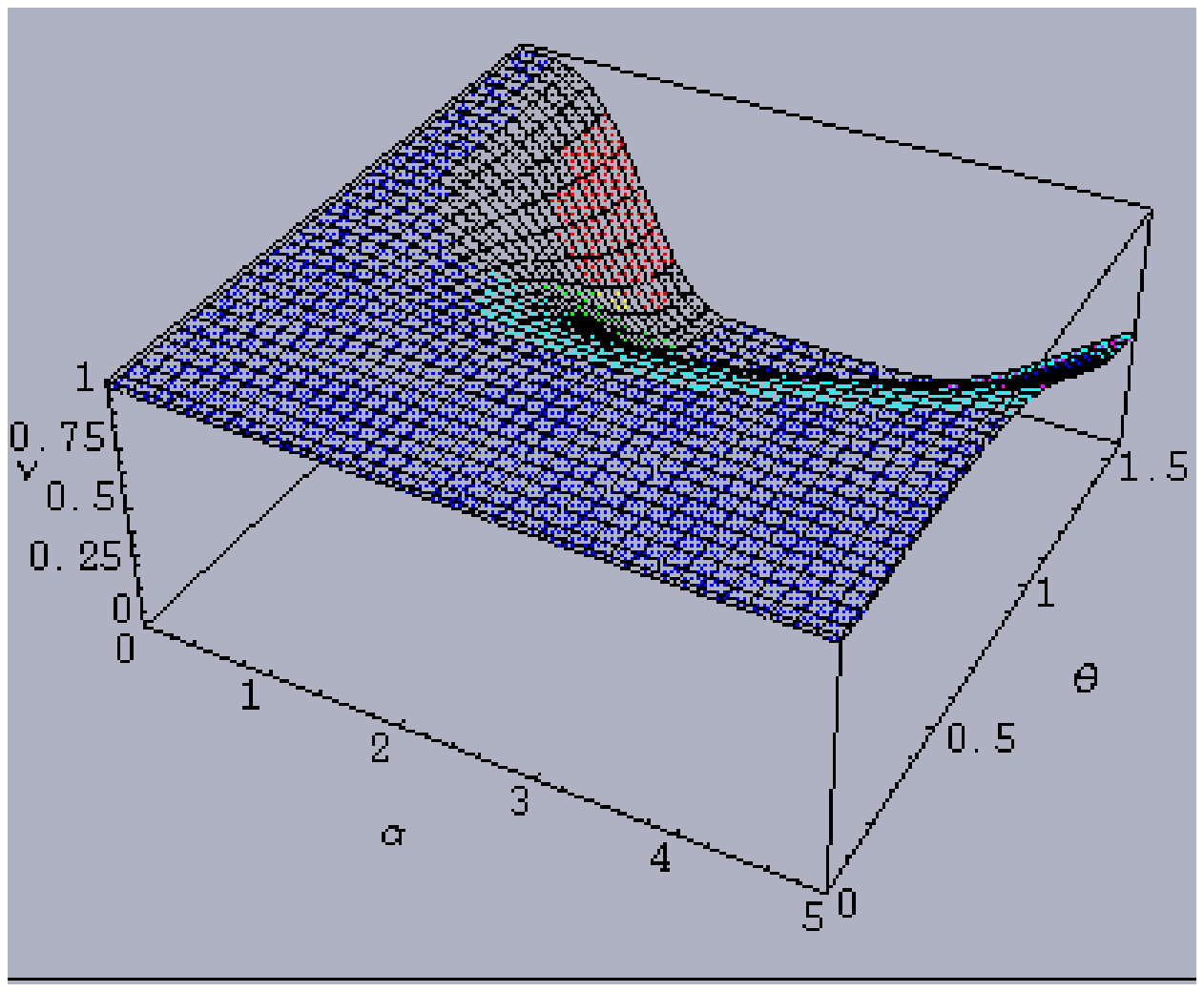}}
\caption{Uhlmann visibility for Gibbsian spin-${5 \over 2}$ systems}
\label{v6}
\end{figure}
\begin{figure}
\centerline{\psfig{figure=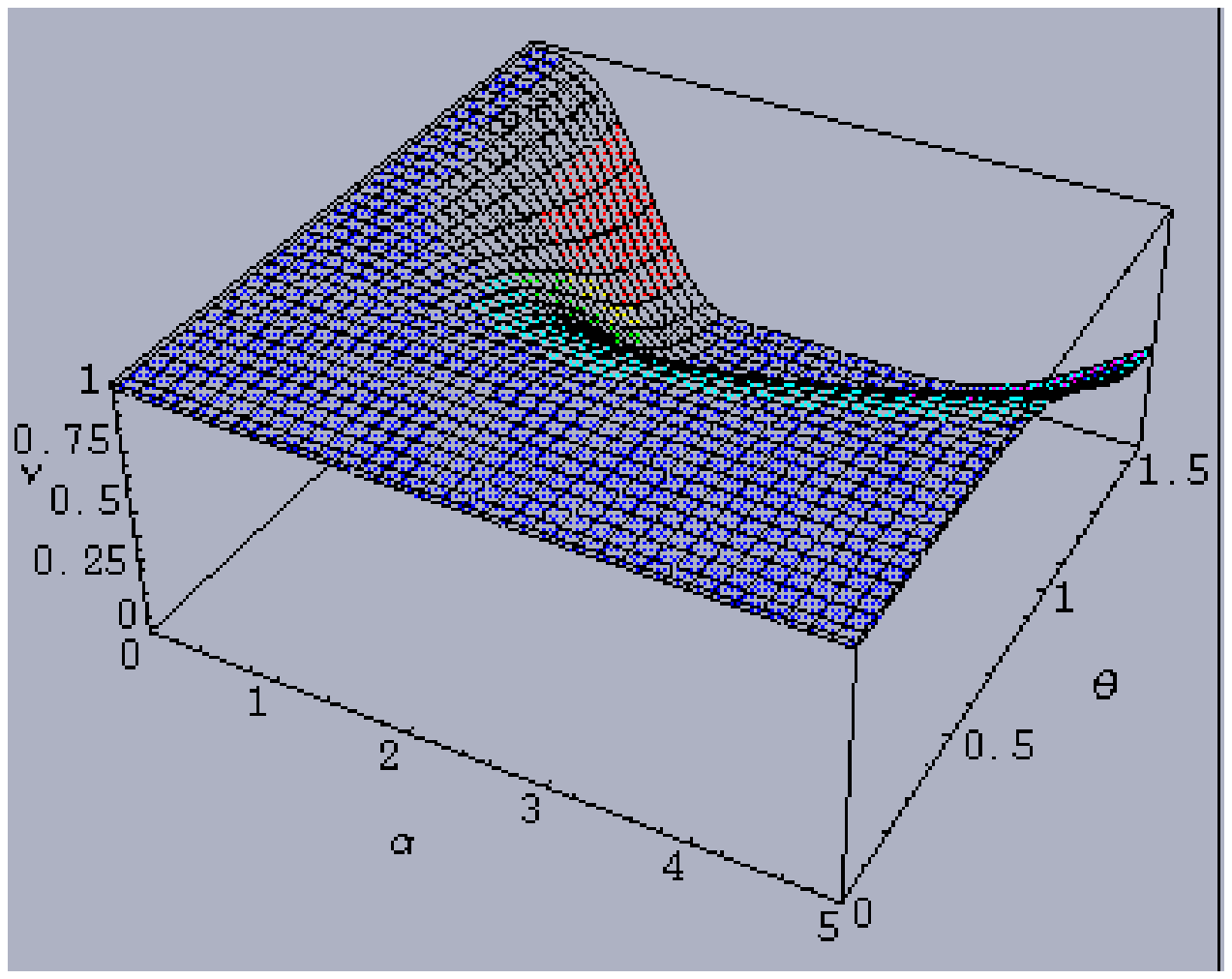}}
\caption{Uhlmann visibility for Gibbsian spin-3 systems}
\label{v7}
\end{figure}
\begin{figure}
\centerline{\psfig{figure=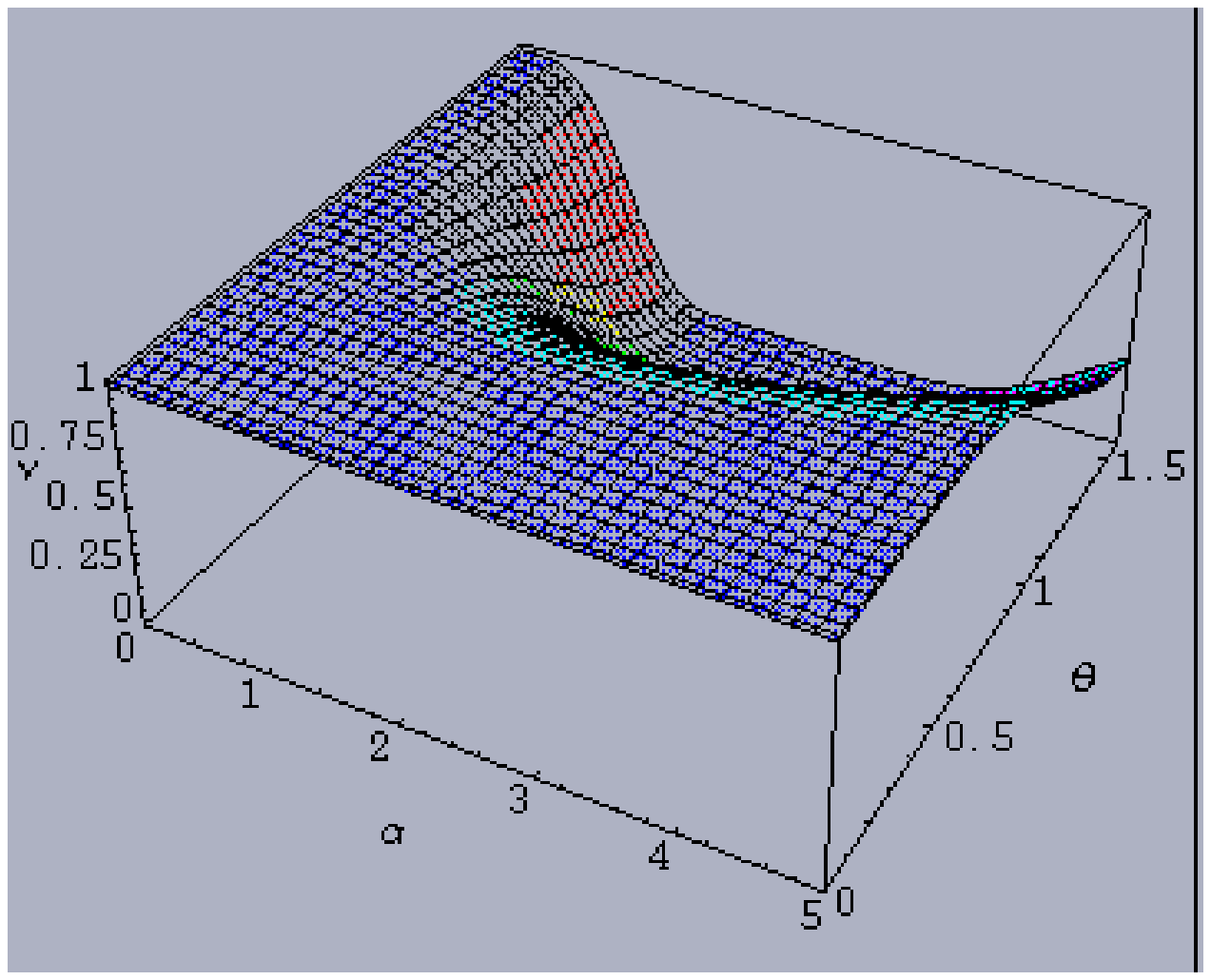}}
\caption{Uhlmann visibility for Gibbsian spin-${7 \over 2}$ systems}
\label{v8}
\end{figure}
\begin{figure}
\centerline{\psfig{figure=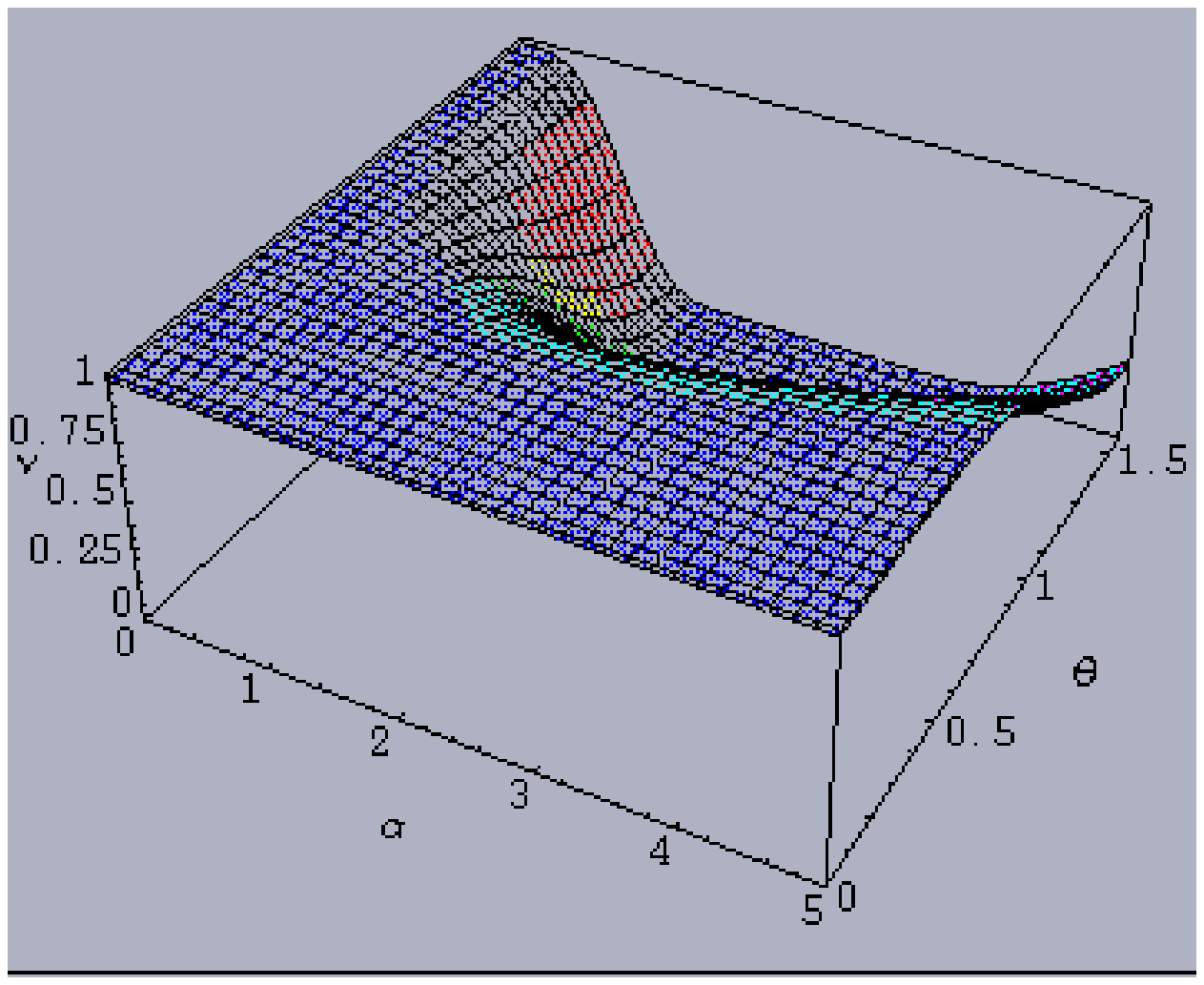}}
\caption{Uhlmann visibility for Gibbsian spin-4 systems}
\label{v9}
\end{figure}
\begin{figure}
\centerline{\psfig{figure=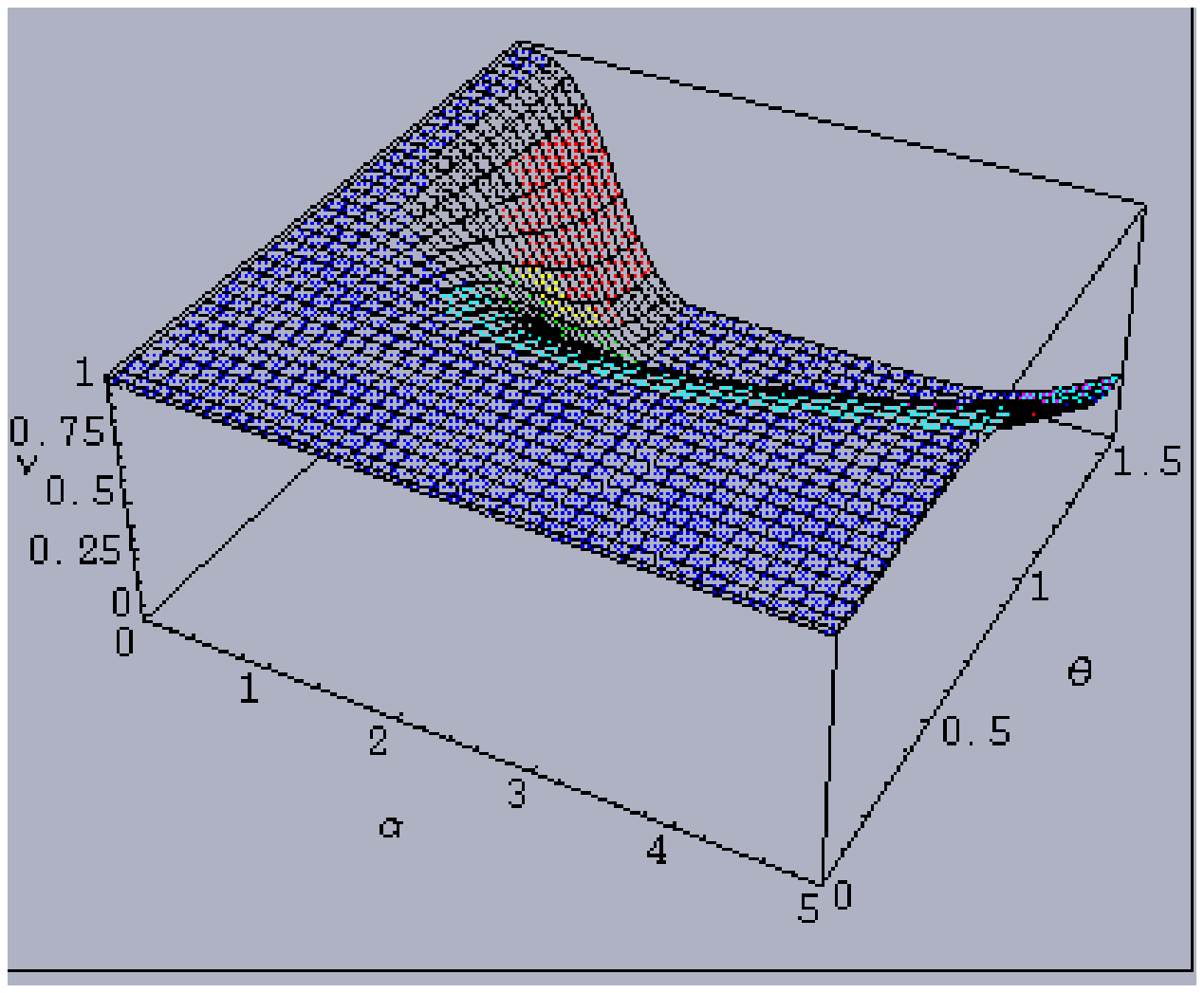}}
\caption{Uhlmann visibility for Gibbsian spin-${9 \over 2}$-systems}
\label{v10}
\end{figure}
\begin{figure}
\centerline{\psfig{figure=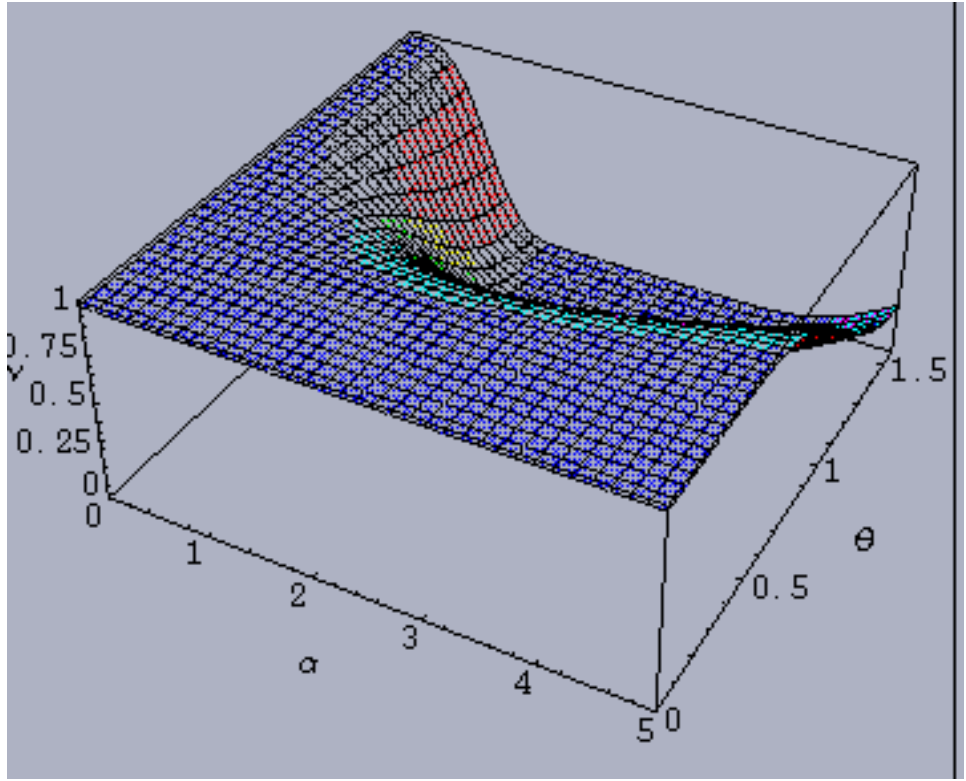}}
\caption{Uhlmann visibility for Gibbsian spin-5 systems}
\label{v11}
\end{figure}
\section{Comparisons across 
Gibbsian $n$-level systems of 
Uhlmann geometric phases and visibilities} \label{r4}
In the spin-${1 \over 2}$ case (Figs.~\ref{g2} and \ref{v2}), we have that
the geometric phase $\gamma_{1 \over 2}$ is the arctangent of ${y \over x}$, 
taking into account which
quadrant the point $(x,y)$ lies in, where
\begin{equation}
x=-\cosh{ i \pi \kappa}, \qquad y = - {\cos{\theta} \sin{\pi \kappa} 
\tanh{\alpha \over 2} \over  \kappa}
\end{equation}
Also, the visibility in this spin-${1 \over 2}$ case is given by
\begin{equation}
\nu_{1 \over 2} = \sqrt{1 + {4 \sinh^{2}{i \pi \kappa} \over \zeta}},
\end{equation}
where 
\begin{equation}
\zeta=  
3 -\cos{2 \theta} + 2 \cos^{2}{\theta} \cosh{\alpha},
\end{equation}
and $\kappa$  is as defined in (\ref{KAP}).
The comparable results in the spin-1 case (Figs.~\ref{g3} and \ref{v3})
are similarly based on
\begin{equation}
x= {e^{-2 i \pi \kappa} \Big(2 (e^{\alpha} + 2 e^{2 i \pi \kappa}+ 
e^{\alpha +4 i \pi \kappa})
\cos^{2}{\theta} + (2 + 2 e^{4 i \pi \kappa} + e^{\alpha} 
(1 + e^{2 i \pi \kappa})^{2}) \mbox{sech}^{2}{{\alpha \over 2}}
\sin^{2}{\theta} \Big) \over 4 \kappa^{2} (1 +2 \cosh{\alpha})},
\end{equation}
and
\begin{equation}
y={e^{\alpha} \cos{\theta} \sin{2 \pi \kappa} \over \kappa + 2 
\kappa \cos{\alpha}}.
\end{equation}
One feature distinguishing the (strikingly similar) 
visibility plots (Figs.~\ref{v2} - \ref{v11}) from one another
is that as $j$ increases, $\nu_{j}$ decreases at the 
(boundary) point 
$\alpha =5, \theta= {\pi \over 2}$. For instance this value is
.871618 in Fig.~\ref{v2} and .498776 in Fig.~\ref{v6}.

\begin{figure}
\centerline{\psfig{figure=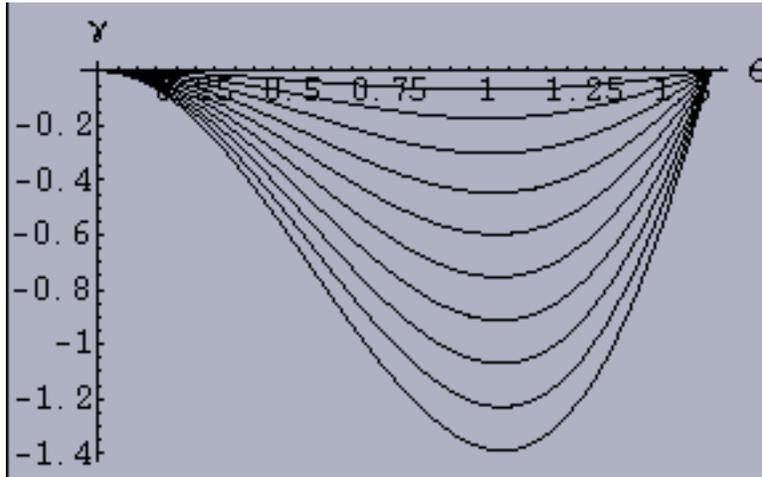}}
\caption{Uhlmann geometric phases for $n$-level Gibbsian systems 
($n=2,\ldots,11$) holding $\alpha=1$. The curve for $n=2$ dominates
that for $n=3$, which dominates that for $n=4,\ldots$}
\label{s1}
\end{figure}

\begin{figure}
\centerline{\psfig{figure=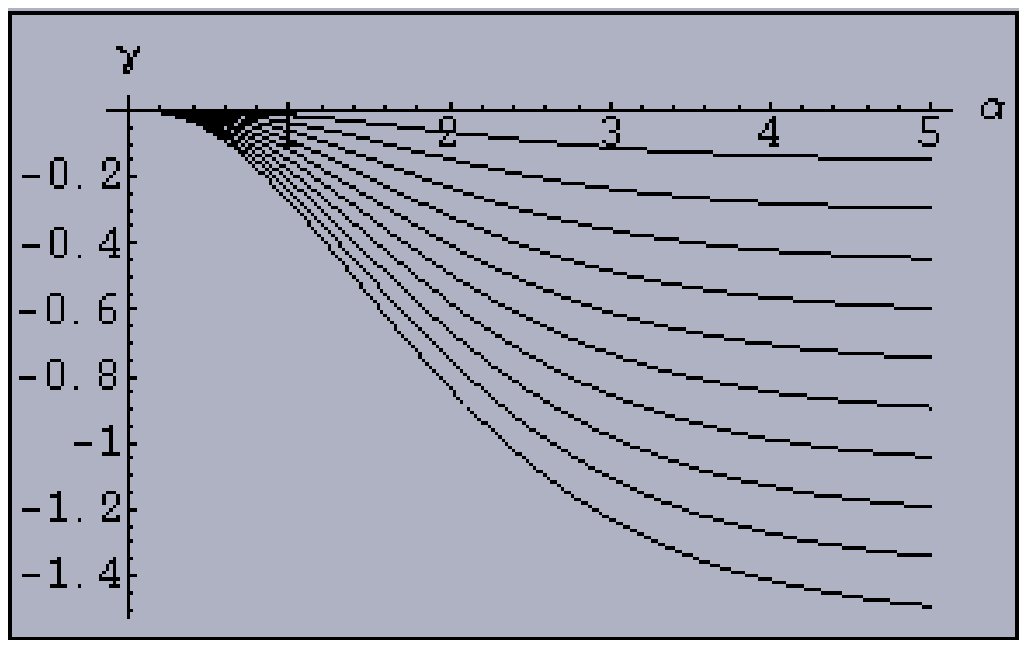}}
\caption{Uhlmann geometric phases for $n$-level Gibbsian systems
($n=2,\ldots,11$) holding $\theta$ fixed at ${\pi \over 10}$. The curve for
$n=2$ dominates that for $n=3$, which dominates that for $n=4, \ldots$}
\label{s2}
\end{figure}

\begin{figure}
\centerline{\psfig{figure=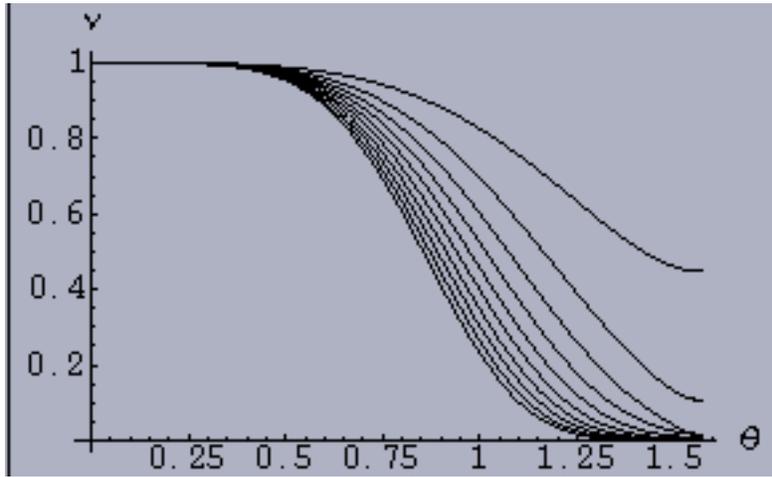}}
\caption{Uhlmann visibilities for $n$-level Gibbsian systems ($n=2,\ldots,11$) holding 
$\alpha$ fixed at 2. The curve for $n=2$ dominates that for $n=3$, which
dominates that for $n=4,\ldots$}
\label{s3}
\end{figure}

\begin{figure}
\centerline{\psfig{figure=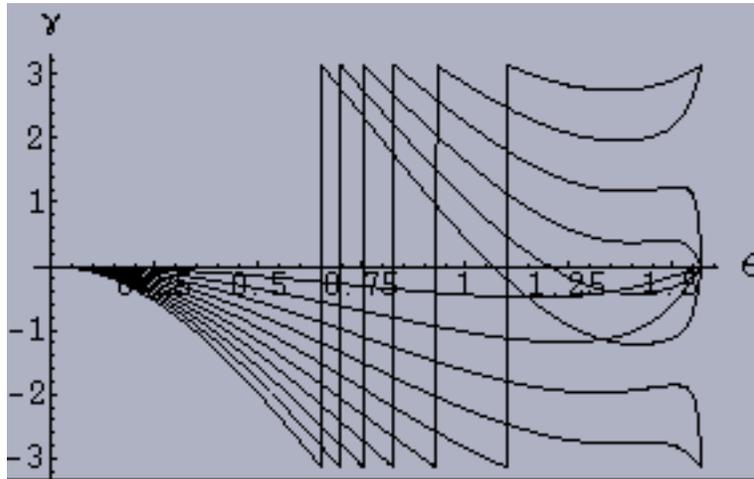}}
\caption{Uhlmann geometric phases for $n$-level Gibbsian systems $(n=2,\ldots,11)$ 
holding $\alpha =2$. Only below $\theta \approx .65$ does the curve 
for $n=2$ dominate that for $n=3$, which dominates that for $n=4,\ldots$}
\label{s4}
\end{figure}

\begin{figure}
\centerline{\psfig{figure=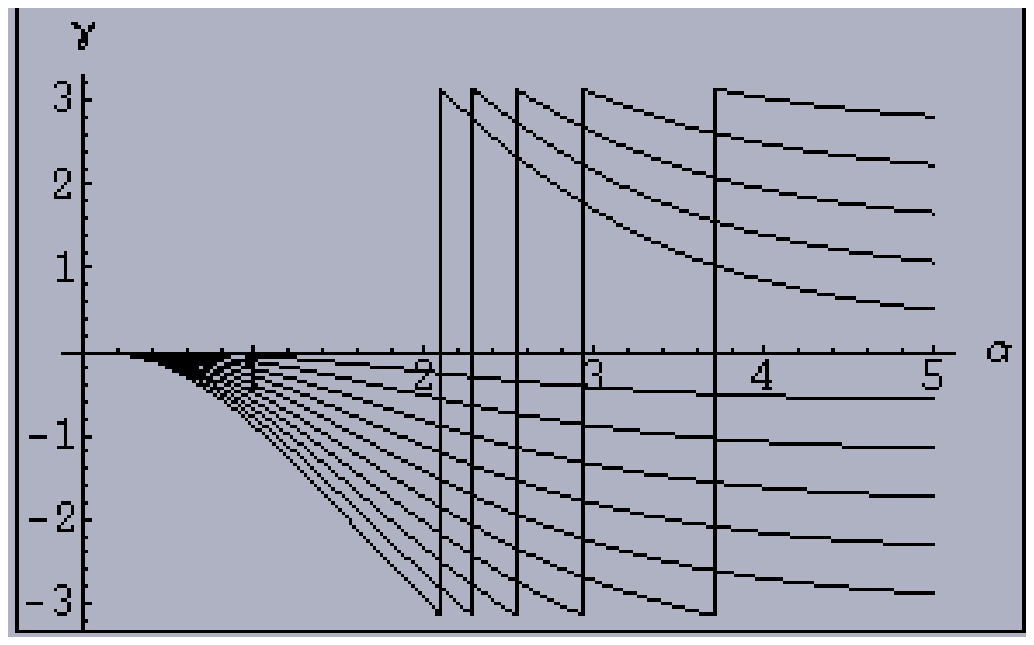}}
\caption{Uhlmann geometric phases for $n$-level Gibbsian systems ($n=2,\ldots,11$)
holding $\theta = {\pi \over 5}$. Only below $\alpha \approx 2$ does 
the curve for $n=2$ dominate that for $n=3$, which dominates that
for $n=4,\ldots$}
\label{s5}
\end{figure}

\begin{figure}
\centerline{\psfig{figure=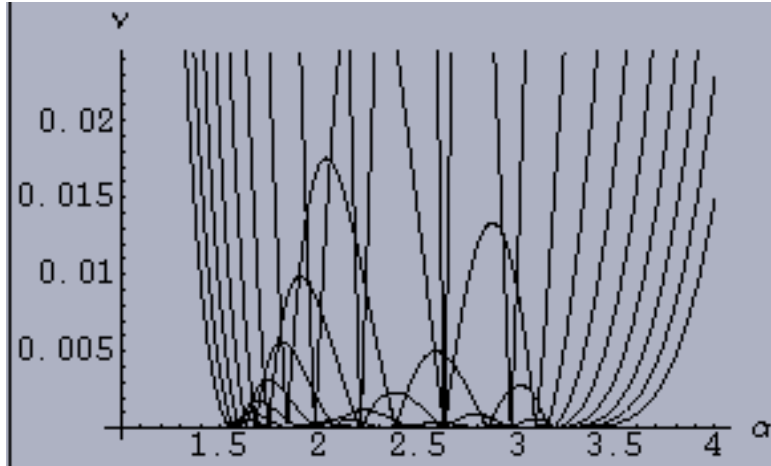}}
\caption{Uhlmann visibilities for $n$-level Gibbsian systems 
($n=2,\ldots,11$) fixing $\theta = {\pi \over 2}$. At $\alpha=2.1$,
the values of $\nu$ monotonically decline as $n$ increases. The highest 
peak belongs to $n=2$.}
\label{s6}
\end{figure}

In Fig.~\ref{s1}, we plot $\gamma_{j}$ ($j ={1 \over 2},\ldots , $)
versus $\theta$, holding $\alpha$ fixed at 1. Curves for lower-dimensional
Gibbsian systems 
here strictly dominate those for higher-dimensional systems.
(As $\alpha$ increases, however, above certain thresholds for
$\theta$, this simple monotonic behavior vanishes 
[cf. Fig.~\ref{s4}].)
In Fig.~\ref{s2}, we ``reverse'' this scenario, now 
holding $\theta$ fixed at
${\pi \over 10}$ and letting $\alpha$ vary over [0,5].
The monotonicity of the ten curves is completely analogous to that in
Fig.~\ref{s1}, with curves for lower $j$ dominating those for higher $j$.
In Fig.~\ref{s3}, we hold $\alpha$ at 2 and plot the {\it visibilities}
for the ten spin scenarios.
Again, as in Figs.~\ref{s1} and \ref{s2}, curves for lower values of $j$
dominate those for higher values.
While in Fig.~\ref{s1} the inverse temperature parameter 
$\alpha$ was fixed at 1, in Fig.~\ref{s4}, it is
held at 2. Now the same simple monotonicity with $j$  observed in
the preceding three figures, holds below
$\theta \approx .65$. However, 
it is lost at higher values of $\theta$, with the curves for the 
higher spin states oscillating from -${\pi \over 2}$ to ${\pi \over 2}$ at 
lower values of
$\theta$ than do the curves for some of the lower spin
states.
In Fig.~\ref{s5}, in which $\theta$ is fixed at ${\pi \over 5}$, similar
behavior to that observed in Fig.~\ref{s4} takes place. 
Below $\alpha \approx 1.43$, the simple monotonicity with $j$ holds, then
the curves begin to jump from -${\pi \over 2}$ to ${\pi \over 2}$ 
with the curves for higher
$j$ jumping first.
In Fig.~\ref{s6}, holding $\theta = {\pi \over 2}$, we plot the various
visibilities over the range $\alpha \in [1,4]$. At $\alpha=2.2$,
the values of $\nu_{j}$ monotonically decline from $j={1 \over 2}$ to
${9 \over 2}$.
\section{Higher-order Uhlmann holonomy invariants} \label{r5}
\subsection{Traces of powers of Uhlmann holonomy invariants}
\begin{figure}
\centerline{\psfig{figure=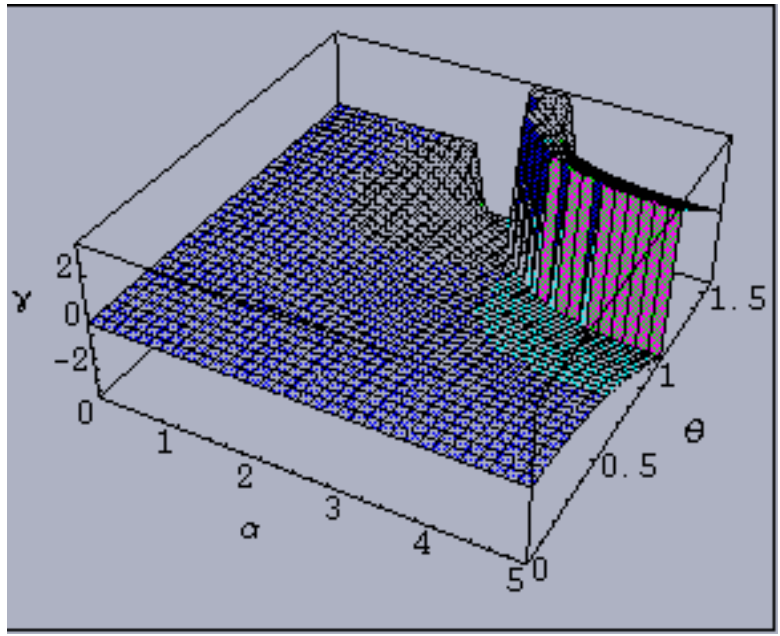}}
\caption{Argument of the trace of the {\it second} power of the holonomy
invariant (\ref{hi}) for the two-level Gibbsian systems ($n=2$)}
\label{ve1}
\end{figure}

\begin{figure}
\centerline{\psfig{figure=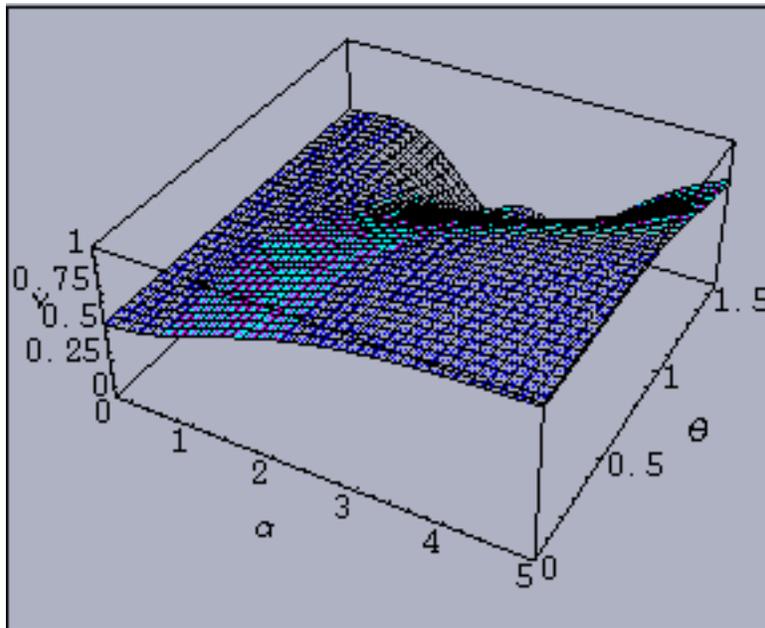}}
\caption{Absolute value of the trace of the {\it second} power of the
holonomy invariant (\ref{hi}) for the {\it two}-level Gibbsian systems
($n=2$)}
\label{ve2}
\end{figure}

\begin{figure}
\centerline{\psfig{figure=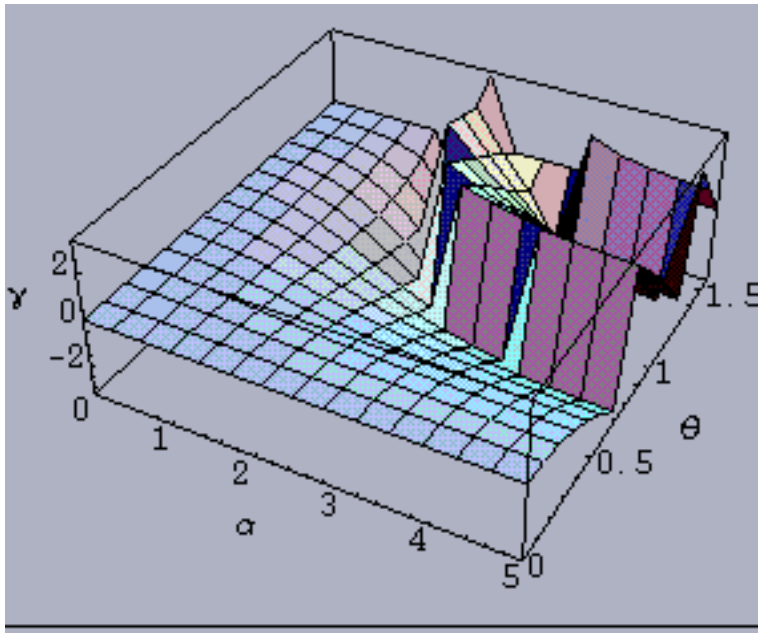}}
\caption{Argument of the trace of the {\it second} power of the holonomy
invariant (\ref{hi}) for the {\it three}-level Gibbsian systems
($n=3$)}
\label{ve3}
\end{figure}

\begin{figure}
\centerline{\psfig{figure=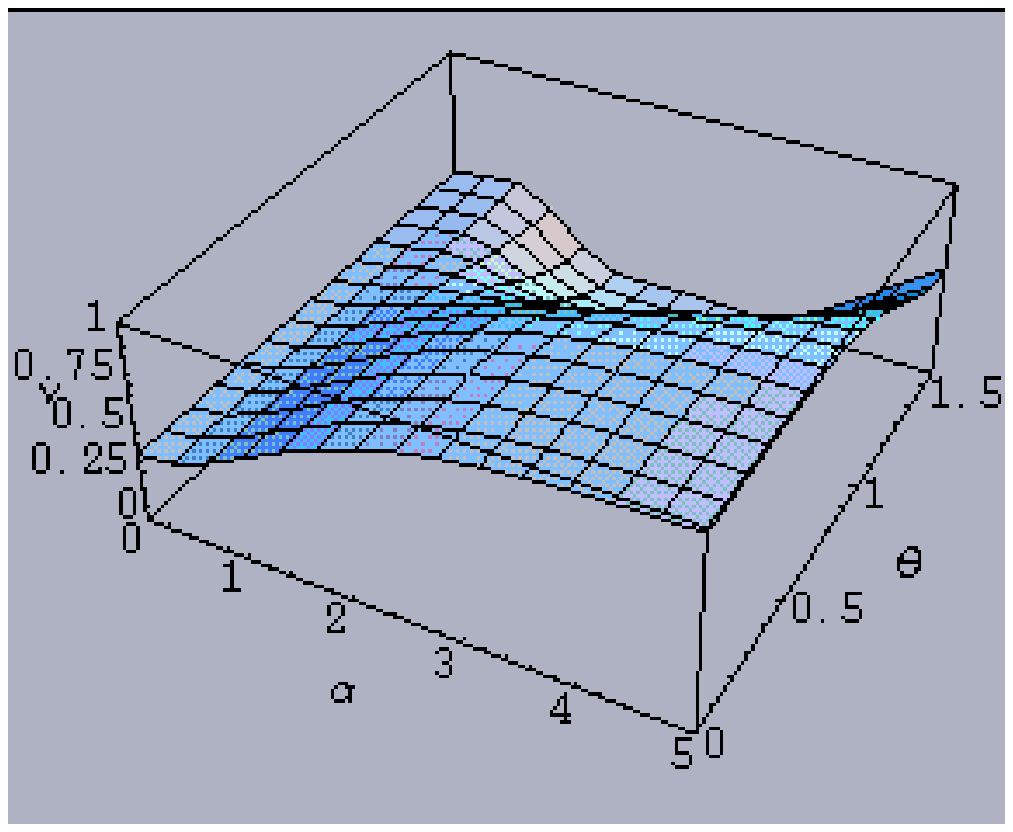}}
\caption{Absolute value of the trace of the {\it second} power of the
holonomy invariant (\ref{hi}) for the {\it three}-level Gibbsian 
systems ($n=3$)}
\label{ve4}
\end{figure}

All the analyses above have been conducted on the basis of the trace of
the holonomy invariant (\ref{hi}). In addition, however, the traces of 
the higher powers of (\ref{hi}) are also invariants \cite[eq. (102)]{dittuhl}
(all reducing, however, for pure states, to simple powers of the 
``Berry phase'' \cite[eq. (112.5)]{dittuhl}) (cf. \cite{jackiw}).
In Fig.~\ref{ve1}, we show the argument of the 
{\it second} power of the holonomy
invariant for the spin-${1 \over 2}$ case ($n=2$), which we express in the
form of
the arctangent of ${y \over x}$, taking into account
which quadrant the point ($x,y)$ lies in, where
\begin{equation}
x ={1 \over 2 \zeta} \Big( \zeta (-1 + (1 +\cosh{\alpha}) 
\cosh{2 i \pi \kappa}) \mbox{sech}^{2}{\alpha \over 2} 
-8 \sinh^{2}{i \pi \kappa} \Big), \quad y = -
{2 i \kappa \cos{\theta} \sinh{\alpha} \sinh{2 i \pi \kappa} \over \zeta}.
\end{equation}
The corresponding absolute value (Fig.~\ref{ve2}) is the square root of
\begin{equation}
x^2+y^2 = 
1 - {2 \cosh{2 i \pi \kappa} \over 1+\cosh{\alpha}} + {1 \over 4}
\mbox{sech}^{4}{\alpha \over 2} + {4 (2 +\cosh{\alpha}) 
\mbox{sech}^{2}{\alpha 
\over 2} \sinh^{2}{i \pi \kappa} \over \zeta} + {16 \sinh^{4}{i \pi \kappa} 
\over \zeta^{2}}.
\end{equation}
(In some of the later 
[computationally-intensive] figures in this section,
it proved essentially necessary to omit
the [otherwise obvious] axes labels, due to some 
quite distinct peculiarities of the graphics
facilities for the local installation of MATHEMATICA on the more 
powerful workstations.)
In Figs.~\ref{ve3} and \ref{ve4} are shown the analogous quantities
for Gibbsian spin-1 systems.
\begin{figure}
\centerline{\psfig{figure=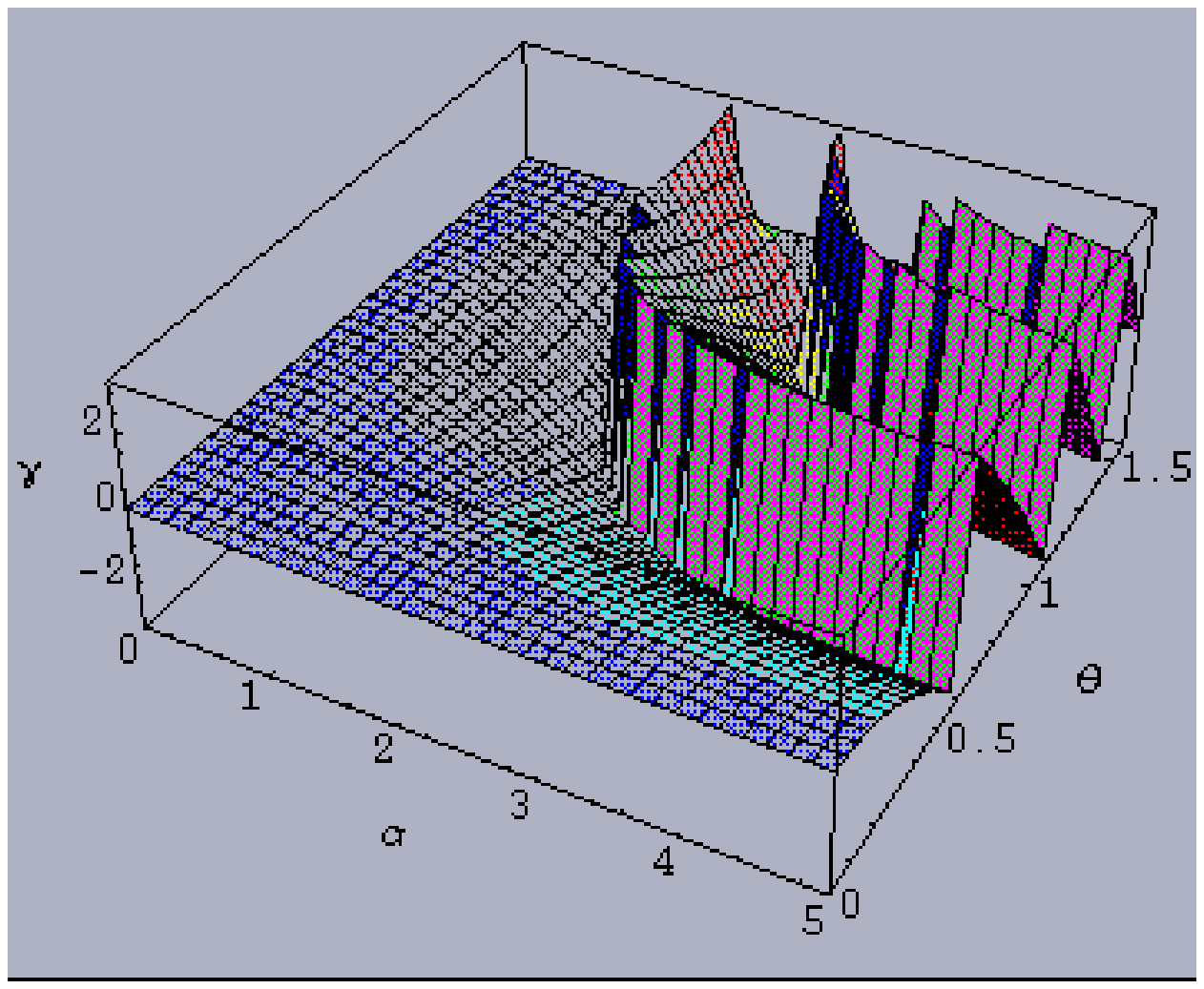}}
\caption{Argument of the trace of the {\it third} power of the holonomy
invariant (\ref{hi}) for the {\it three}-level Gibbsian systems ($n=3$)}
\label{ve5}
\end{figure}
\begin{figure}
\centerline{\psfig{figure=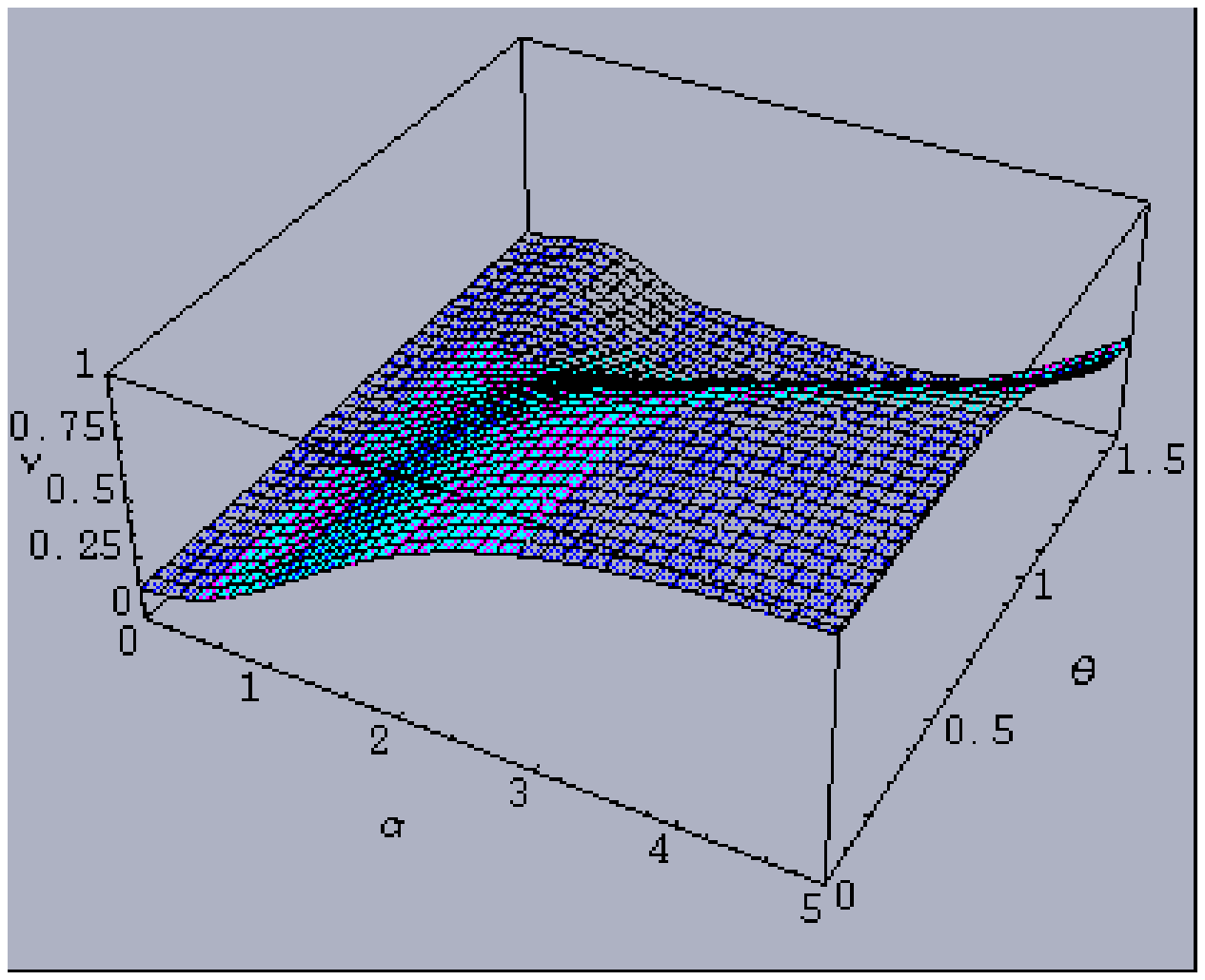}}
\caption{Absolute value of  the trace of the {\it third} power of the 
holonomy invariant (\ref{hi})  for the {\it three}-level Gibbsian
systems ($n=3$)}
\label{ve6}
\end{figure}
\begin{figure}
\centerline{\psfig{figure=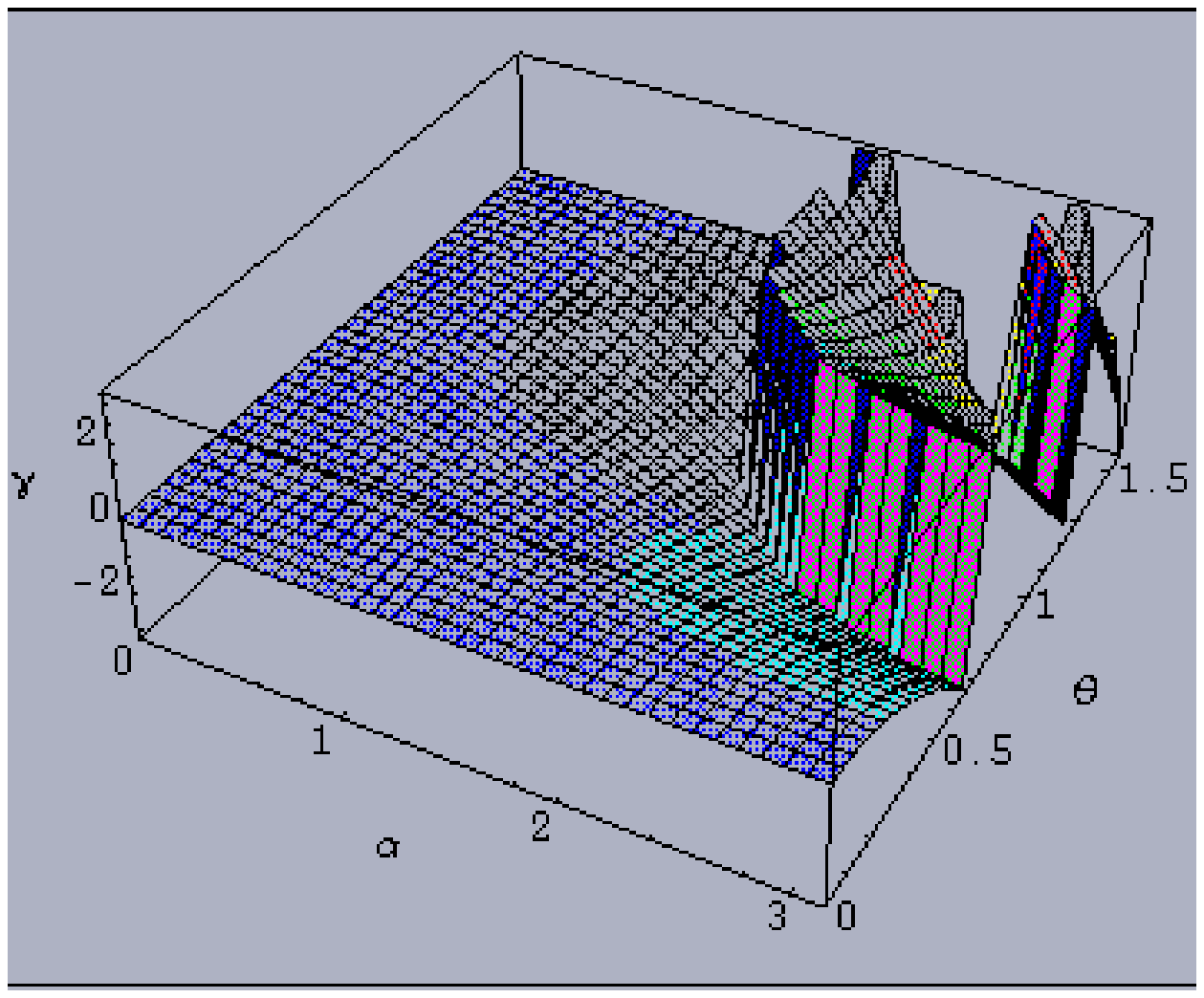}}
\caption{Argument of the trace of the {\it second} power of the holonomy
invariant (\ref{hi}) for the {\it four}-level Gibbsian systems ($n=4$)}
\label{ve7}
\end{figure}
\begin{figure}
\centerline{\psfig{figure=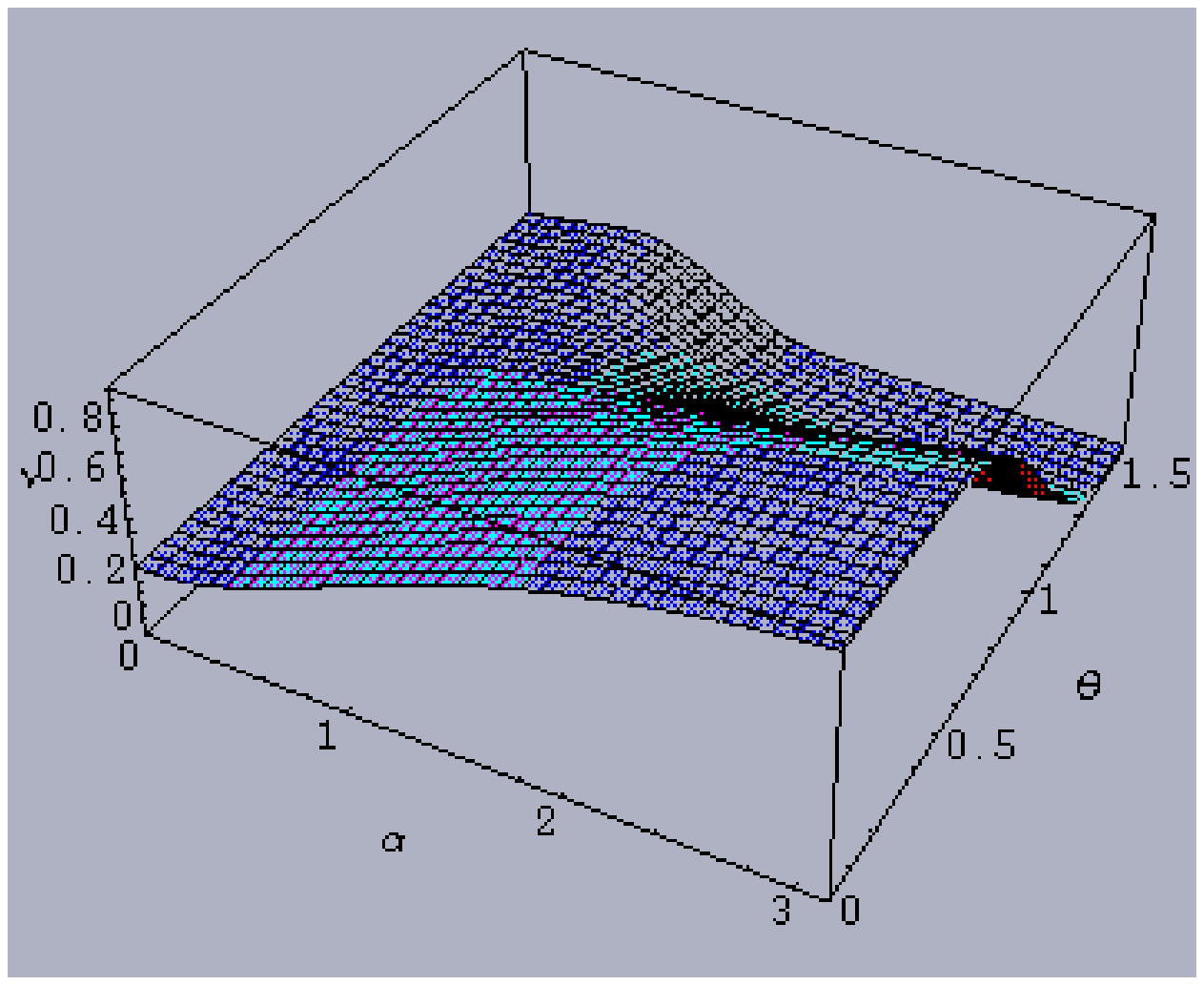}}
\caption{Absolute value of the trace of the {\it second} power of the
holonomy invariant (\ref{hi}) for the {\it four}-level Gibbsian systems
($n=4$)}
\label{ve8}
\end{figure}
\begin{figure}
\centerline{\psfig{figure=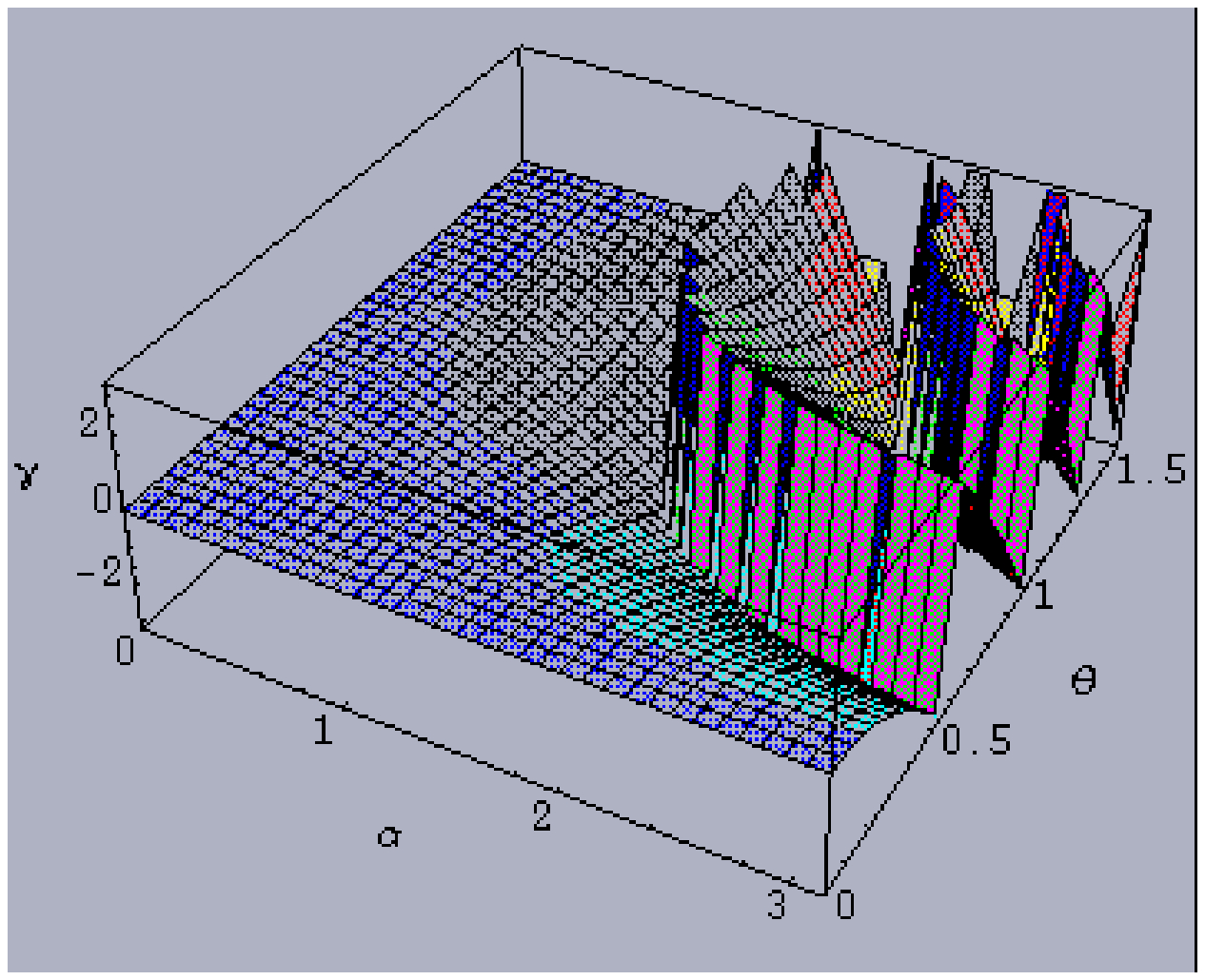}}
\caption{Argument of the trace of the {\it third} power of the holonomy
invariant (\ref{hi}) for the {\it four}-level Gibbsian systems ($n=4$)}
\label{ve9}
\end{figure}
\begin{figure}
\centerline{\psfig{figure=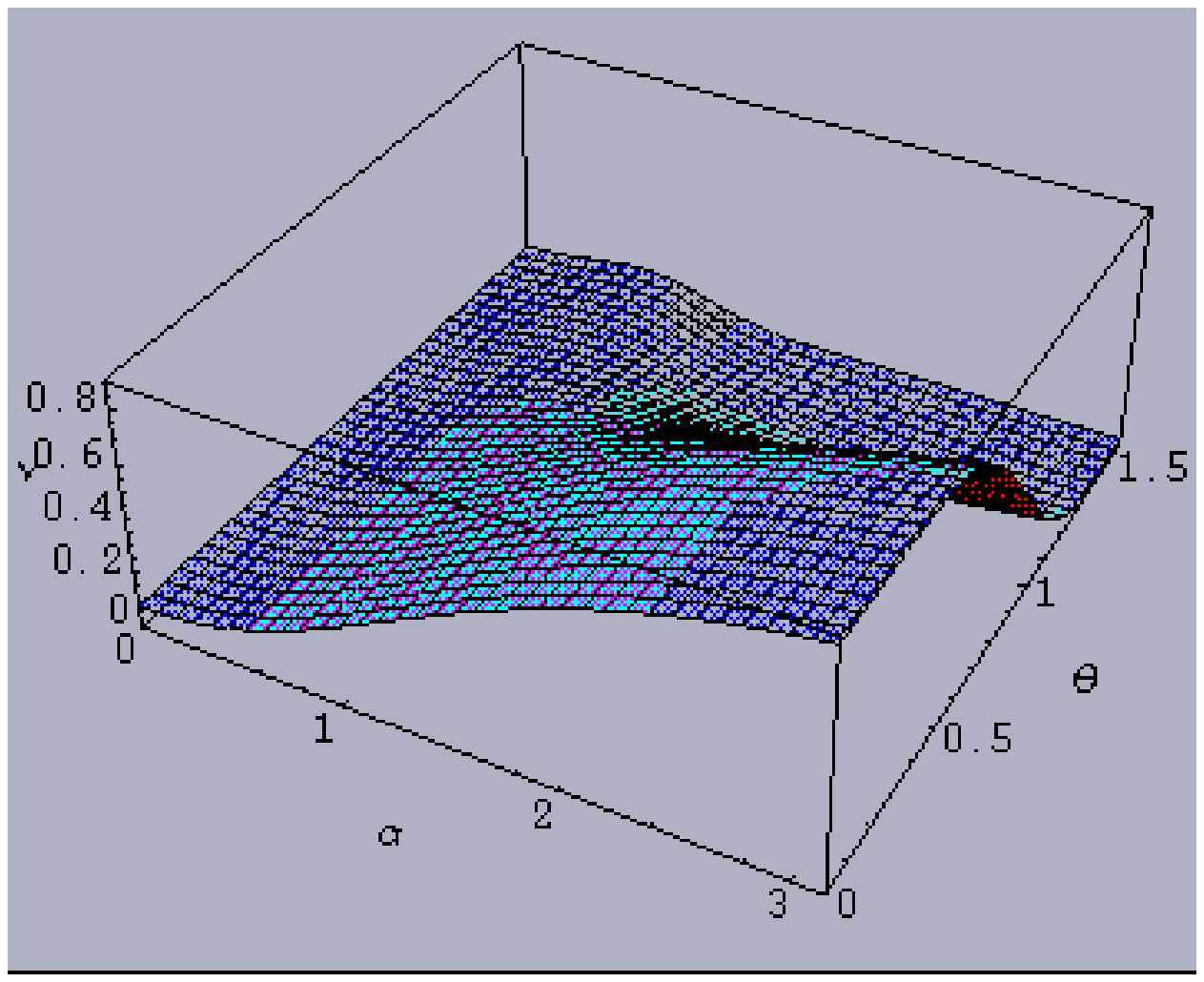}}
\caption{Absolute value of the trace of the {\it third} power of the holonomy 
invariant (\ref{hi}) for the {\it four}-level Gibbsian systems
($n=4$)}
\label{ve10}
\end{figure}
\begin{figure}
\centerline{\psfig{figure=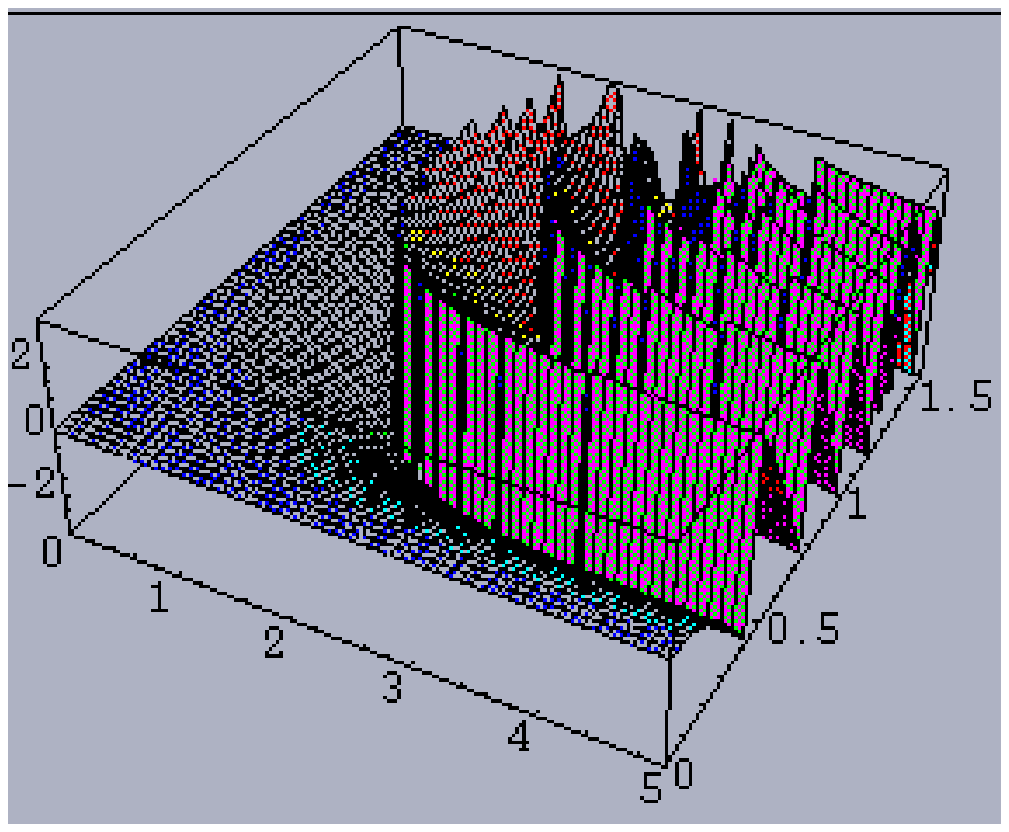}}
\caption{Argument of the trace of the {\it fourth} power of the holonomy 
invariant (\ref{hi}) for the {\it four}-level Gibbsian systems 
($n=4$)}
\label{ve11}
\end{figure}
\begin{figure}
\centerline{\psfig{figure=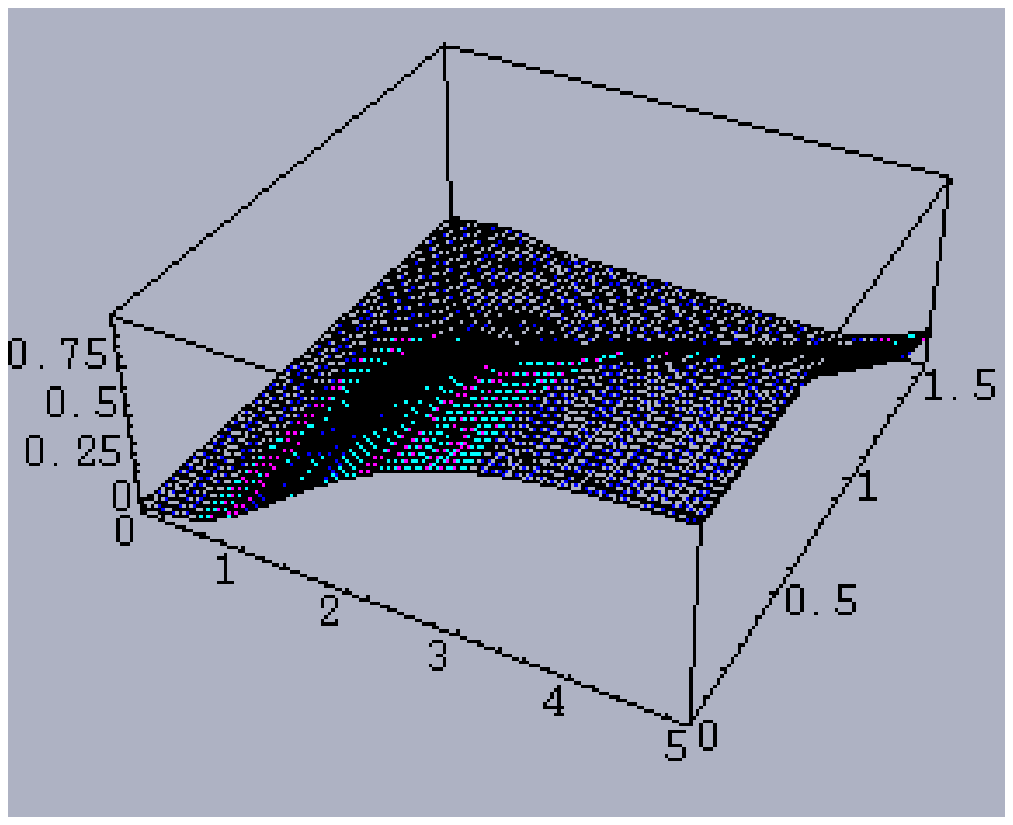}}
\caption{Absolute value of the trace of the {\it fourth} power of the
holonomy invariant (\ref{hi}) for the {\it four}-level Gibbsian systems
($n=4$)}
\label{ve12}
\end{figure}
\begin{figure}
\centerline{\psfig{figure=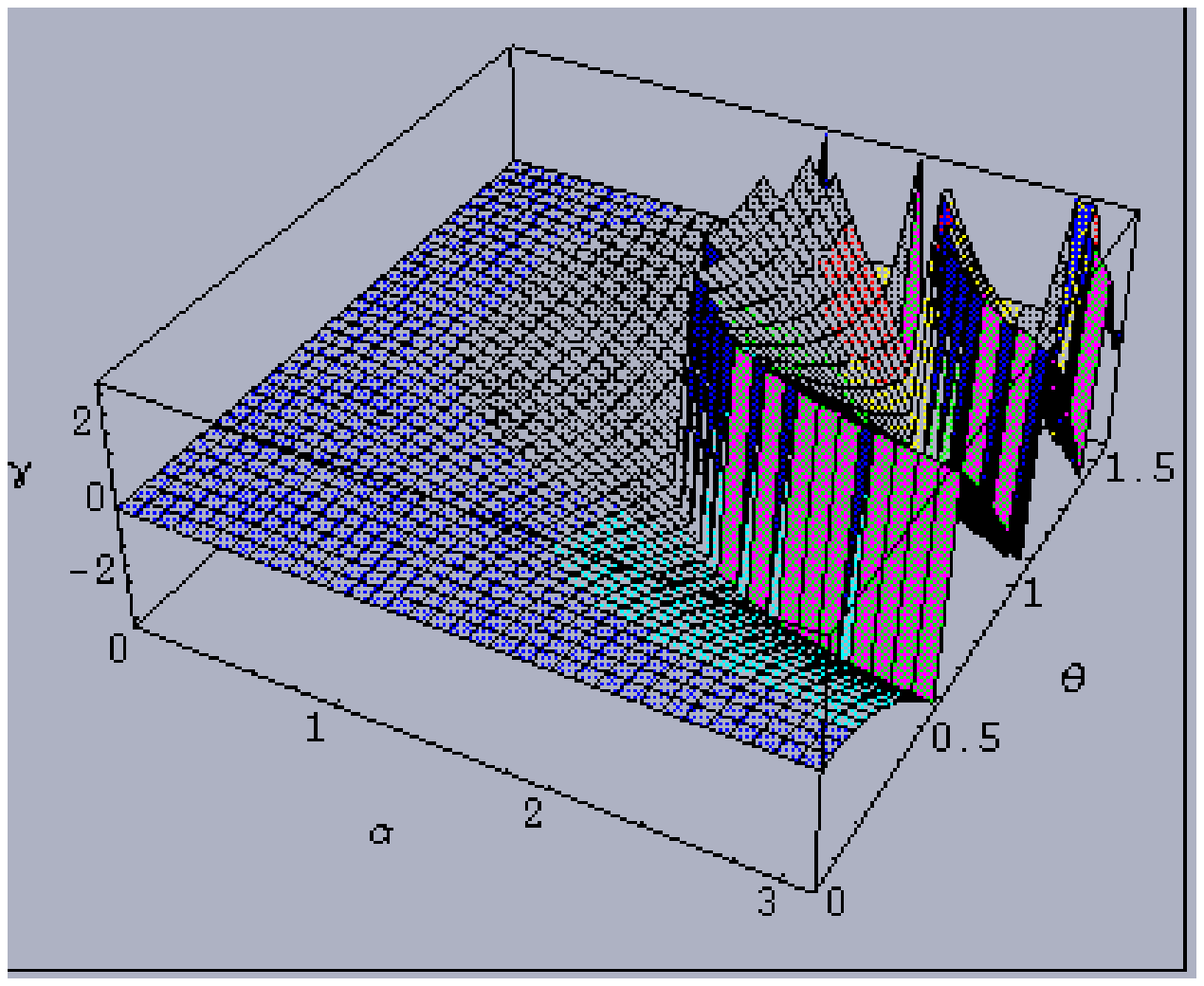}}
\caption{Argument of the trace of the {\it second} power of the holonomy
invariant (\ref{hi}) for the {\it five}-level Gibbsian systems ($n=5$)}
\label{ve13}
\end{figure}
\begin{figure}
\centerline{\psfig{figure=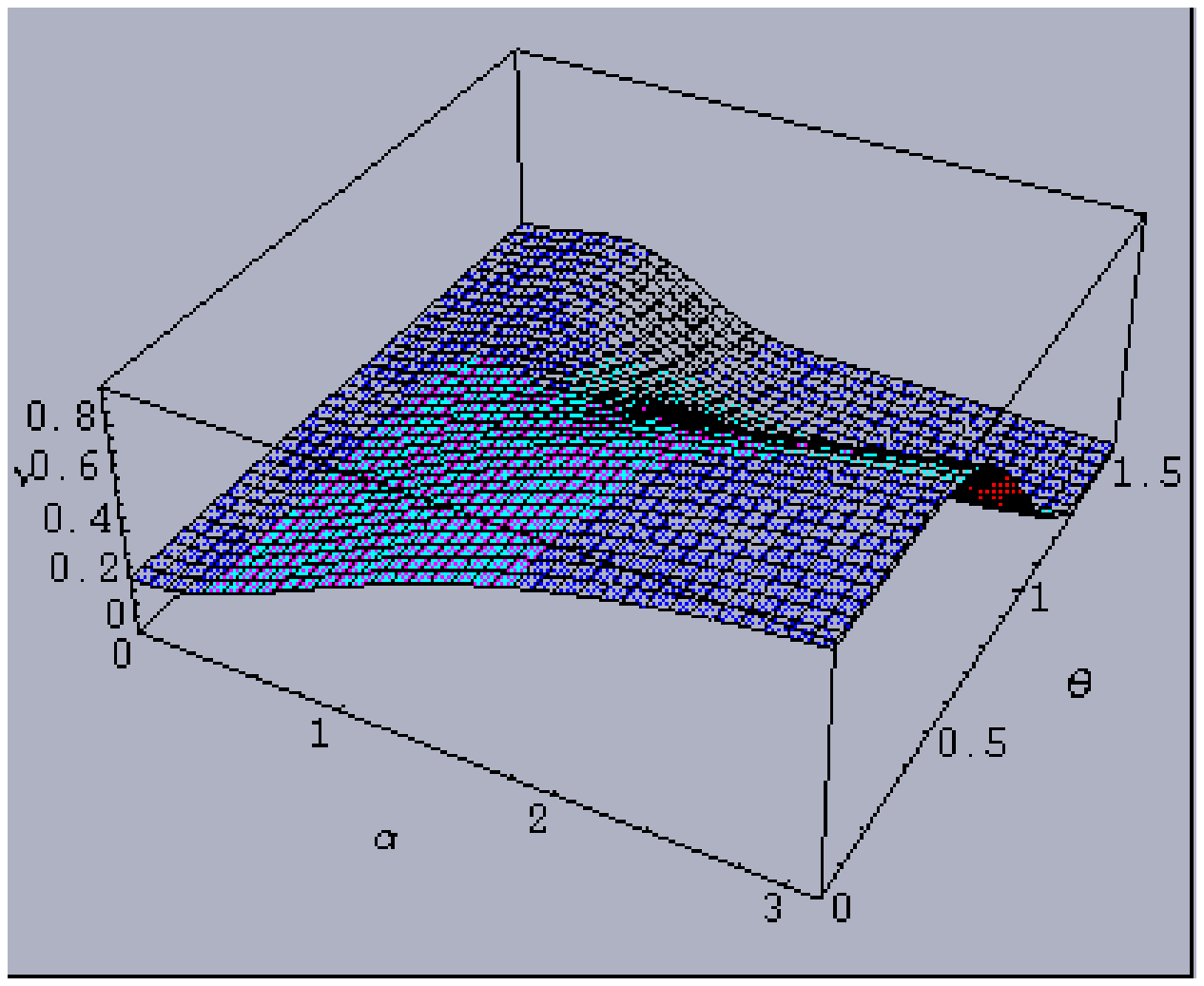}}
\caption{Absolute value of the trace of the {\it second} power of the holonomy 
invariant (\ref{hi}) for the {\it five}-level Gibbsian systems ($n=5$)}
\label{ve14}
\end{figure}
\begin{figure}
\centerline{\psfig{figure=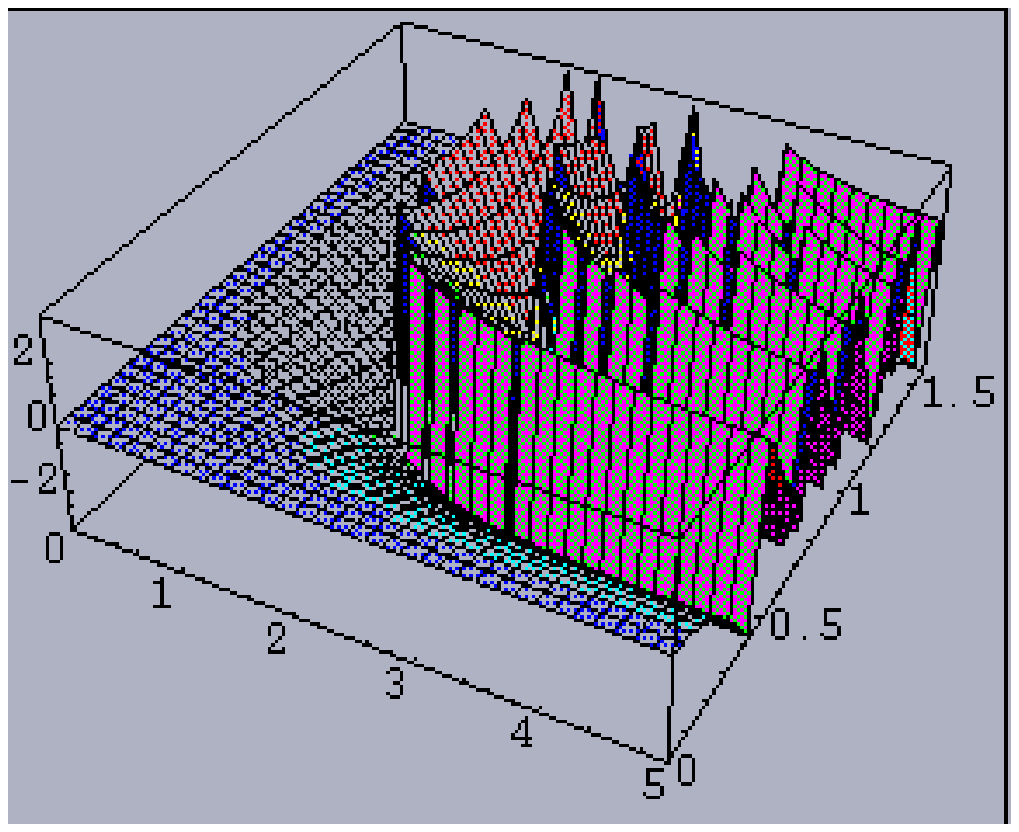}}
\caption{Argument of the trace of the {\it third} power of the holonomy
invariant (\ref{hi}) for the {\it five}-level Gibbsian systems ($n=5$)}
\label{ve15}
\end{figure}
\begin{figure}
\centerline{\psfig{figure=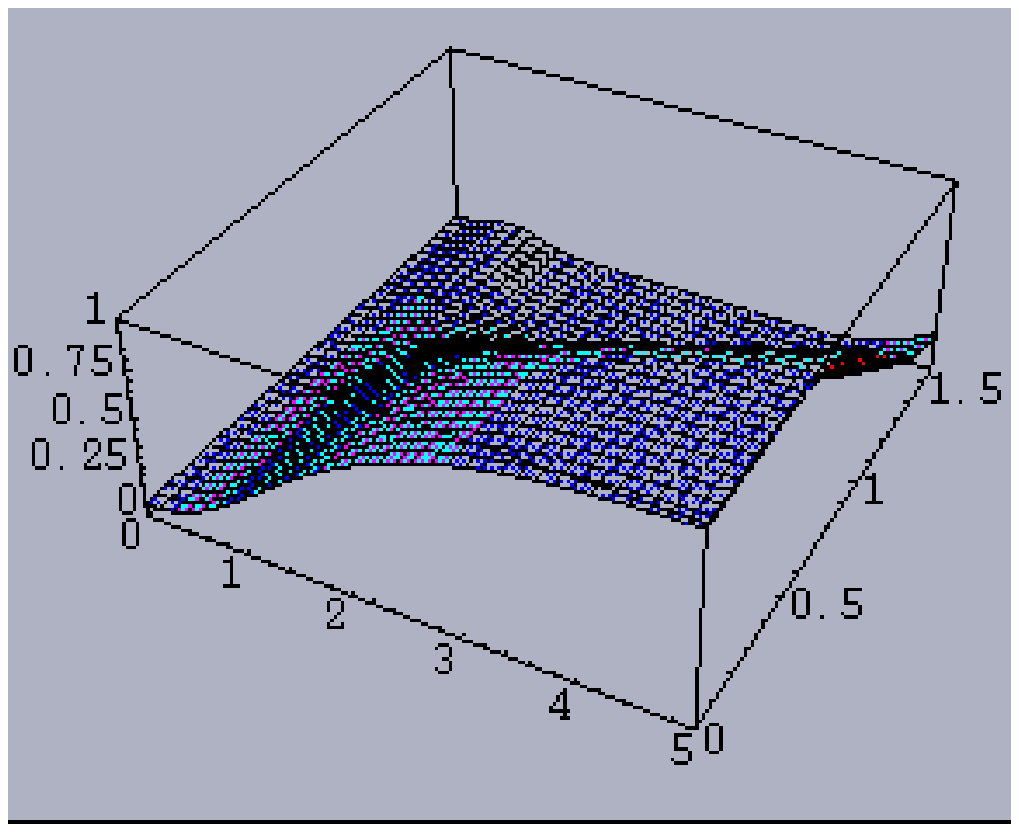}}
\caption{Absolute value of the trace of the {\it third} power of the holonomy
invariant (\ref{hi}) for the {\it five}-level Gibbsian systems ($n=5$)}
\label{ve16}
\end{figure}
\begin{figure}
\centerline{\psfig{figure=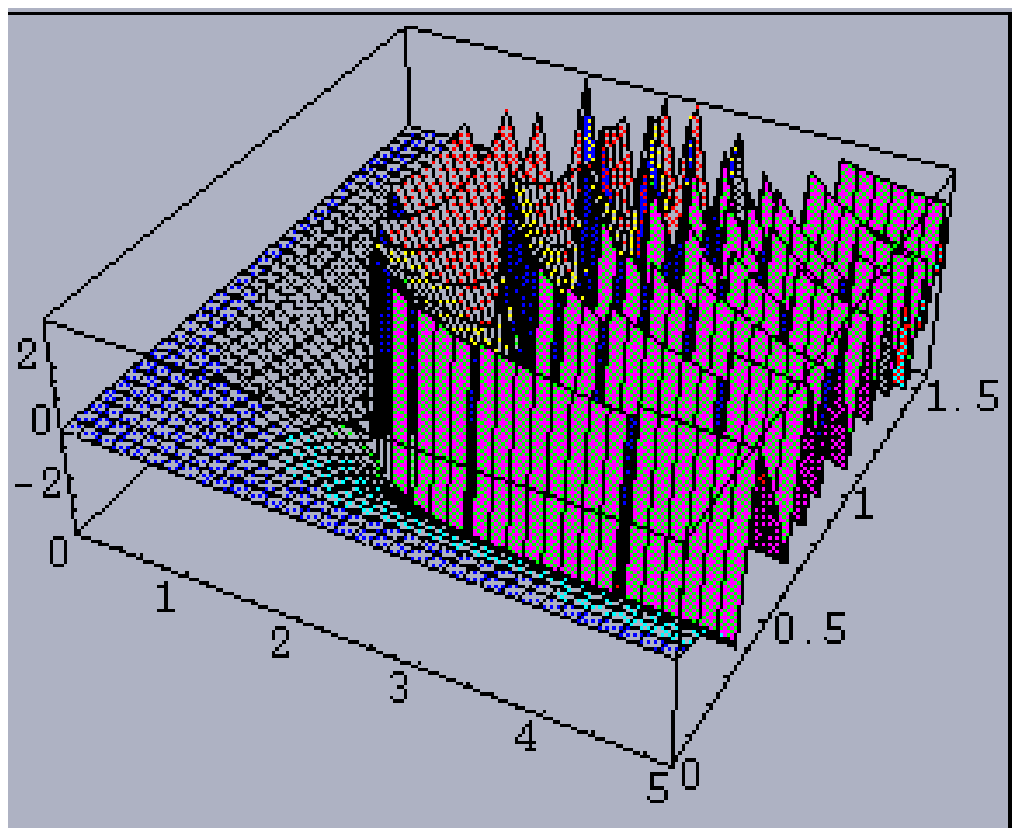}}
\caption{Argument of the trace of the {\it fourth} power of the holonomy 
invariant (\ref{hi}) for the {\it five}-level Gibbsian systems
($n=5$)}
\label{ve17}
\end{figure}
\begin{figure}
\centerline{\psfig{figure=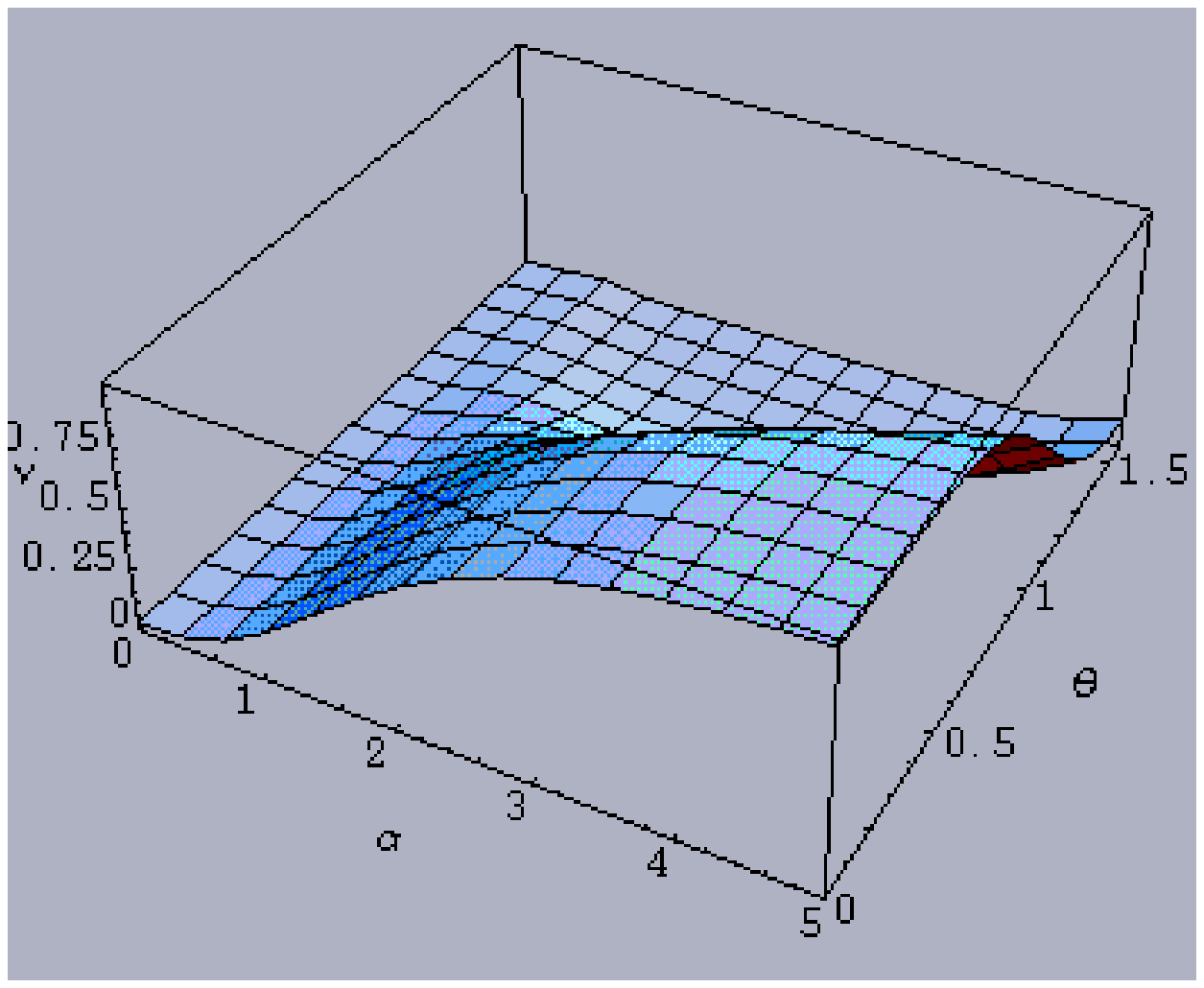}}
\caption{Absolute value of the trace of the {\it fourth} power of the holonomy
invariant (\ref{hi}) for the {\it five}-level Gibbsian systems 
($n=5$)}
\label{ve18}
\end{figure}
\begin{figure}
\centerline{\psfig{figure=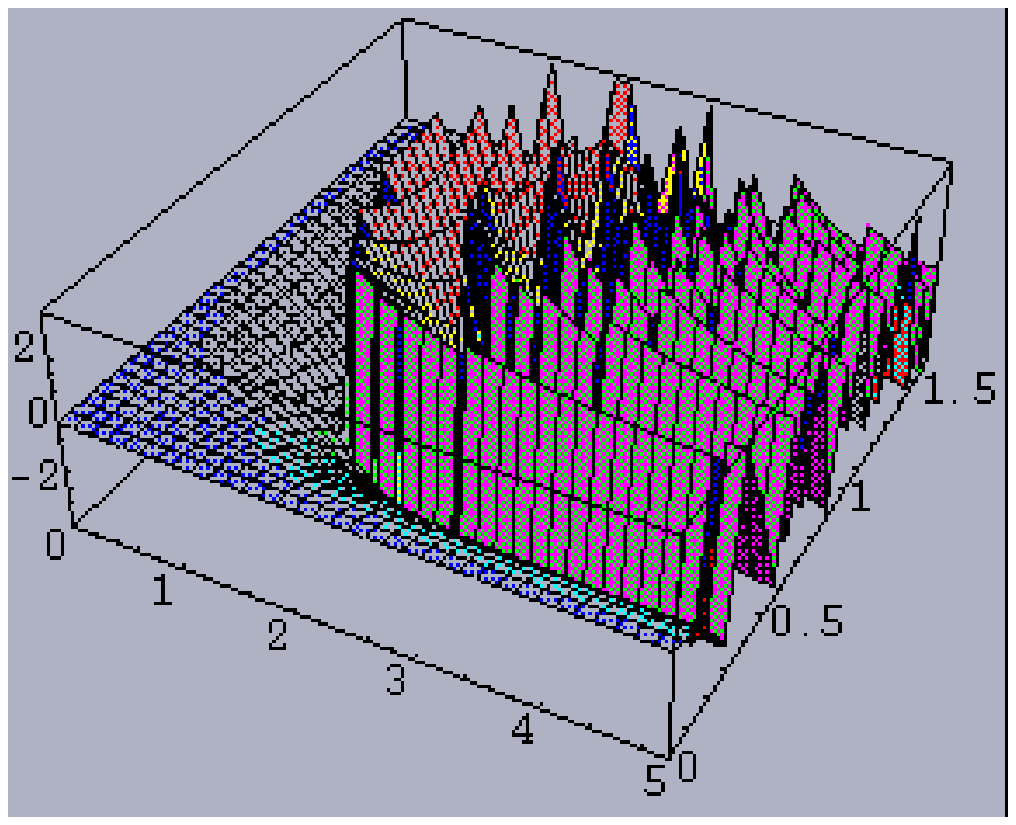}}
\caption{Argument of the trace of the {\it fifth} power of the holonomy
invariant (\ref{hi}) for the {\it five}-level Gibbsian systems ($n=5$)}
\label{ve19}
\end{figure}
\begin{figure}
\centerline{\psfig{figure=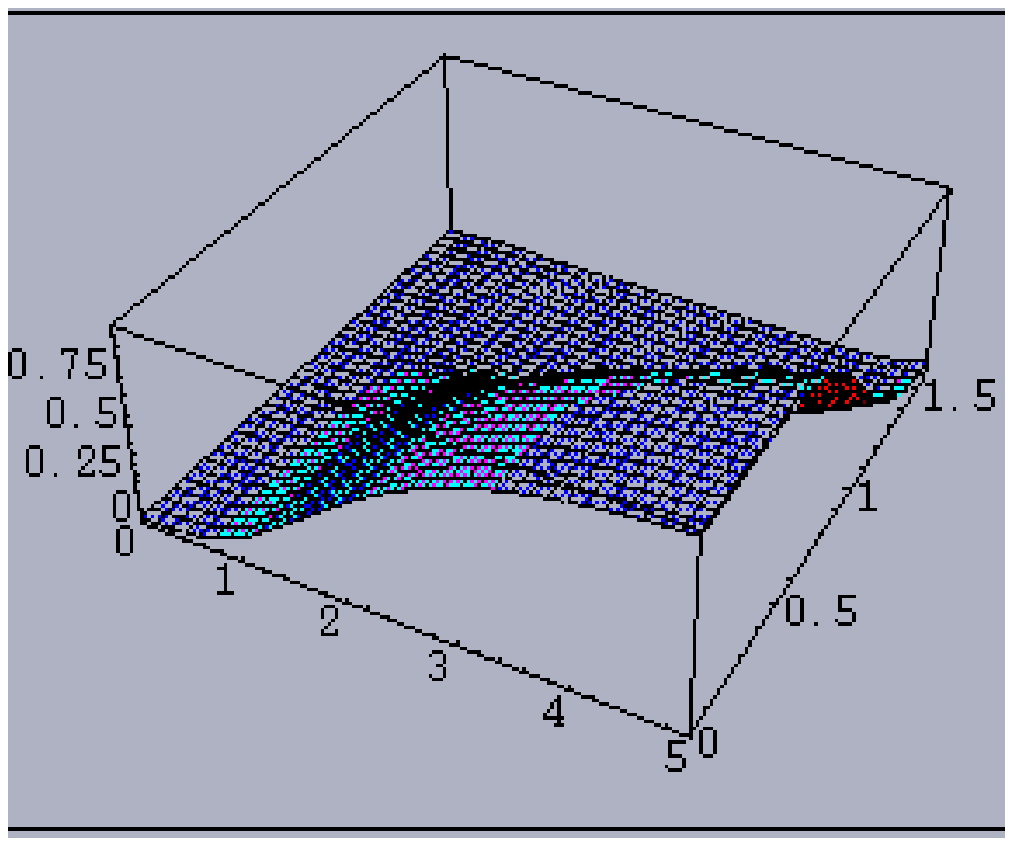}}
\caption{Absolute value of the trace of the {\it fifth} power of the 
holonomy invariant (\ref{hi}) for the {\it five}-level Gibbsian systems
($n=5$)}
\label{ve19b}
\end{figure}
\begin{figure}
\centerline{\psfig{figure=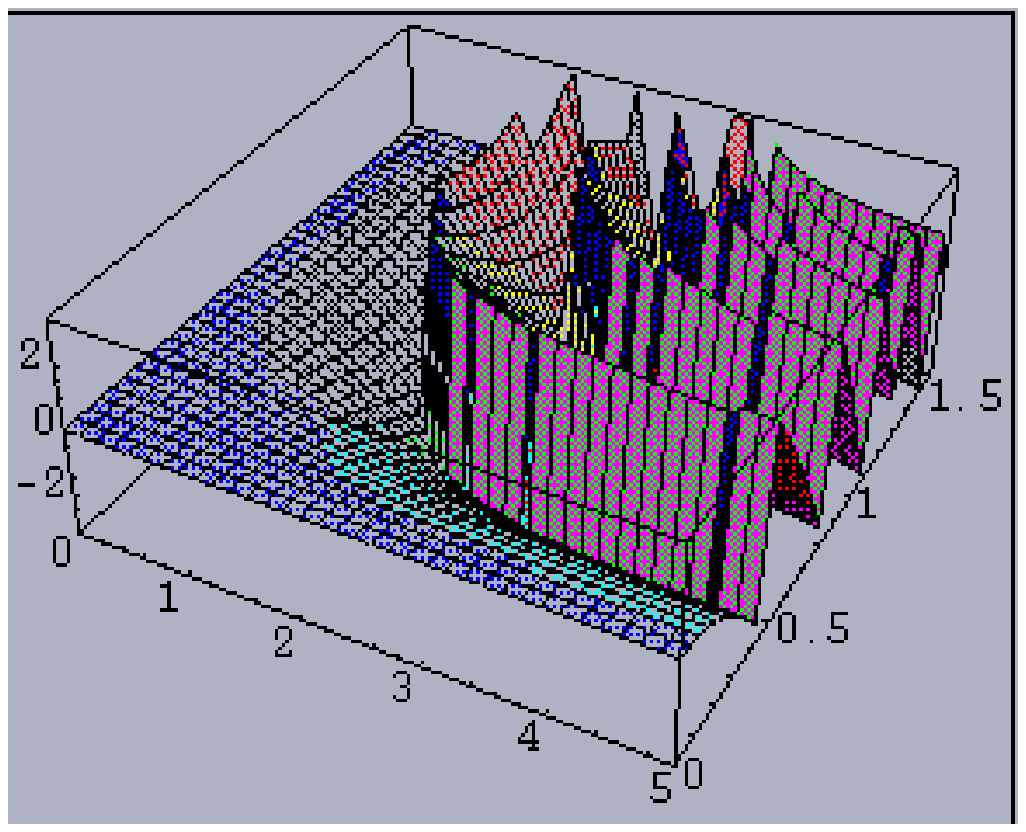}}
\caption{Argument of the trace of the {\it second} power of the holonomy
invariant (\ref{hi}) for the {\it six}-level Gibbsian systems ($n=6$)}
\label{ve20}
\end{figure}
\begin{figure}
\centerline{\psfig{figure=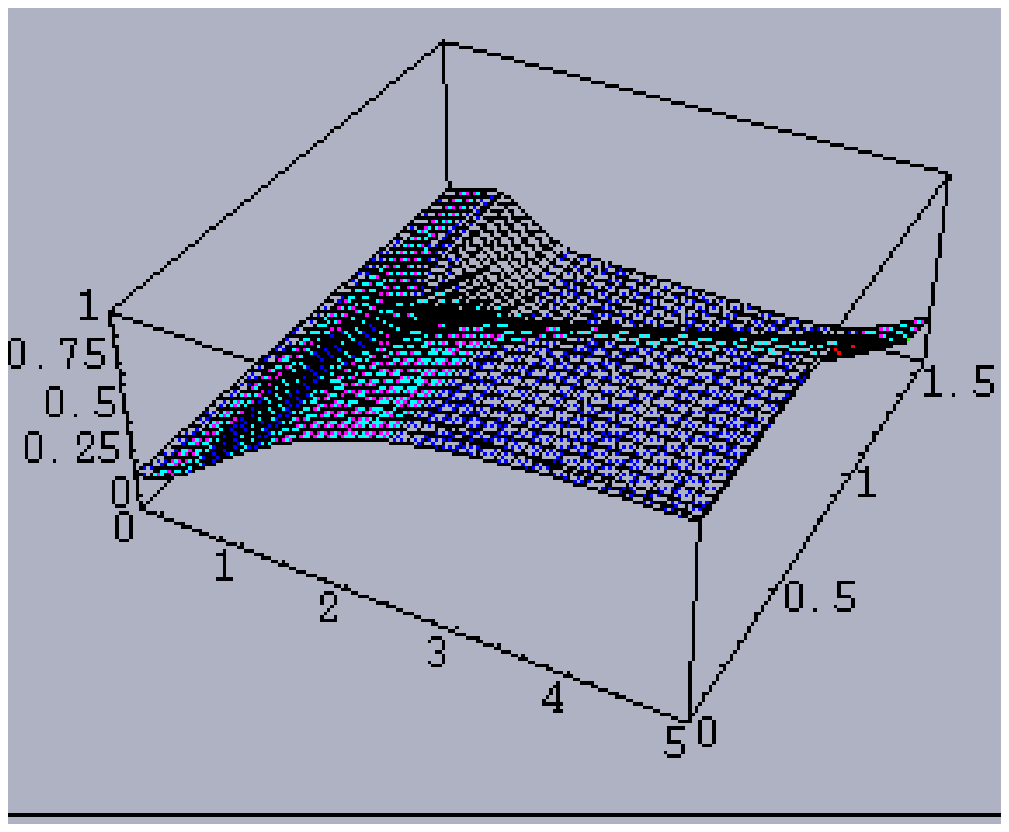}}
\caption{Absolute value of the trace of the {\it second} power of the holonomy
invariant (\ref{hi}) for the {\it six}-level Gibbsian systems ($n=6$)}
\label{ve21}
\end{figure}
\begin{figure}
\centerline{\psfig{figure=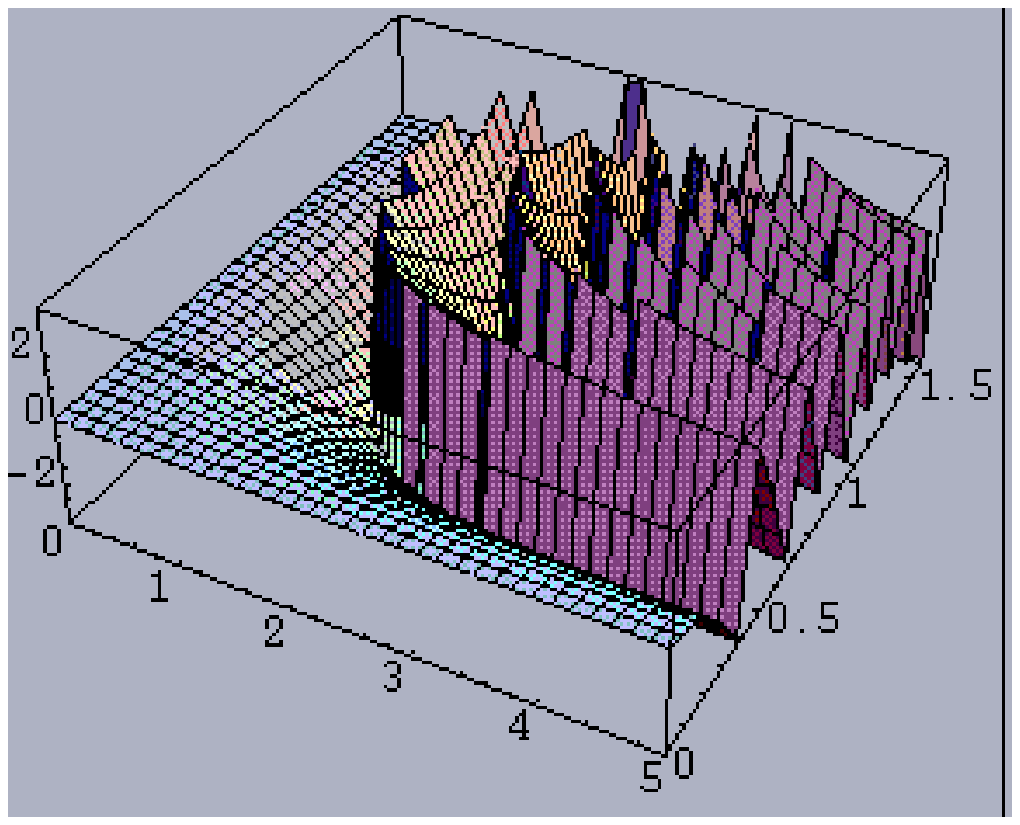}}
\caption{Argument of the trace of the {\it third} power of the holonomy
invariant (\ref{hi}) for the {\it six}-level Gibbsian systems ($n=6$)}
\label{ve22}
\end{figure}
\begin{figure}
\centerline{\psfig{figure=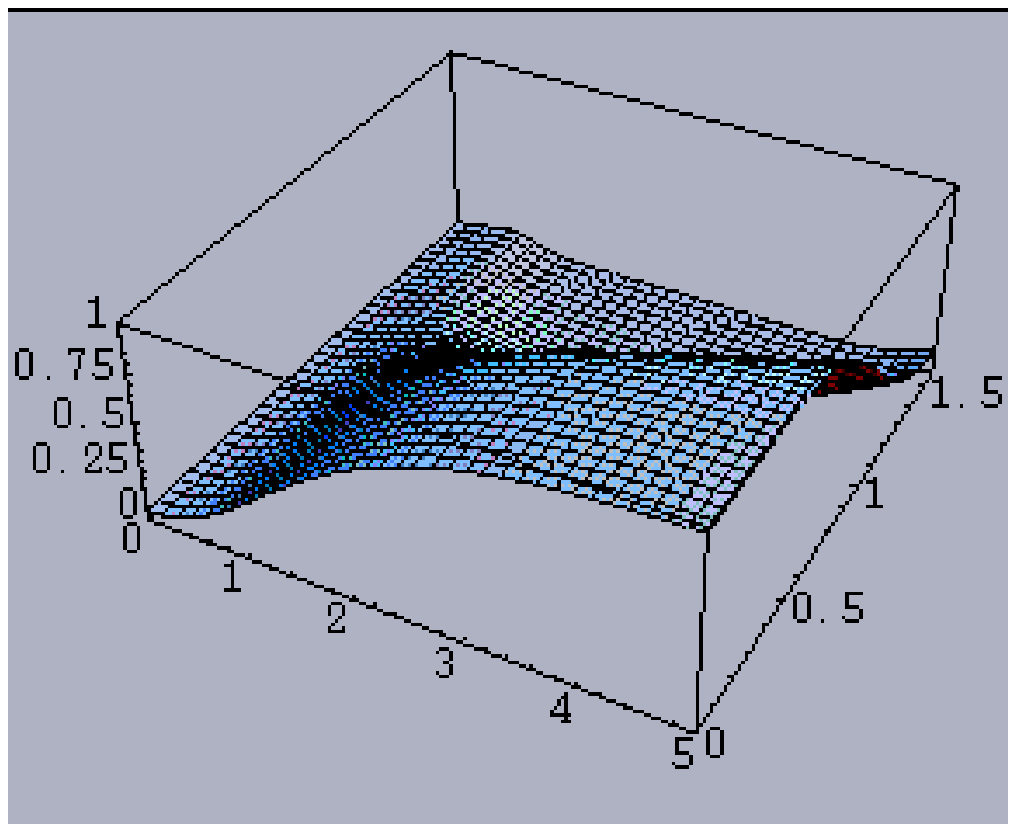}}
\caption{Absolute value of the trace of the {\it third} power of the
holonomy invariant (\ref{hi}) for the {\it six}-level Gibbsian systems
($n=6$)}
\label{ve25}
\end{figure}
\begin{figure}
\centerline{\psfig{figure=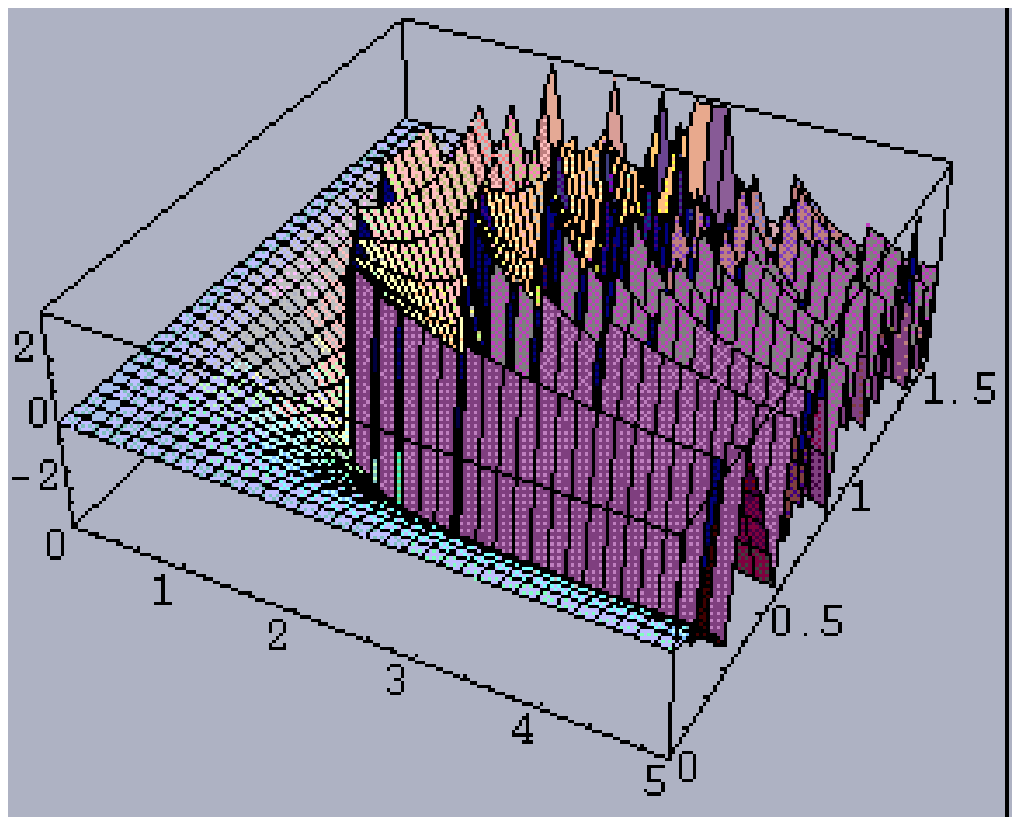}}
\caption{Argument of the trace of the {\it fourth} power of the holonomy
invariant (\ref{hi}) for the {\it six}-level Gibbsian systems ($n=6$)}
\label{ve26}
\end{figure}
\begin{figure}
\centerline{\psfig{figure=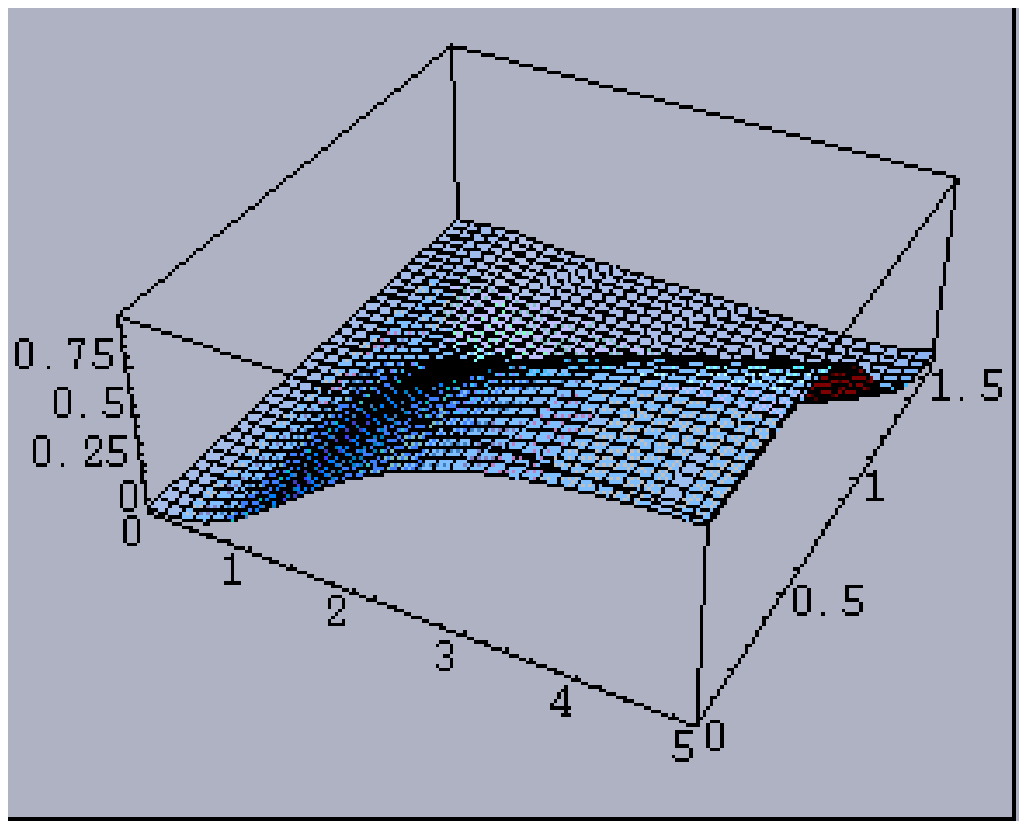}}
\caption{Absolute value of the trace of the {\it fourth} power of the
holonomy invariant (\ref{hi}) for the {\it six}-level Gibbsian systems
($n=6$)}
\label{ve27}
\end{figure}
\begin{figure}
\centerline{\psfig{figure=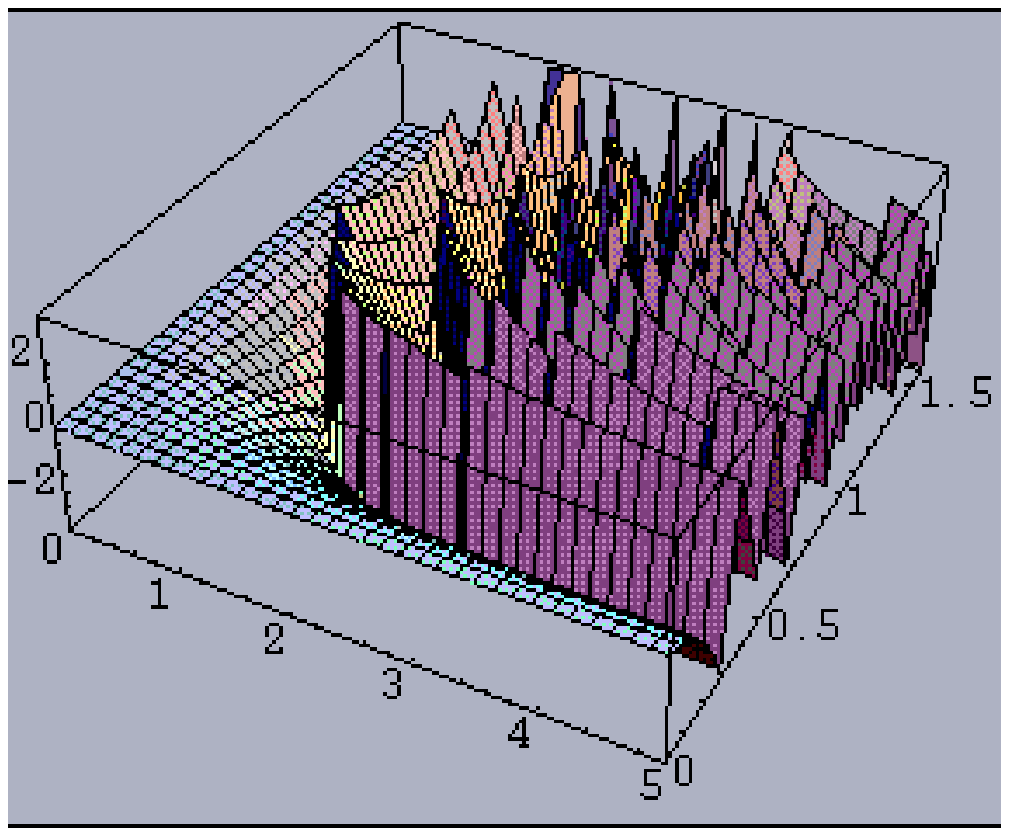}}
\caption{Argument of the trace of the {\it fifth} power of the holonomy
invariant (\ref{hi}) for the {\it six}-level Gibbsian systems ($n=6$)}
\label{ve28}
\end{figure}
\begin{figure}
\centerline{\psfig{figure=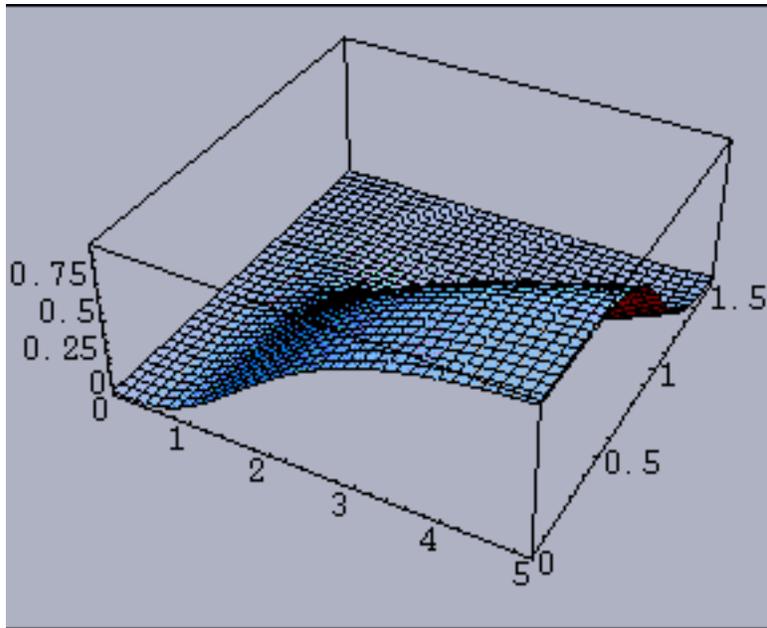}}
\caption{Absolute value of the trace of the {\it fifth} power of the
holonomy invariant (\ref{hi}) for the {\it six}-level Gibbsian systems
($n=6$)}
\label{ve29}
\end{figure}
\subsection{Cross-comparisons of Uhlmann holonomy invariants}
\begin{figure}
\centerline{\psfig{figure=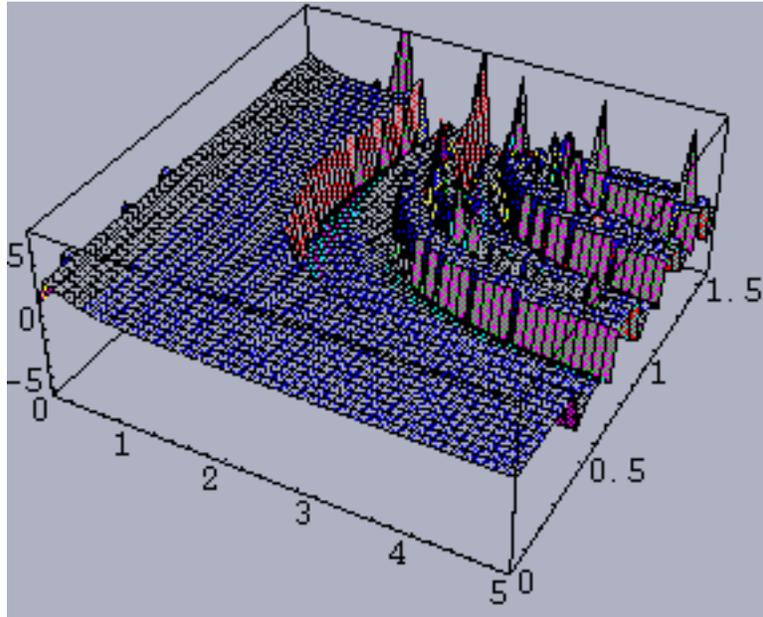}}
\caption{Ratio of the argument of the trace of the second power of the
holonomy invariant (\ref{hi}) to the argument of the 
trace of the first power for
the {\it eight}-level Gibbsian systems}
\label{ve30}
\end{figure}
\begin{figure}
\centerline{\psfig{figure=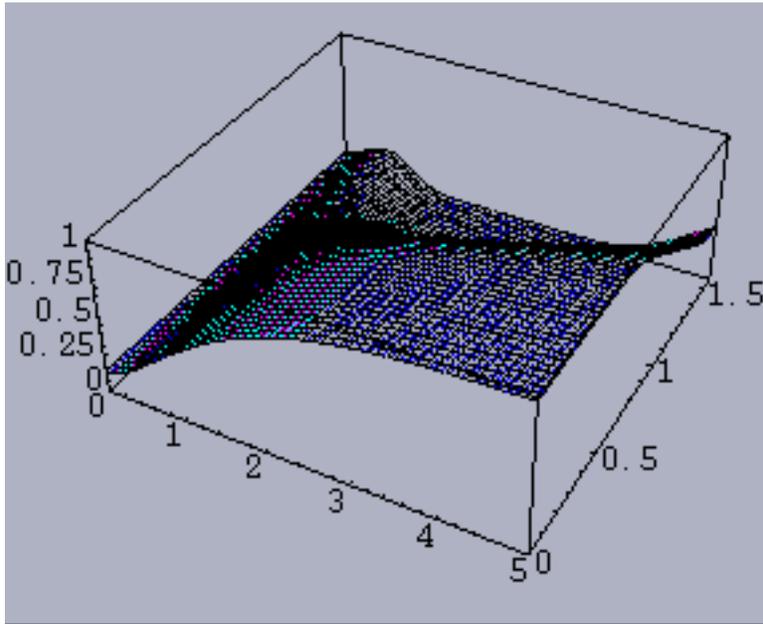}}
\caption{Ratio of the absolute value of the trace of the second power of the
holonomy invariant (\ref{hi}) to the absolute value of the 
trace of the first power for
the {\it eight}-level Gibbsian systems}
\label{ve31}
\end{figure}
\begin{figure}
\centerline{\psfig{figure=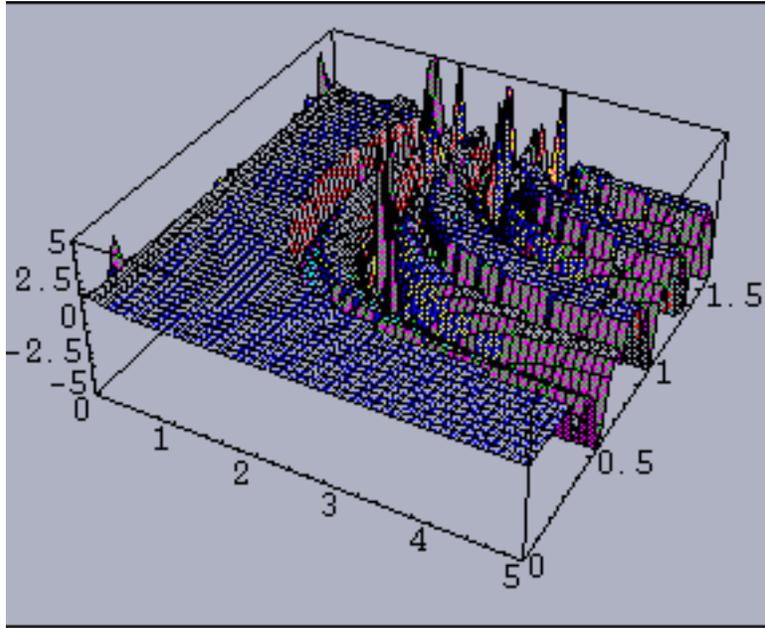}}
\caption{Ratio of the argument of the trace of the third power of the
holonomy invariant (\ref{hi}) to the argument of the trace of the second
power for the {\it eight}-level Gibbsian systems}
\label{ve32}
\end{figure}
\begin{figure}
\centerline{\psfig{figure=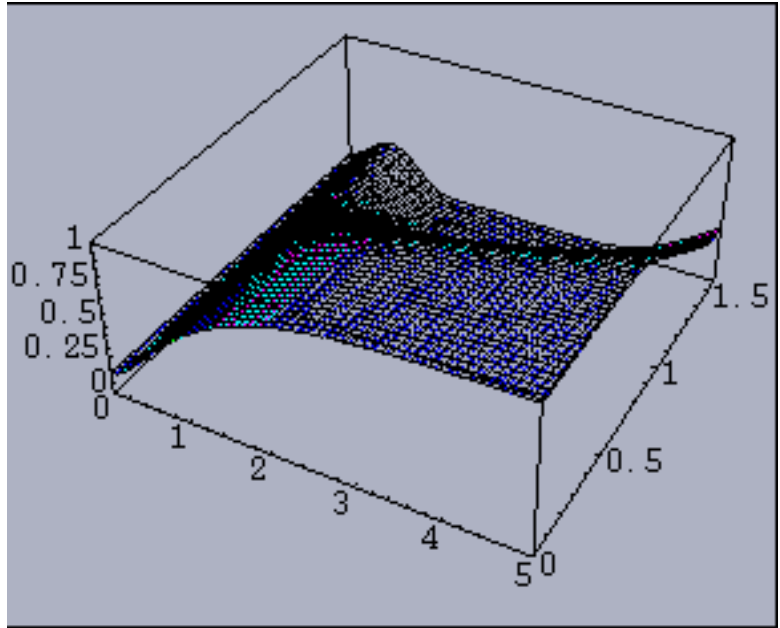}}
\caption{Ratio of the absolute value of the trace of the third power
of the holonomy invariant (\ref{hi}) to the absolute value of the trace of
the second power for the {\it eight}-level Gibbsian systems}
\label{ve33}
\end{figure}
\subsection{Eigenvalues of Uhlmann holonomy invariant for $n=2, 3, 5$} 
\label{reigen}
\begin{figure}
\centerline{\psfig{figure=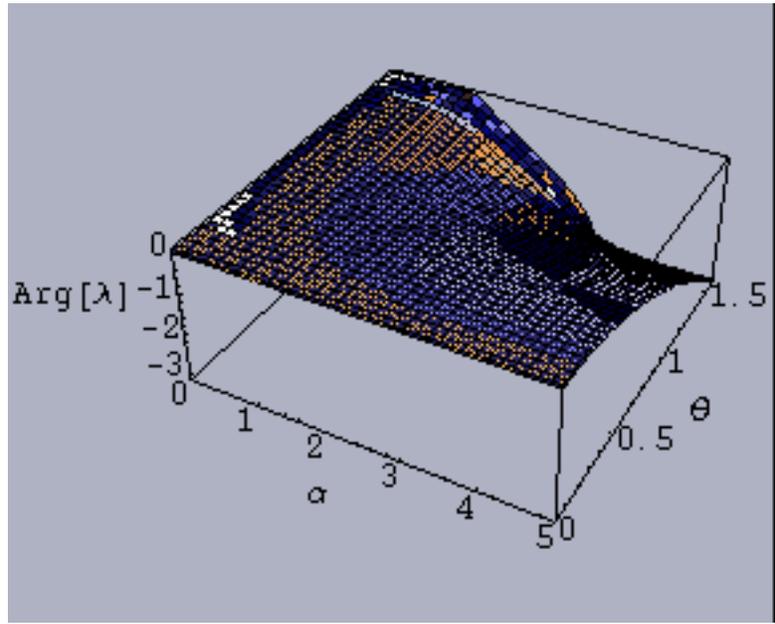}}
\caption{Argument of the dominant eigenvalue of the Uhlmann holonomy 
invariant (\ref{hi}) for the {\it two}-level Gibbsian density matrices}
\label{bh1}
\end{figure}

\begin{figure}
\centerline{\psfig{figure=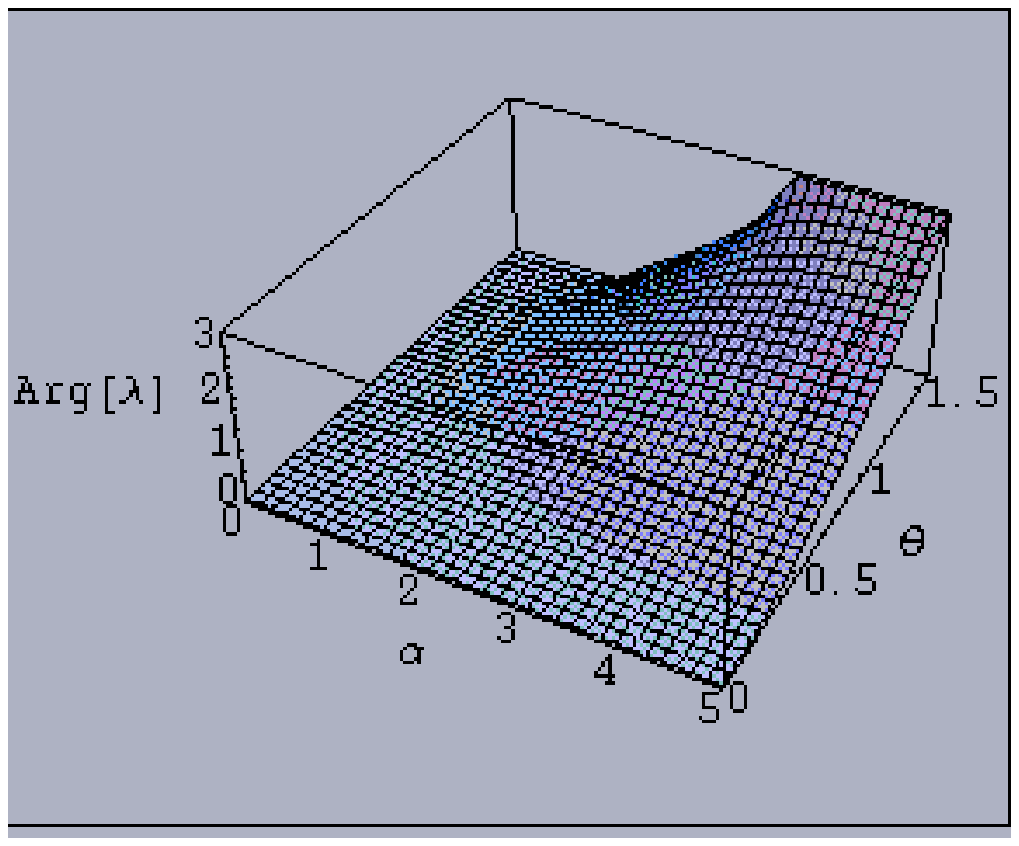}}
\caption{Argument of the subordinate eigenvalue of the Uhlmann holonomy
invariant (\ref{hi}) for the {\it two}-level Gibbsian density matrices}
\label{bh2}
\end{figure}

\begin{figure}
\centerline{\psfig{figure=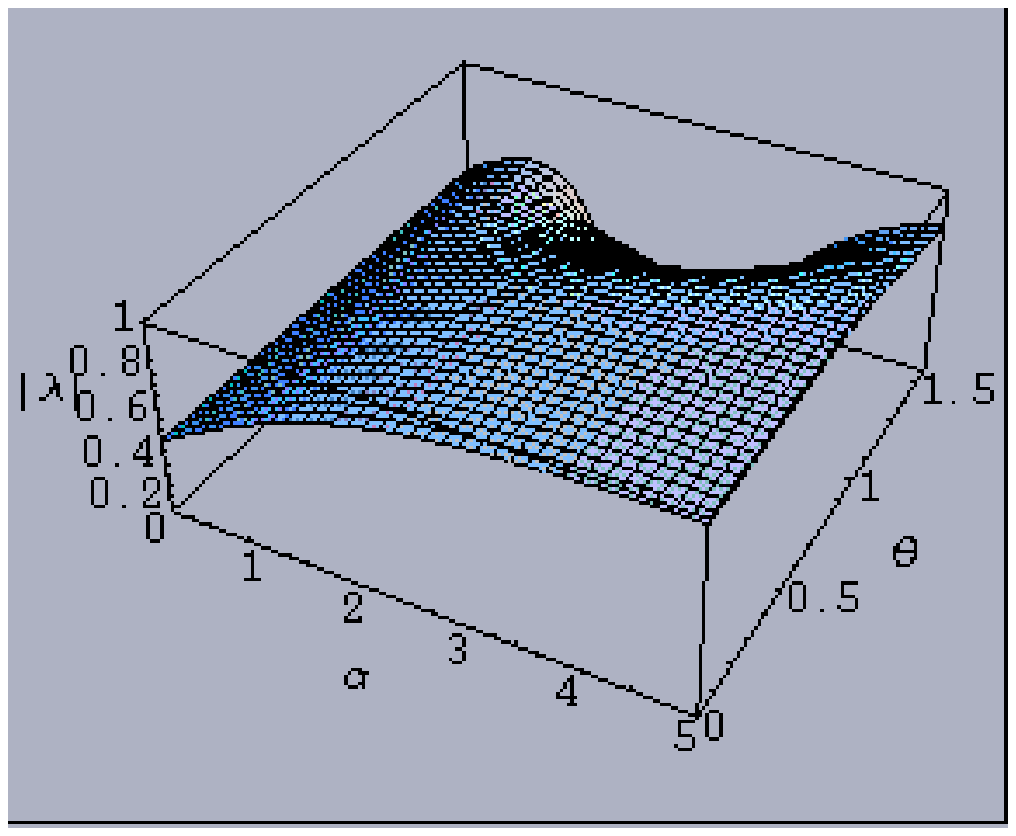}}
\caption{Absolute value of the dominant eigenvalue of the Uhlmann
holonomy invariant (\ref{hi}) for the {\it two}-level Gibbsian
density matrices}
\label{bh3}
\end{figure}

\begin{figure}
\centerline{\psfig{figure=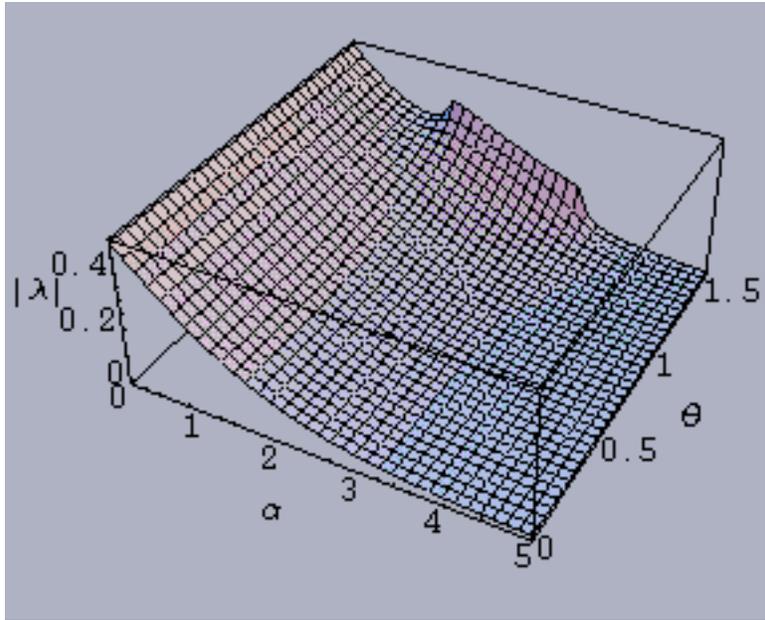}}
\caption{Absolute value of the subordinate eigenvalue of the Uhlmann
holonomy invariant (\ref{hi}) for the {\it two}-level Gibbsian
density matrices}
\label{bh4}
\end{figure}

\begin{figure}
\centerline{\psfig{figure=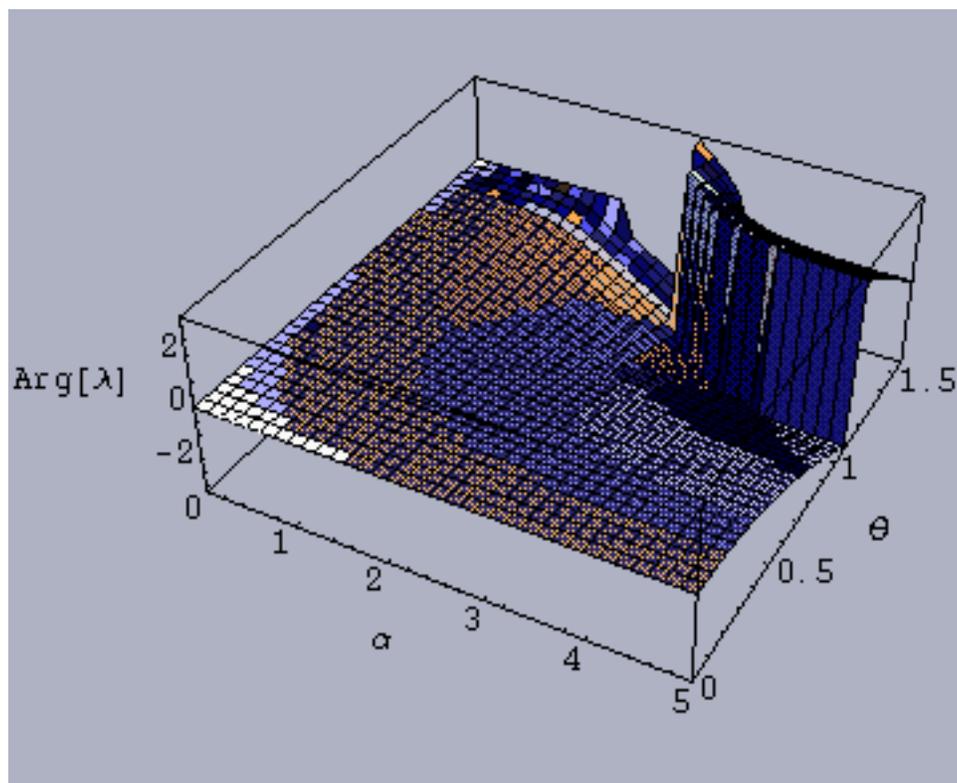}}
\caption{Argument of the dominant eigenvalue of the Uhlmann holonomy
invariant (\ref{hi}) for the {\it three}-level Gibbsian density matrices}
\label{bv1}
\end{figure}

\begin{figure}
\centerline{\psfig{figure=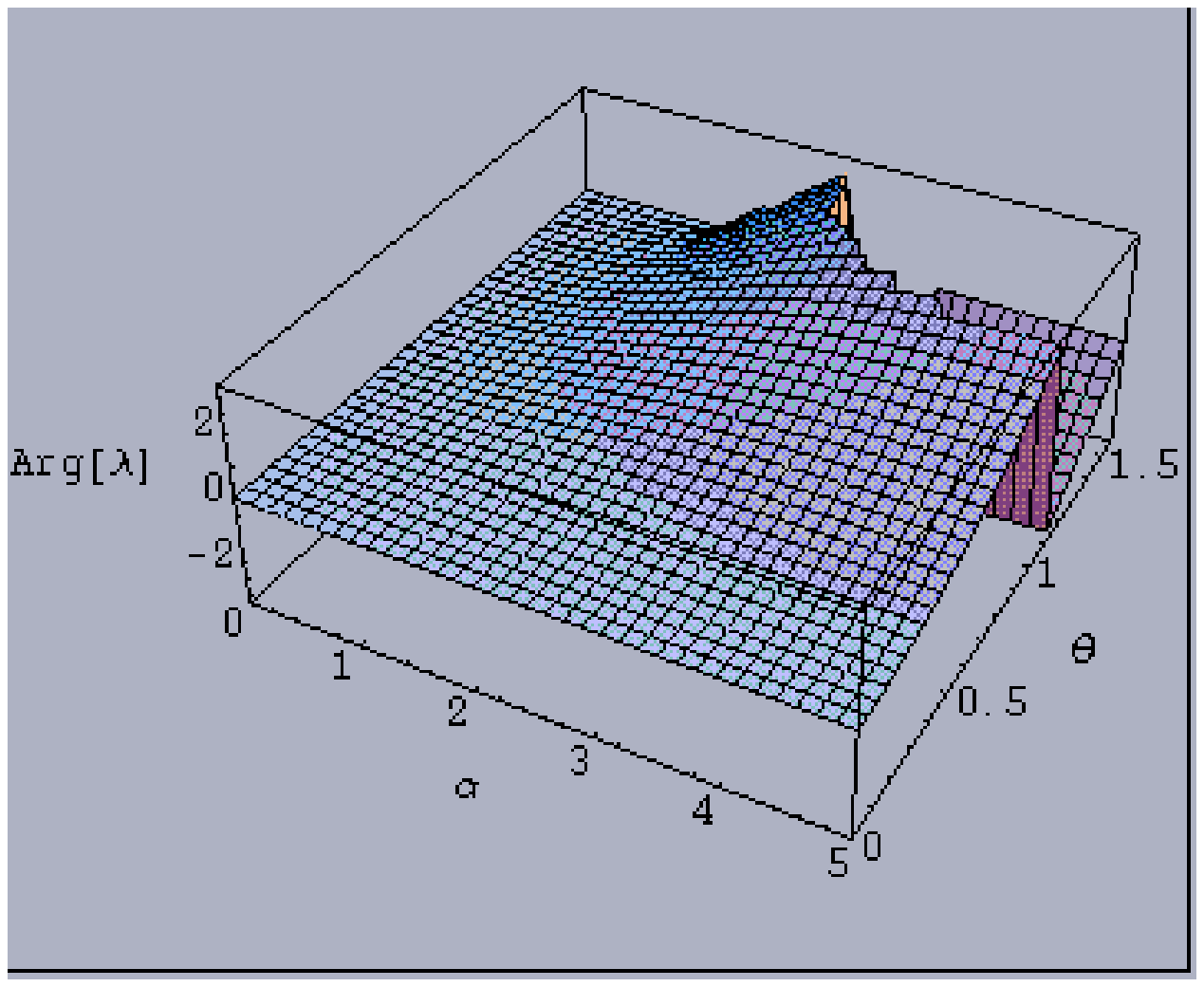}}
\caption{Argument of the subordinate eigenvalue of the Uhlmann holonomy
invariant (\ref{hi}) for the {\it three}-level Gibbsian density matrices}
\label{bv3}
\end{figure}

\begin{figure}
\centerline{\psfig{figure=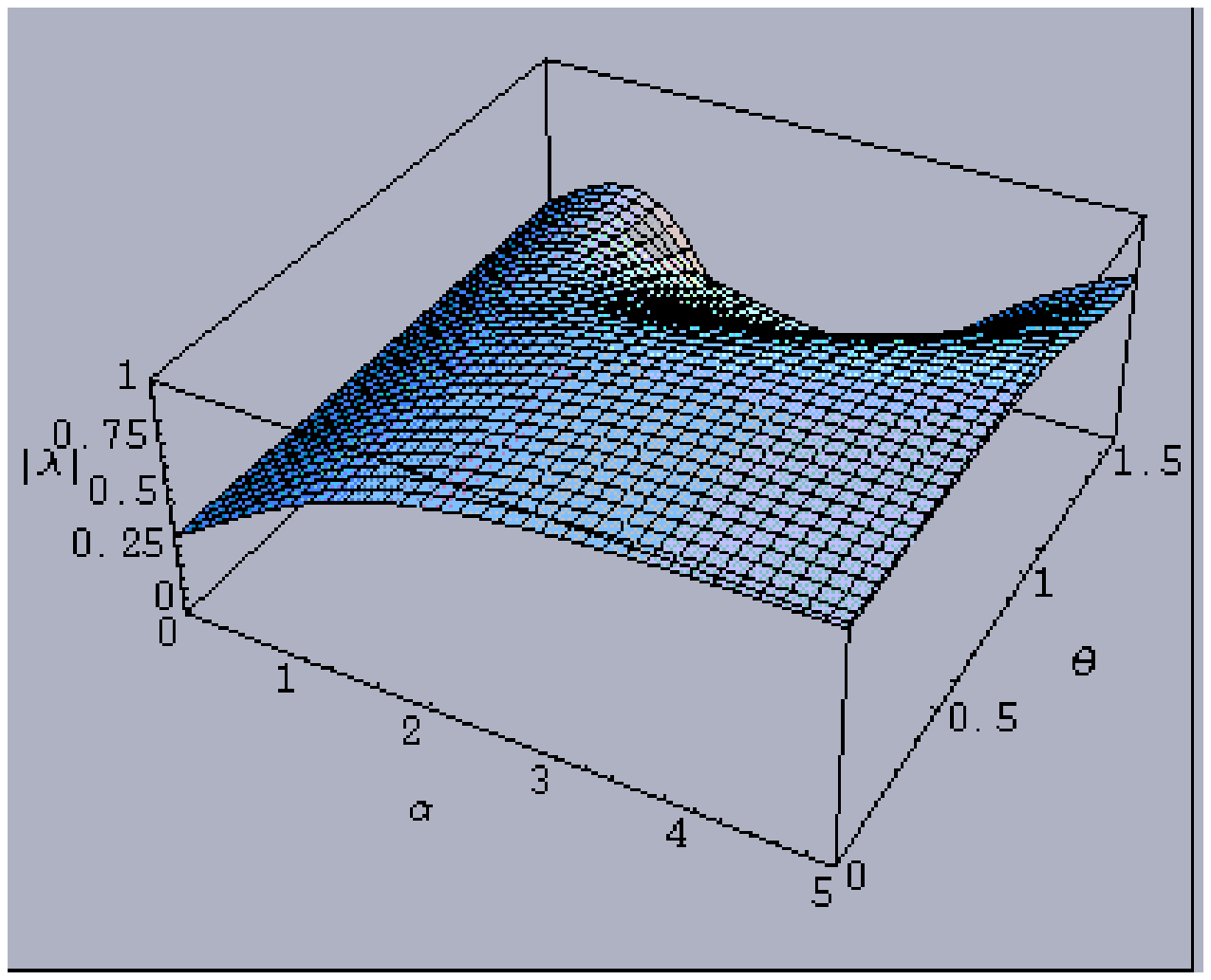}}
\caption{Absolute value of the dominant eigenvalue of the Uhlmann holonomy
invariant (\ref{hi}) for the {\it three}-level Gibbsian density matrices}
\label{bv4}
\end{figure}

\begin{figure}
\centerline{\psfig{figure=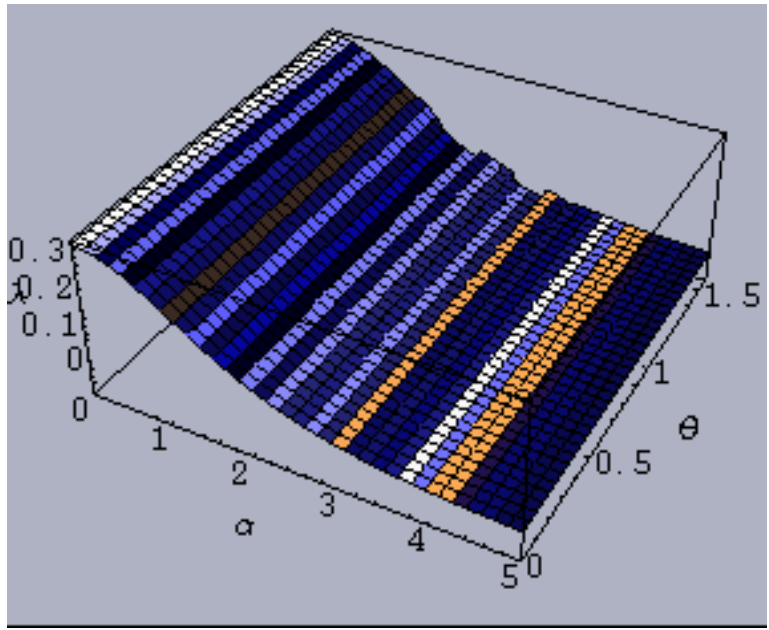}}
\caption{Intermediate eigenvalue of the Uhlmann
holonomy invariant (\ref{hi}) for the {\it three}-level 
Gibbsian density matrices}
\label{bv5}
\end{figure}

\begin{figure}
\centerline{\psfig{figure=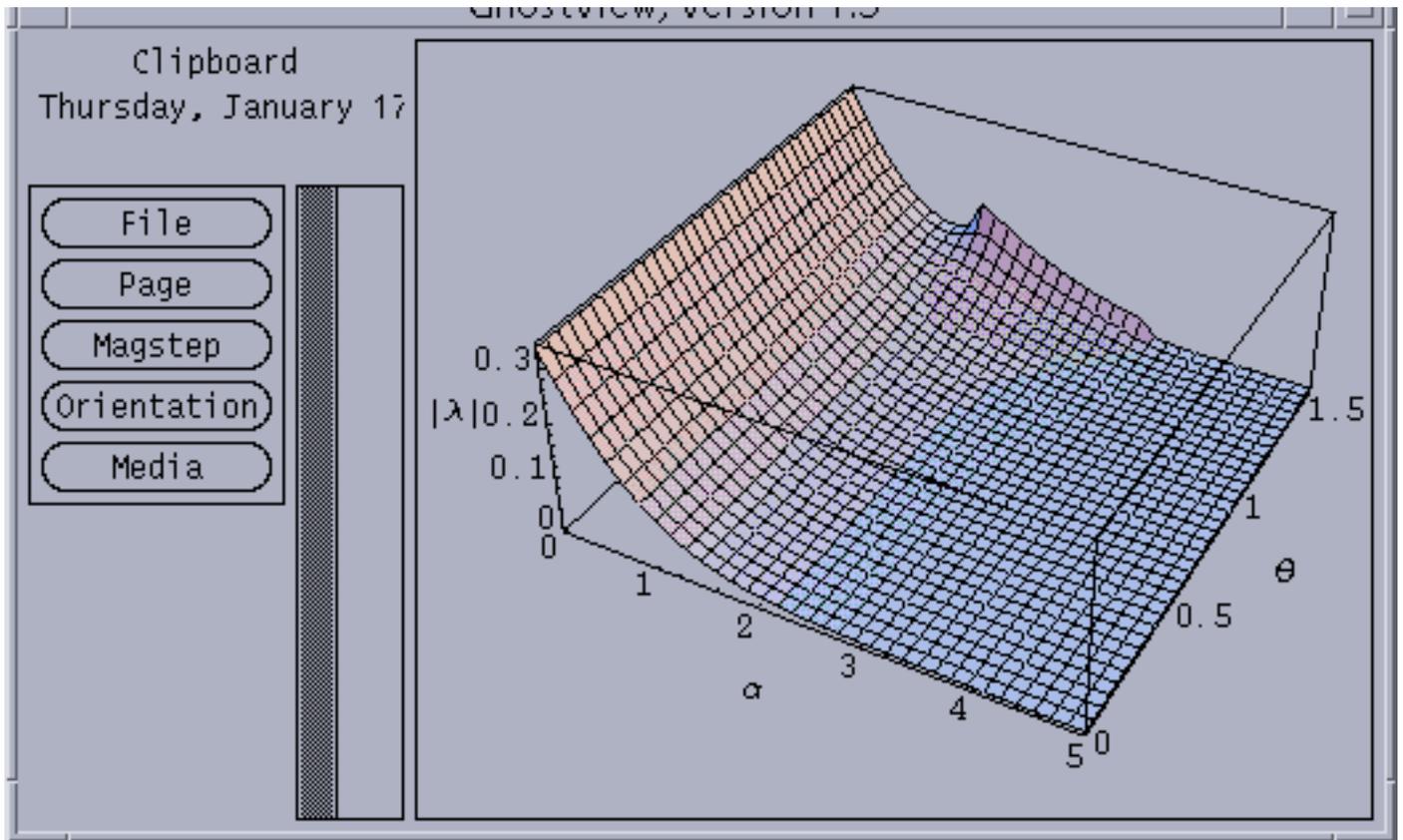}}
\caption{Absolute value of the subordinate eigenvalue of the Uhlmann
holonomy invariant (\ref{hi}) for the {\it three}-level 
Gibbsian density matrices}
\label{bv6}
\end{figure}

\begin{figure}
\centerline{\psfig{figure=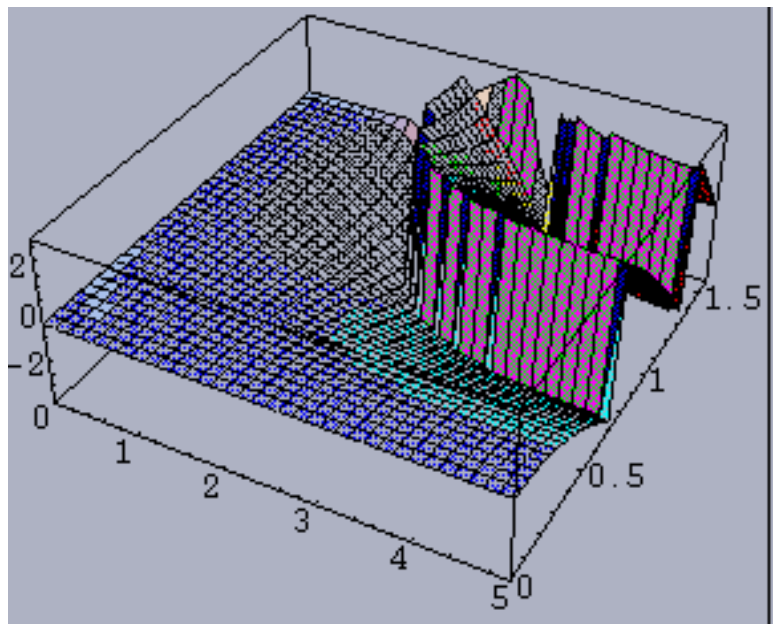}}
\caption{Argument of the leading eigenvalue of the Uhlmann holonomy 
invariant (\ref{hi}) for the {\it five}-level Gibbsian density matrices}
\label{cd1}
\end{figure}

\begin{figure}
\centerline{\psfig{figure=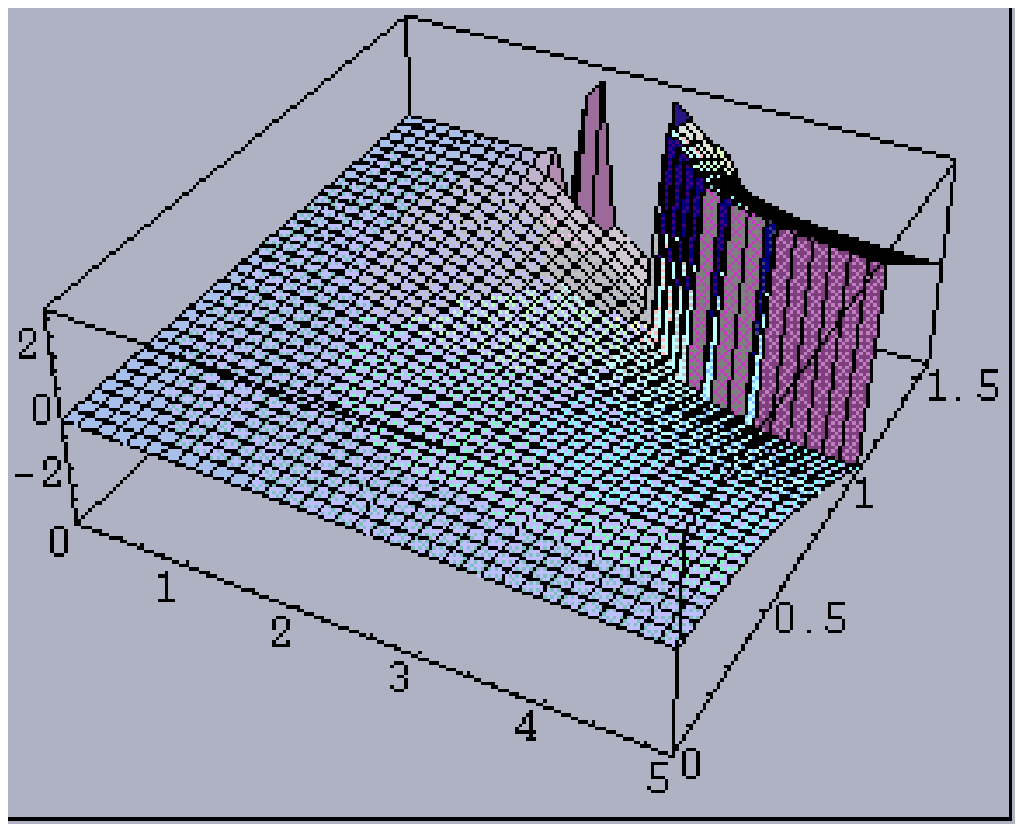}}
\caption{Argument of the second leading eigenvalue of the Uhlmann 
holonomy invariant (\ref{hi}) for the {\it five}-level Gibbsian density matrices}
\label{cd2}
\end{figure}

\begin{figure}
\centerline{\psfig{figure=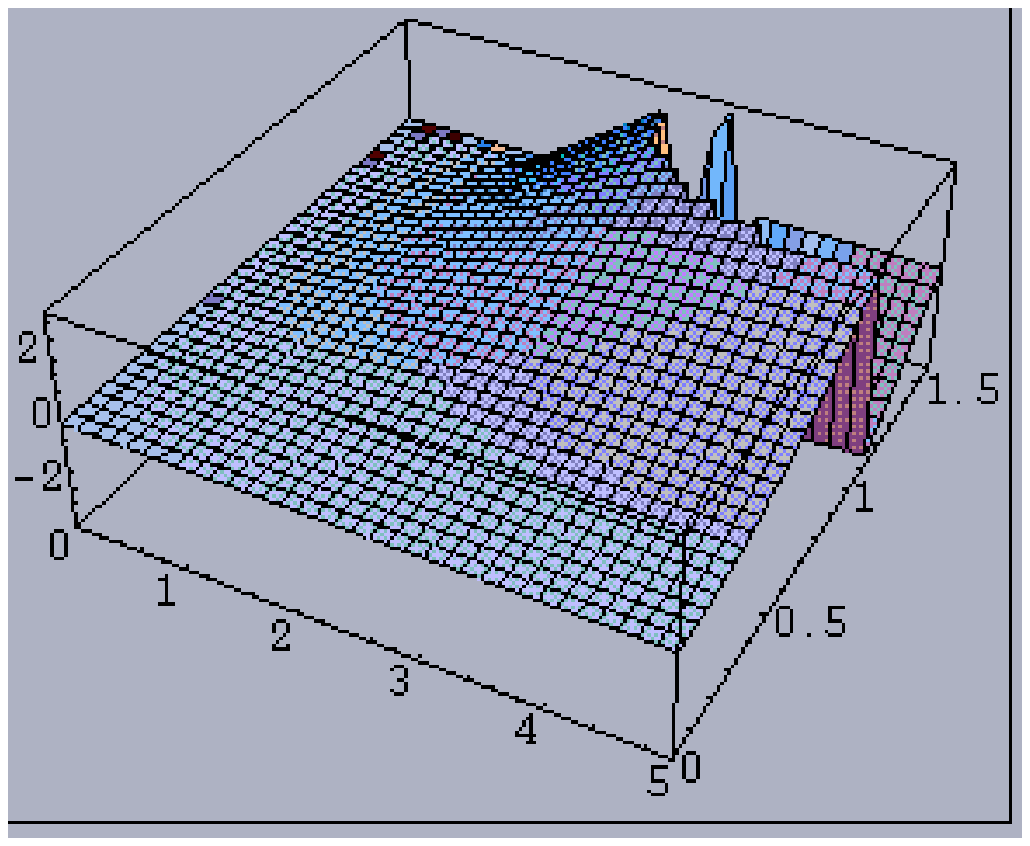}}
\caption{Argument of the second smallest eigenvalue of the Uhlmann holonomy 
invariant (\ref{hi}) for the {\it five}-level Gibbsian density matrices}
\label{cd3}
\end{figure}

\begin{figure}
\centerline{\psfig{figure=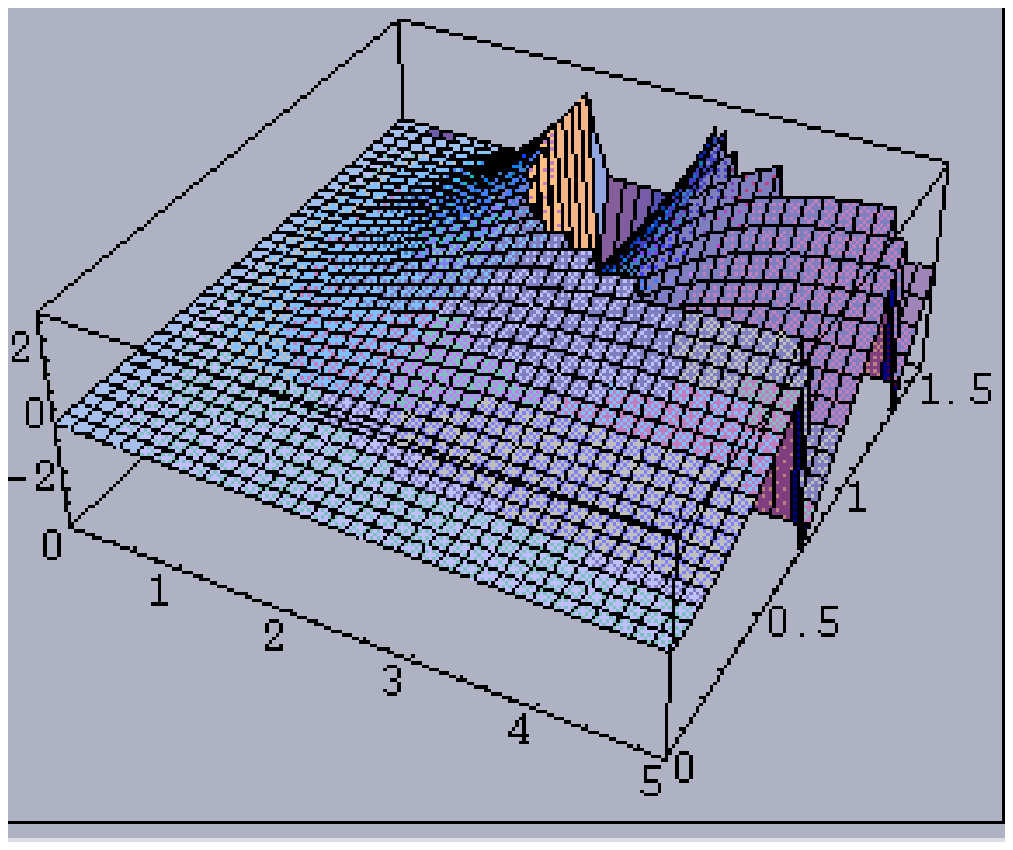}}
\caption{Argument of the last eigenvalue of the Uhlmann holonomy
invariant (\ref{hi}) for the {\it five}-level Gibbsian density matrices}
\label{cd4}
\end{figure}

\begin{figure}
\centerline{\psfig{figure=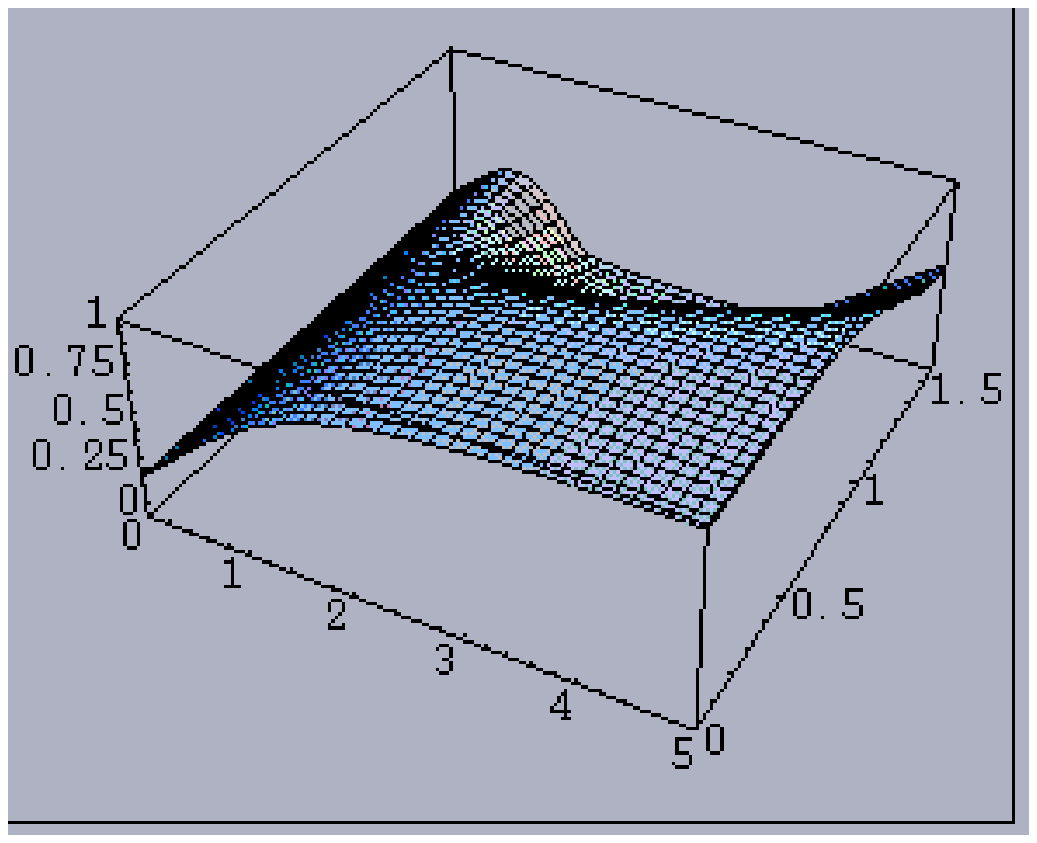}}
\caption{Absolute value of the leading eigenvalue of the Uhlmann holonomy
invariant (\ref{hi}) for the {\it five}-level Gibbsian density matrices}
\label{cd5}
\end{figure}

\begin{figure}
\centerline{\psfig{figure=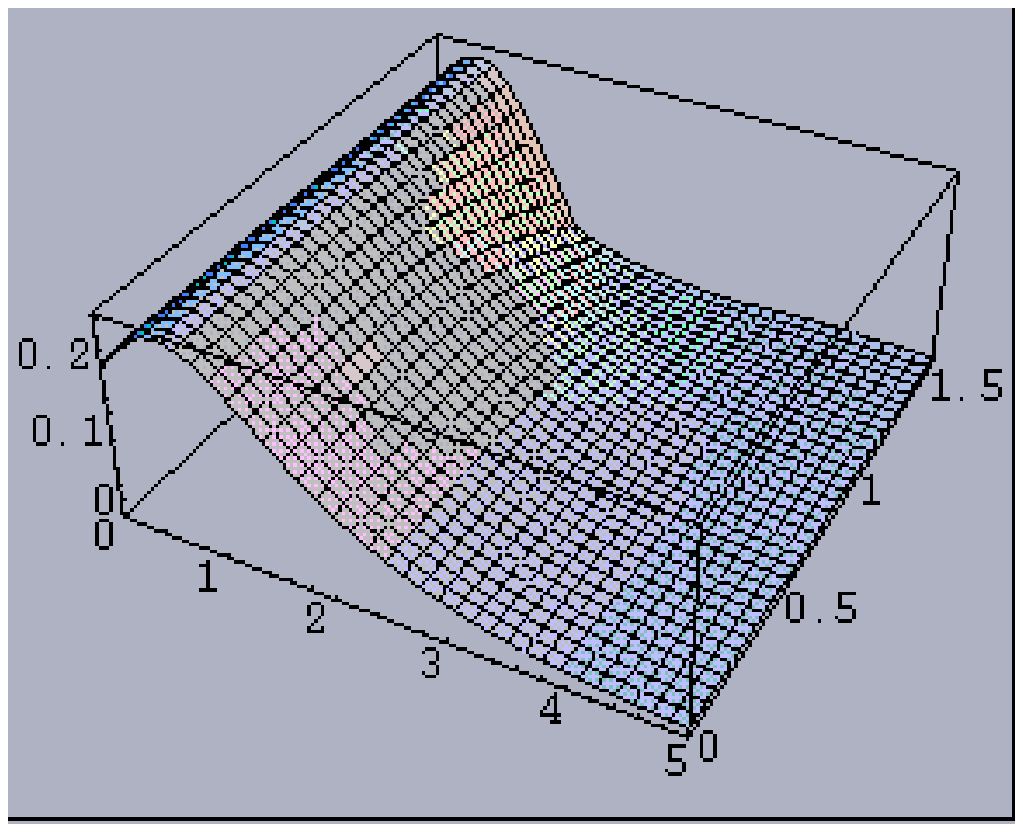}}
\caption{Absolute value of the second leading eigenvalue of the Uhlmann
holonomy invariant (\ref{hi}) for the {\it five}-level Gibbsian density
matrices}
\label{cd6}
\end{figure}

\begin{figure}
\centerline{\psfig{figure=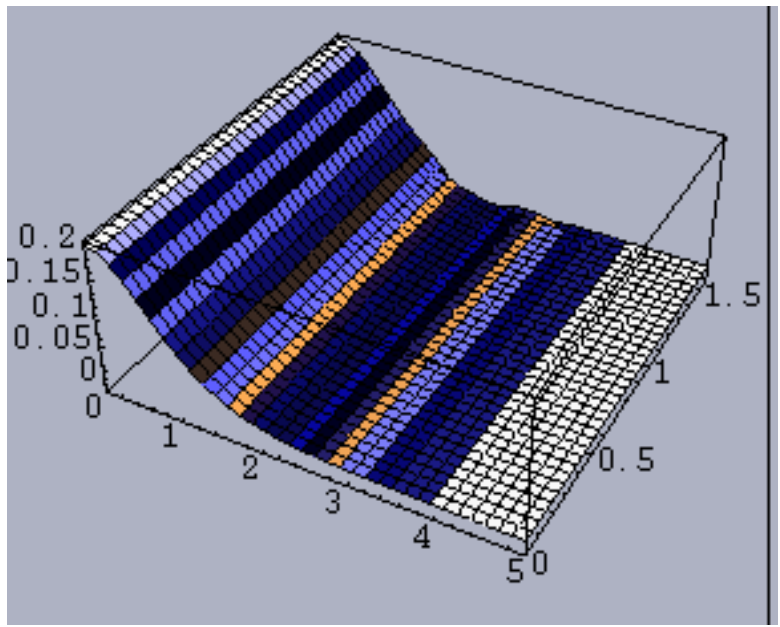}}
\caption{Central eigenvalue of the Uhlmann holonomy
invariant (\ref{hi}) for the {\it five}-level Gibbsian density matrices}
\label{cd7}
\end{figure}

\begin{figure}
\centerline{\psfig{figure=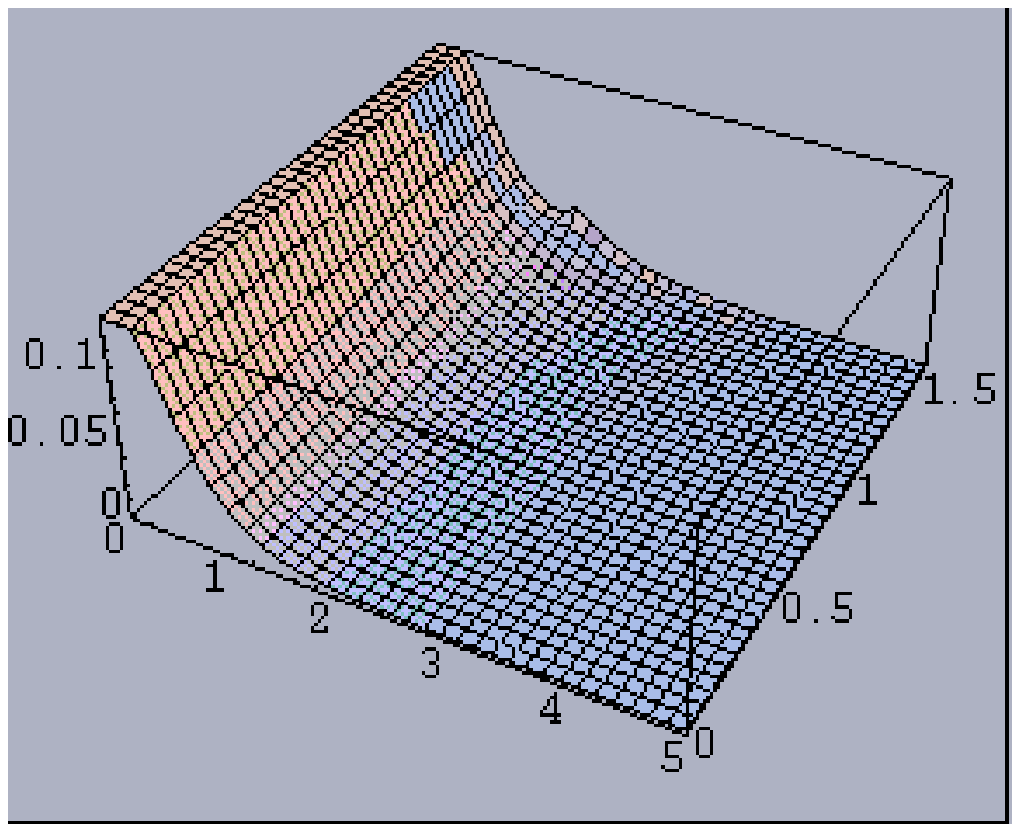}}
\caption{Absolute value of the second smallest eigenvalue of the Uhlmann
holonomy invariant (\ref{hi}) for the {\it five}-level Gibbsian density
matrices}
\label{cd8}
\end{figure}

\begin{figure}
\centerline{\psfig{figure=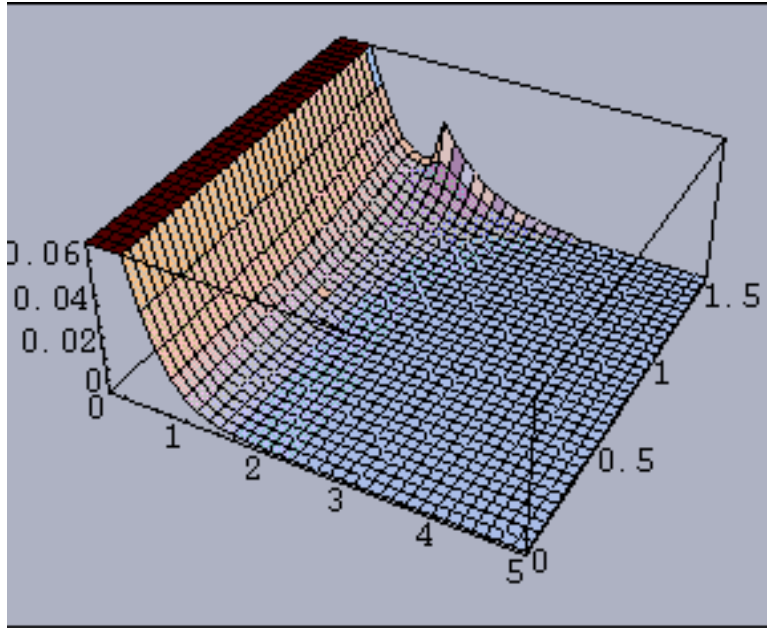}}
\caption{Absolute value of the last eigenvalue of the Uhlmann holonomy
invariant (\ref{hi}) for the {\it five}-level Gibbsian density matrices}
\label{cd9}
\end{figure}

To supplement our consideration above of  the 
{\it traces} of various powers of the Uhlmann
holonomy invariant (\ref{hi}), here we examine the eigenvalues themselves. 
(Both these traces and eigenvalues are themselves invariants, of course.)
In Figs.~\ref{bh1} and \ref{bh2} are shown the arguments of the dominant
and subordinate eigenvalues of the holonomy invariant (\ref{hi}) for the
case $n=2$, and in Figs.~\ref{bh3} and \ref{bh4}, the corresponding absolute
values.
In Figs.~\ref{bv1} and \ref{bv3}, we plot 
the arguments of the dominant and subordinate eigenvalues of the holonomy 
invariant (\ref{hi}) for the case $n=3$ and in
Figs.~\ref{bv4} and \ref{bv6}, the absolute values of the two corresponding 
eigenvalues.
The argument of the intermediate (in absolute value) 
one of the three eigenvalues 
appears to always be identically zero, that is,
this intermediate eigenvalue is real (Fig.~\ref{bv5}).
It also appears numerically, on the basis of further analyses we have 
conducted, 
 that this result is generalizable to the proposition 
that for odd 
$n$, 
 the central/most intermediate 
one (that is, the $\lceil {n \over 2} \rceil$-th) of the $n$
eigenvalues is real. (Also, the centrally located diagonal entry
of the holonomy invariant for odd $n$ appears to be real, in general.)
In Figs.~\ref{cd1}-\ref{cd4}, we show the arguments of the eigenvalues
of the Uhlmann holonomy invariant (\ref{hi}) in the case $n=5$.
(The case $n=4$ proved more intractable in nature.) The third eigenvalue
is simply real. In Figs.~\ref{cd5}, \ref{cd6}, \ref{cd8}, 
\ref{cd9} are shown the corresponding
absolute values. In Fig.~\ref{cd7} is shown the value of the 
(real) intermediate eigenvalue.
\section{Sj\"oqvist {\it et al} geometric phases for 
Gibbsian $n$-level systems
($n=2,\ldots,11$)} \label{r6}
\begin{figure}
\centerline{\psfig{figure=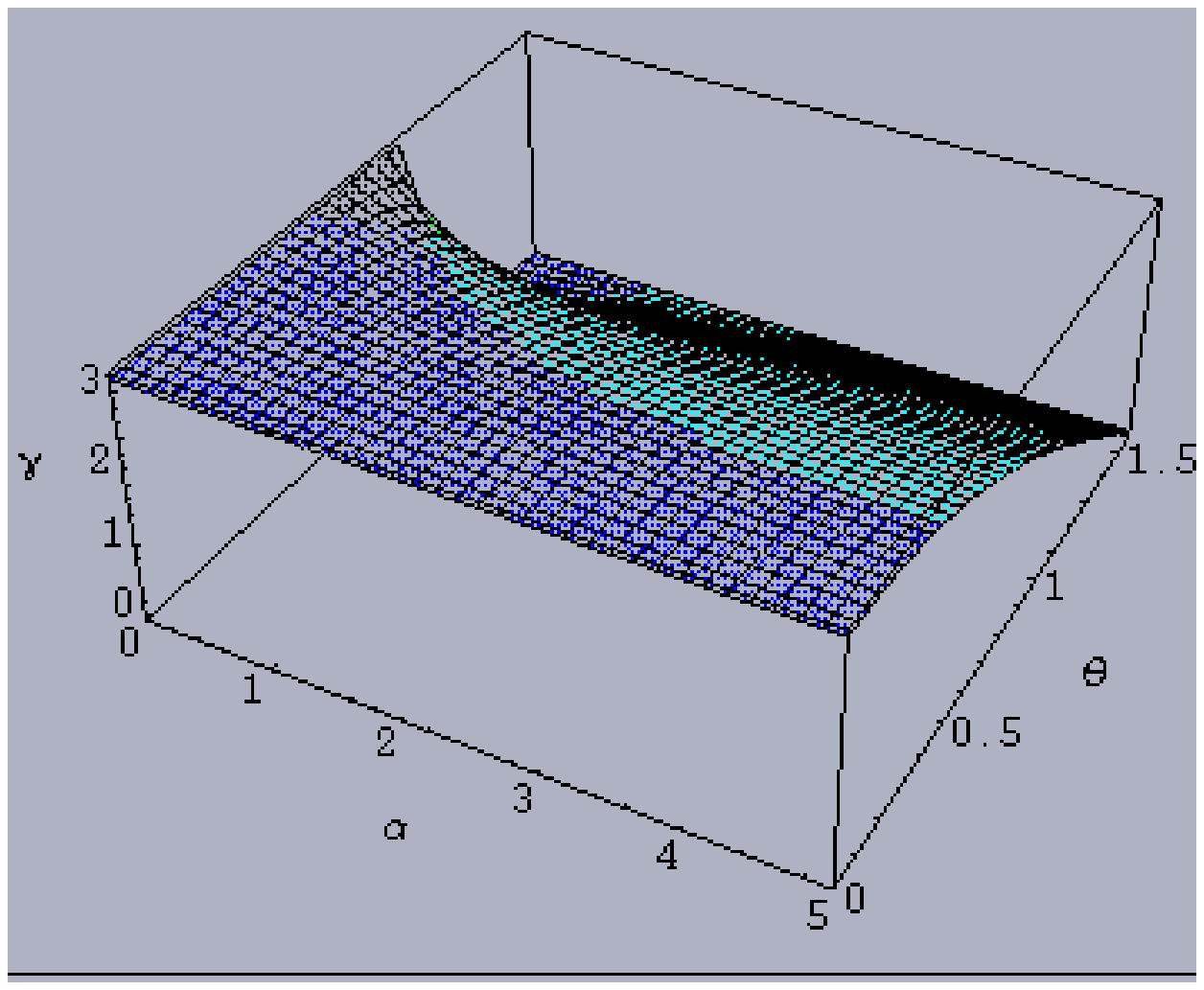}}
\caption{Sj\"oqvist {\it et al} geometric phase for Gibbsian 
spin-${1 \over 2}$ systems}
\label{k2}
\end{figure}

\begin{figure}
\centerline{\psfig{figure=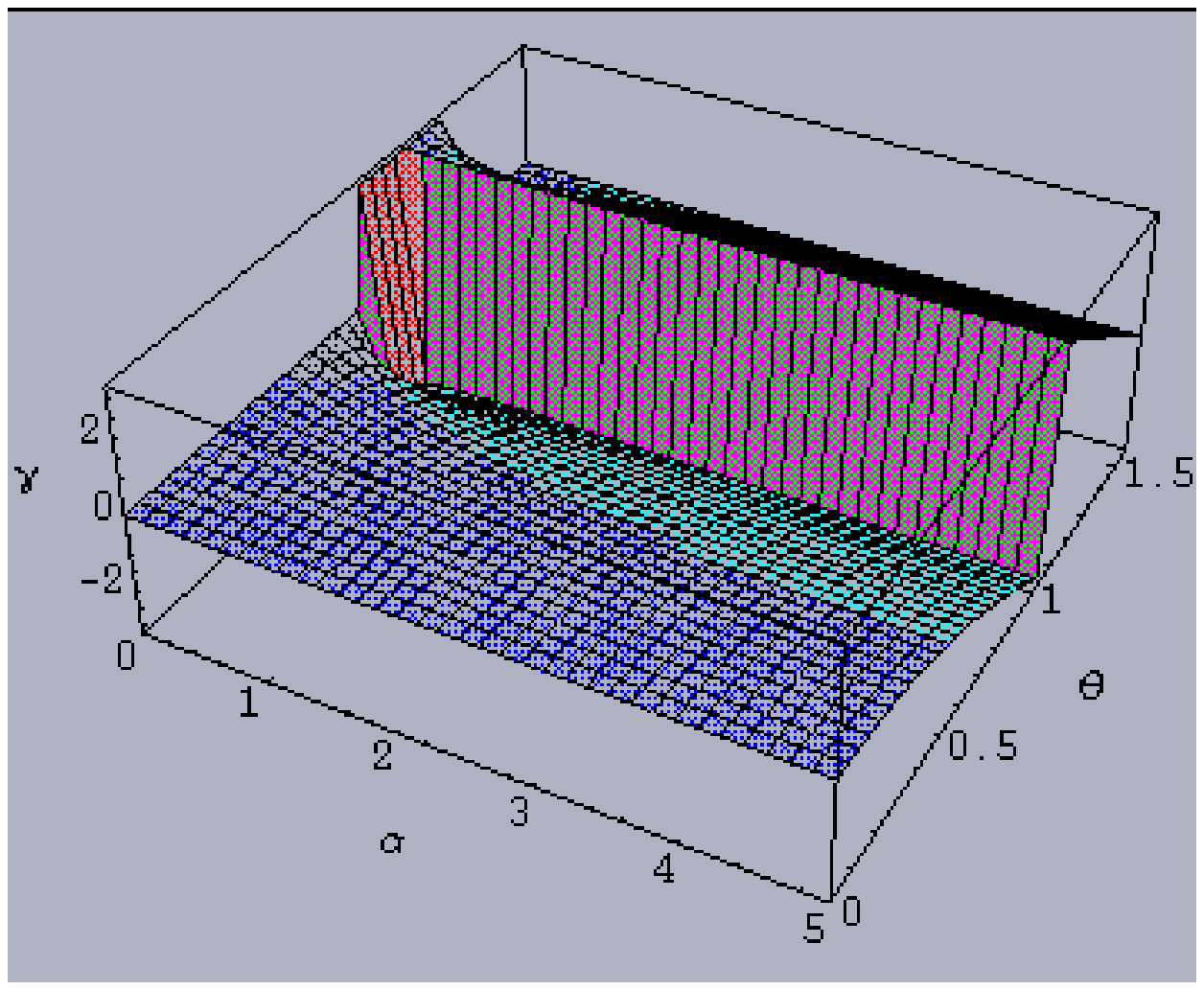}}
\caption{Sj\"oqvist {\it et al} geometric phase for 
Gibbsian spin-1 systems}
\label{k3}
\end{figure}

\begin{figure}
\centerline{\psfig{figure=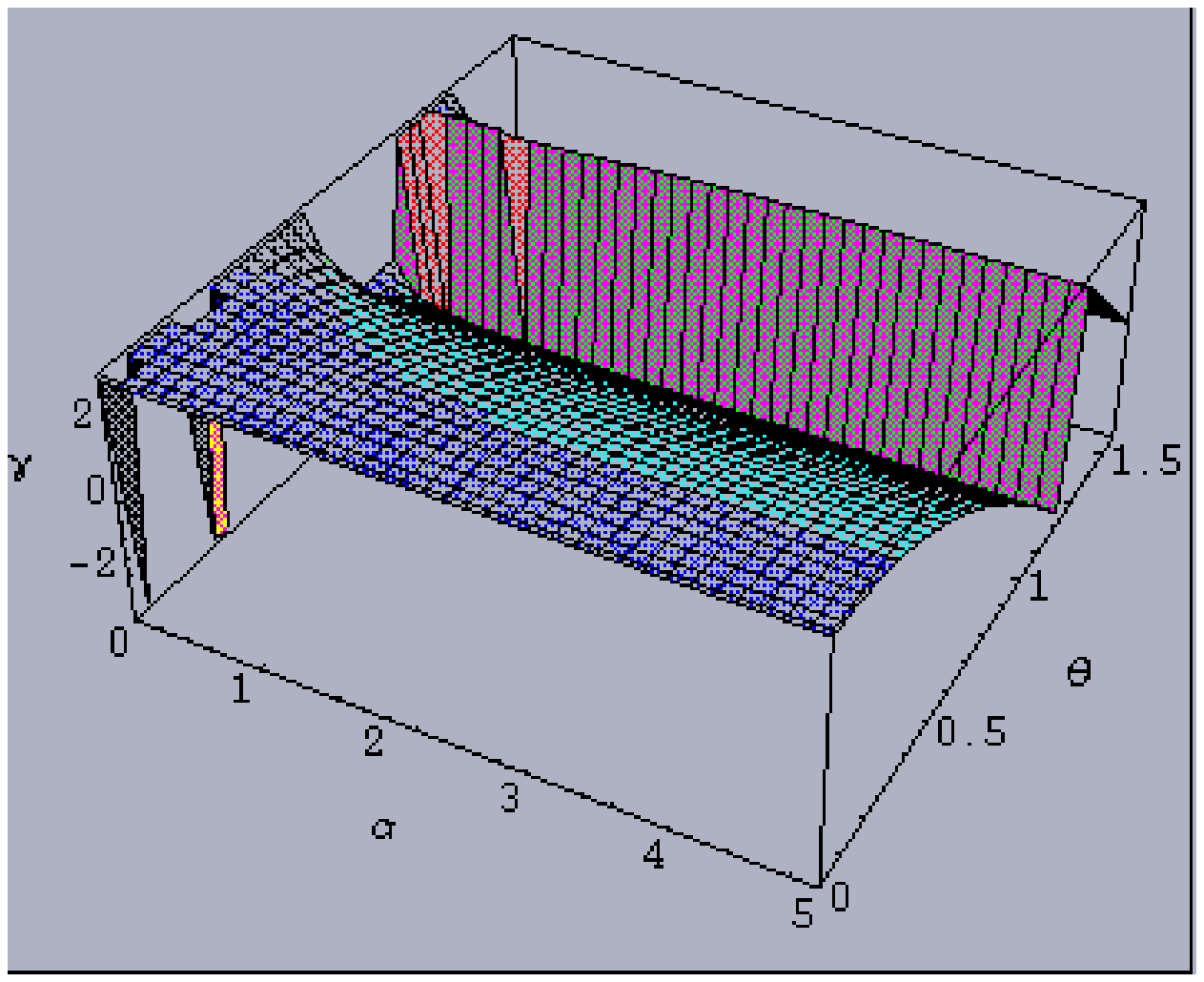}}
\caption{Sj\"oqvist {\it et al} geometric phase 
for Gibbsian spin-${3 \over 2}$ 
systems}
\label{k4}
\end{figure}

\begin{figure}
\centerline{\psfig{figure=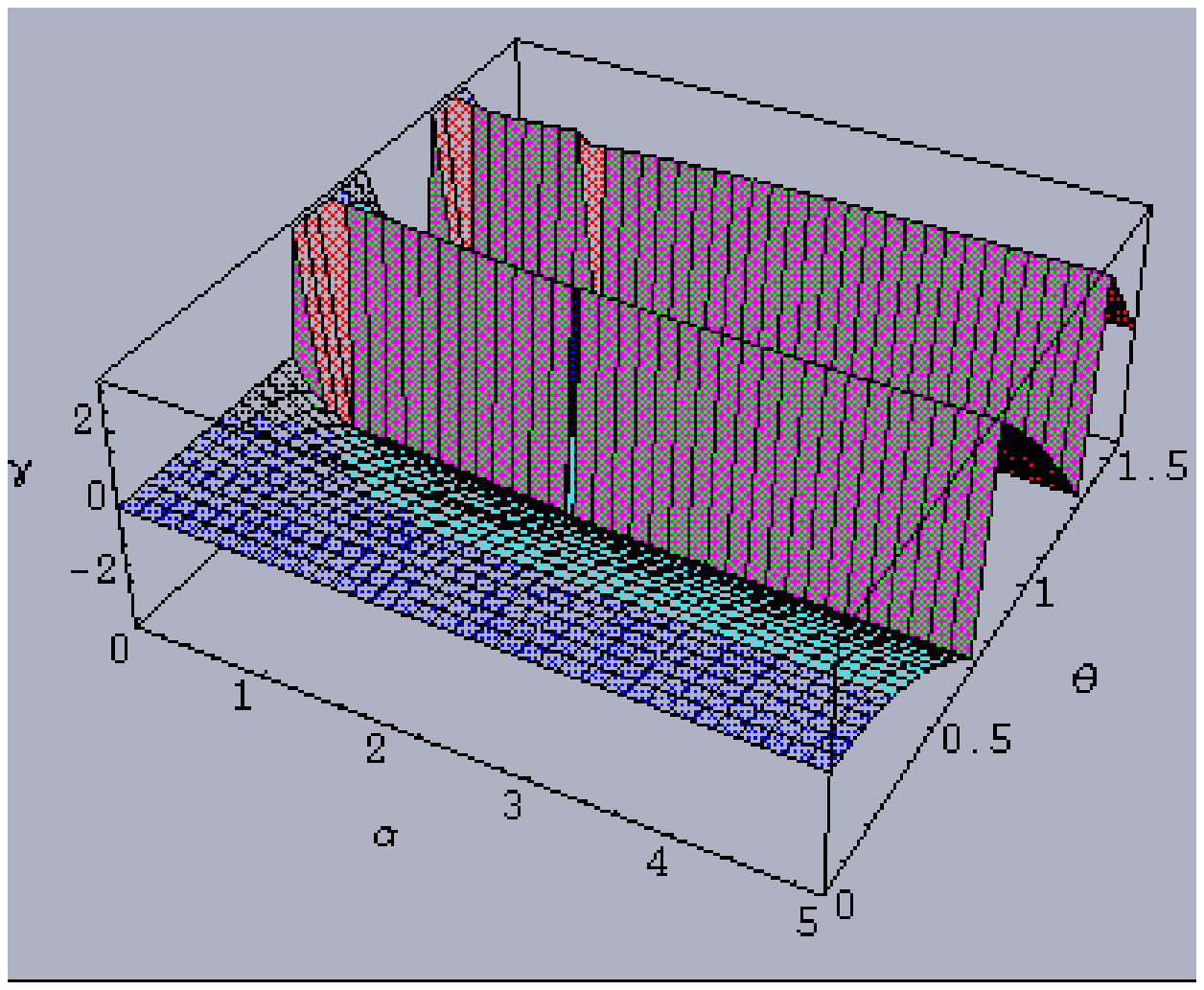}}
\caption{Sj\"oqvist {\it et al} geometric phase for Gibbsian spin-2 systems}
\label{k5}
\end{figure}

\begin{figure}
\centerline{\psfig{figure=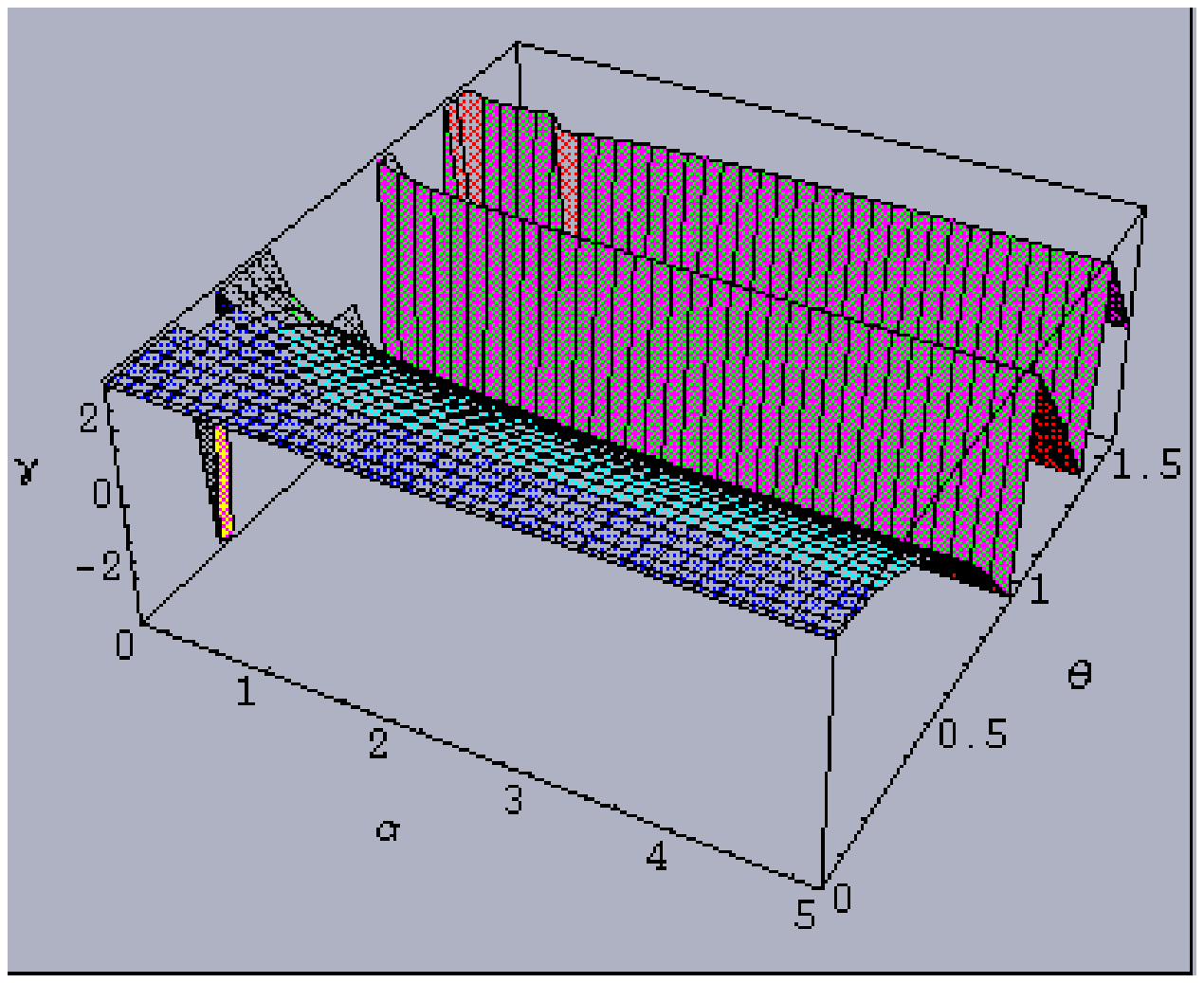}}
\caption{Sj\"oqvist {\it et al} geometric phase for Gibbsian spin-${5 \over 2}$ systems}
\label{k6}
\end{figure}

\begin{figure}
\centerline{\psfig{figure=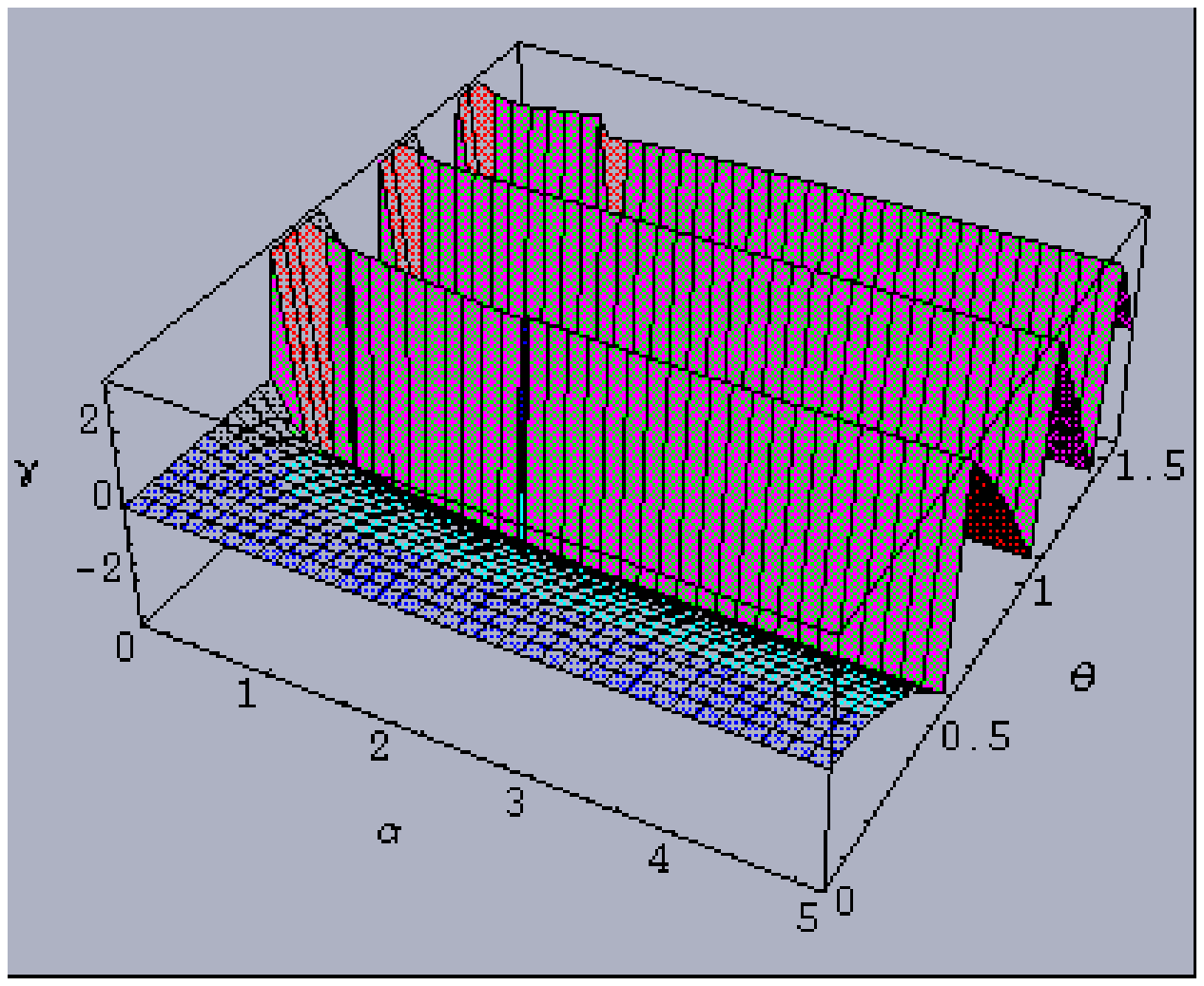}}
\caption{Sj\"oqvist {\it et al} geometric phase for Gibbsian spin-3 systems}
\label{k7}
\end{figure}

\begin{figure}
\centerline{\psfig{figure=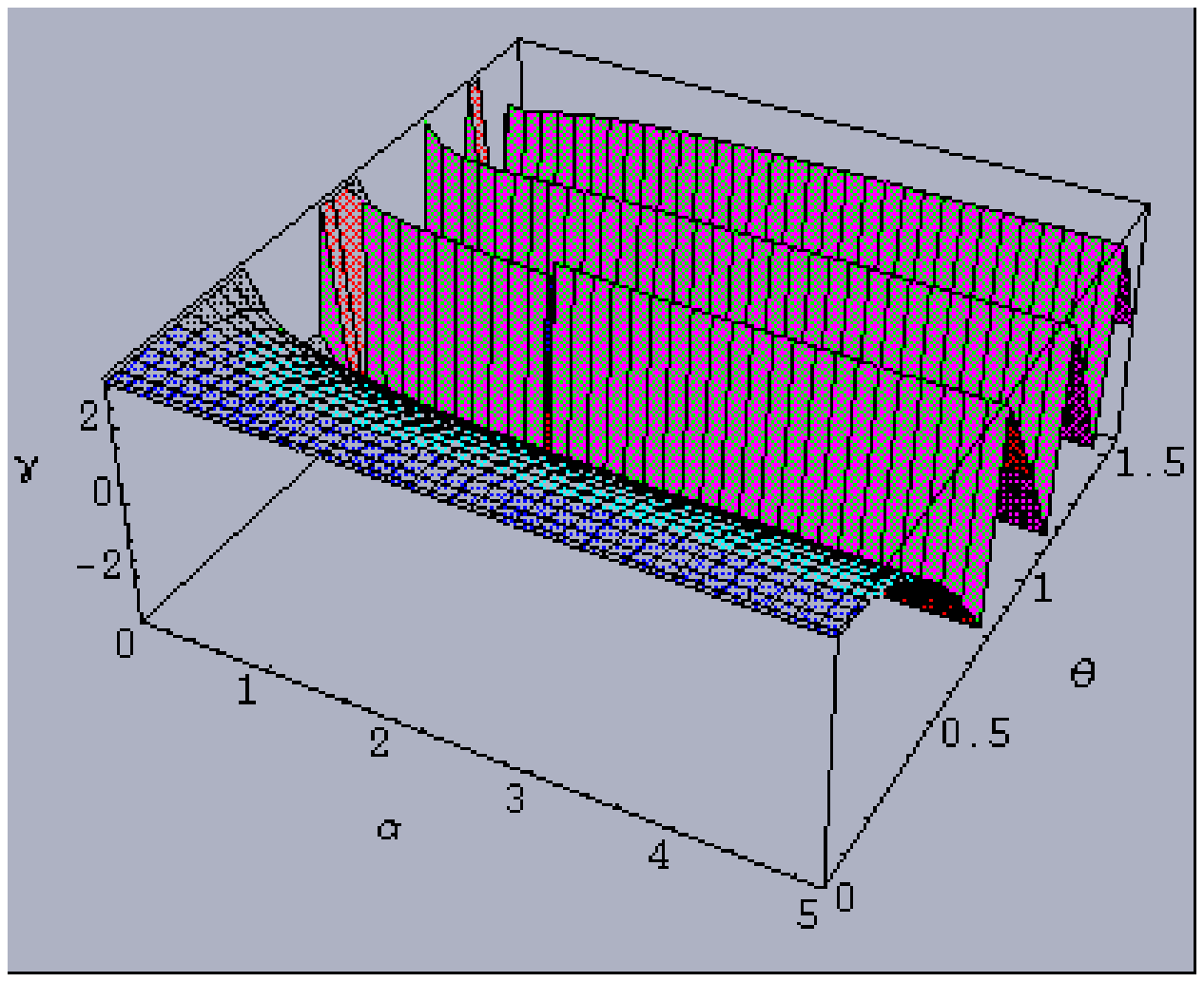}}
\caption{Sj\"oqvist {\it et al} 
geometric phase for Gibbsian spin-${7 \over 2}$ 
systems}
\label{k8}
\end{figure}

\begin{figure}
\centerline{\psfig{figure=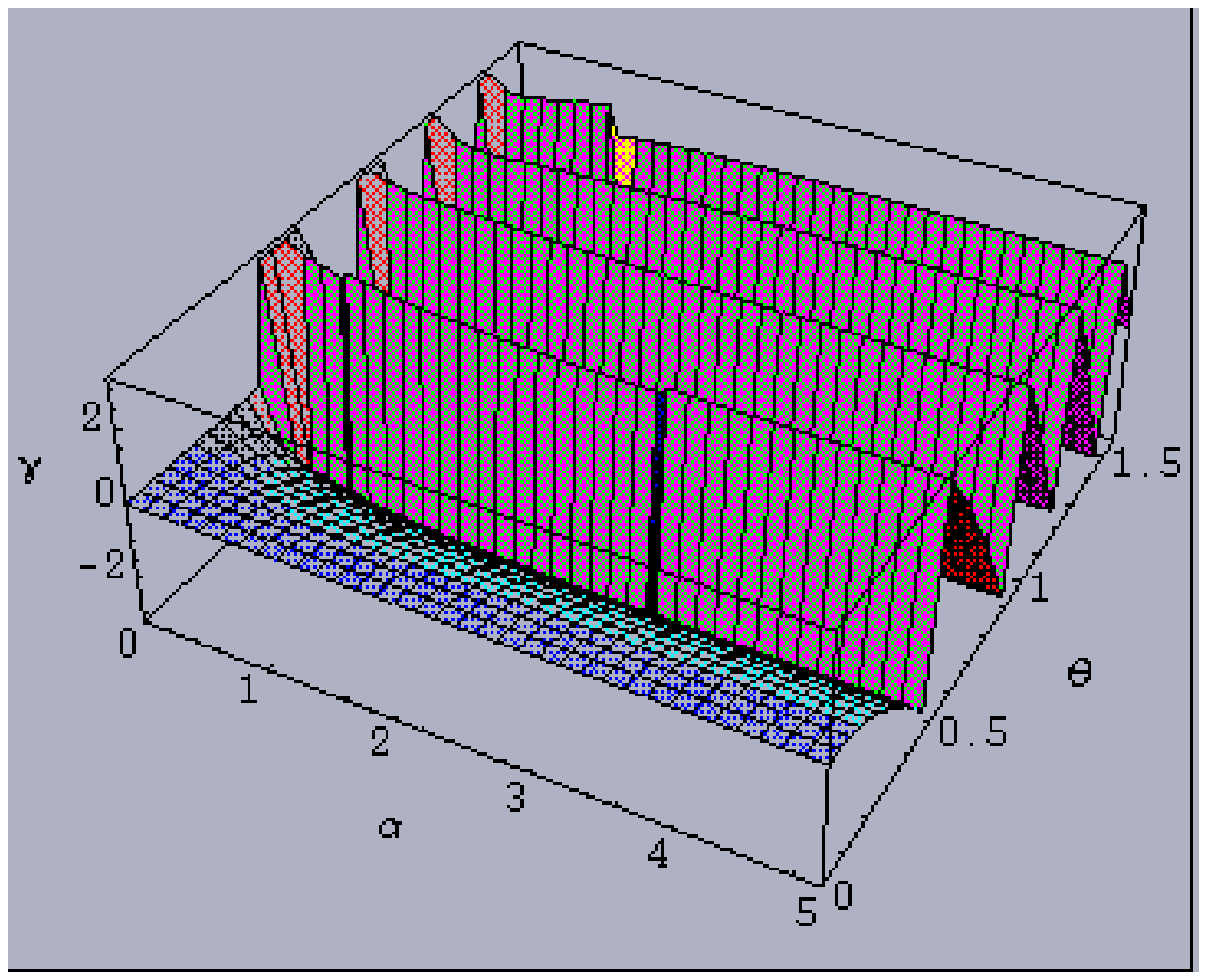}}
\caption{Sj\"oqvist {\it et al} geometric phase for Gibbsian spin-4 systems}
\label{k9}
\end{figure}

\begin{figure}
\centerline{\psfig{figure=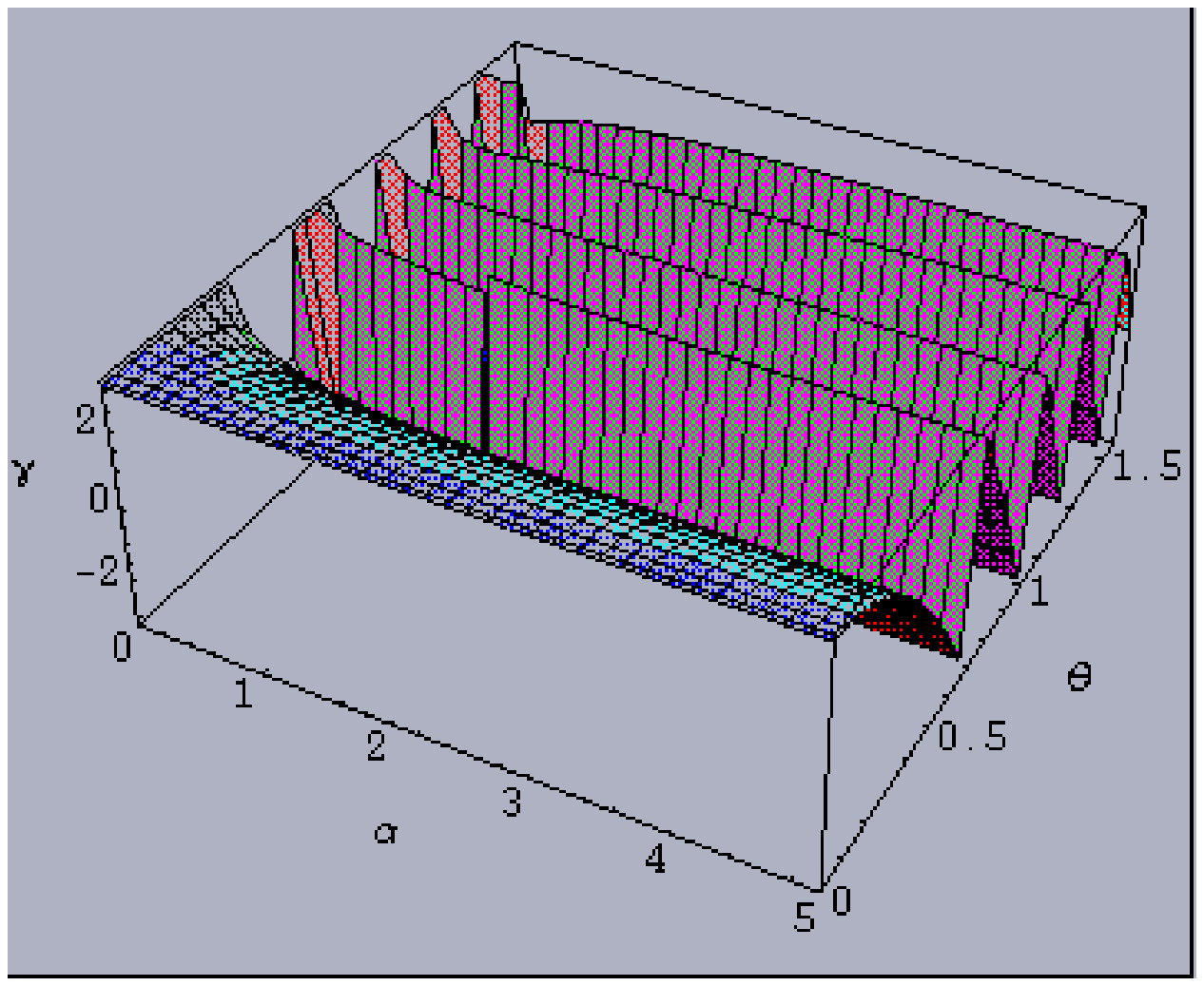}}
\caption{Sj\"oqvist {\it et al} geometric phase for 
Gibbsian spin-${9 \over 2}$ systems}
\label{k10}
\end{figure}

\begin{figure}
\centerline{\psfig{figure=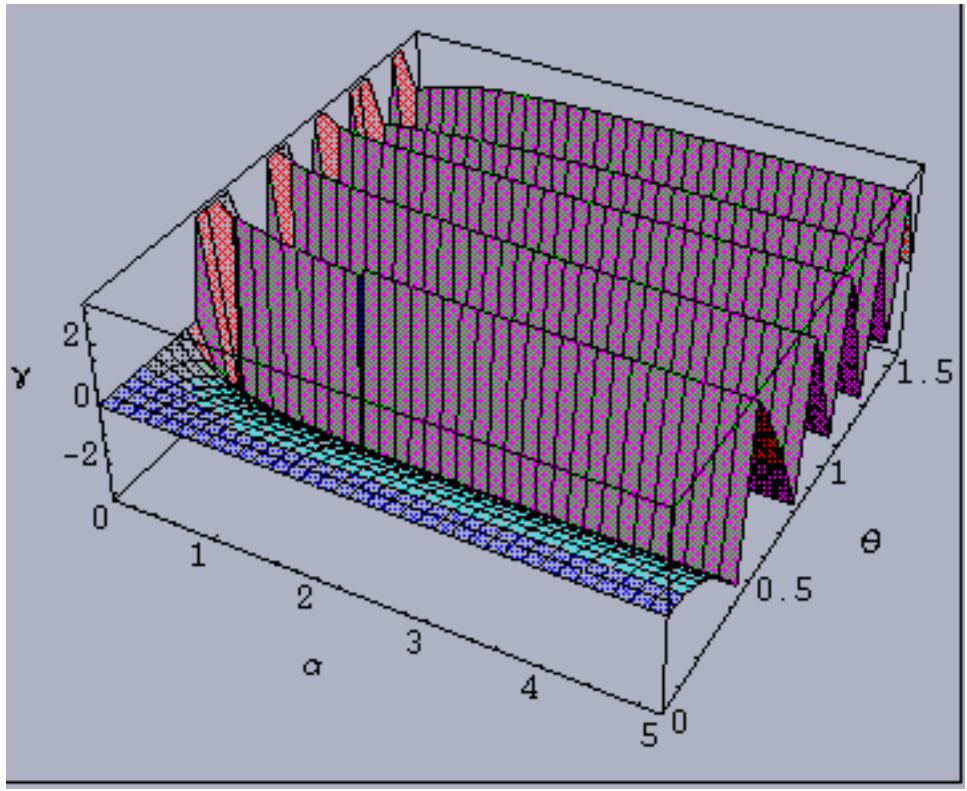}}
\caption{Sj\"oqvist {\it et al} geometric phase for Gibbsian spin-5 systems}
\label{k11}
\end{figure}

\begin{figure}
\centerline{\psfig{figure=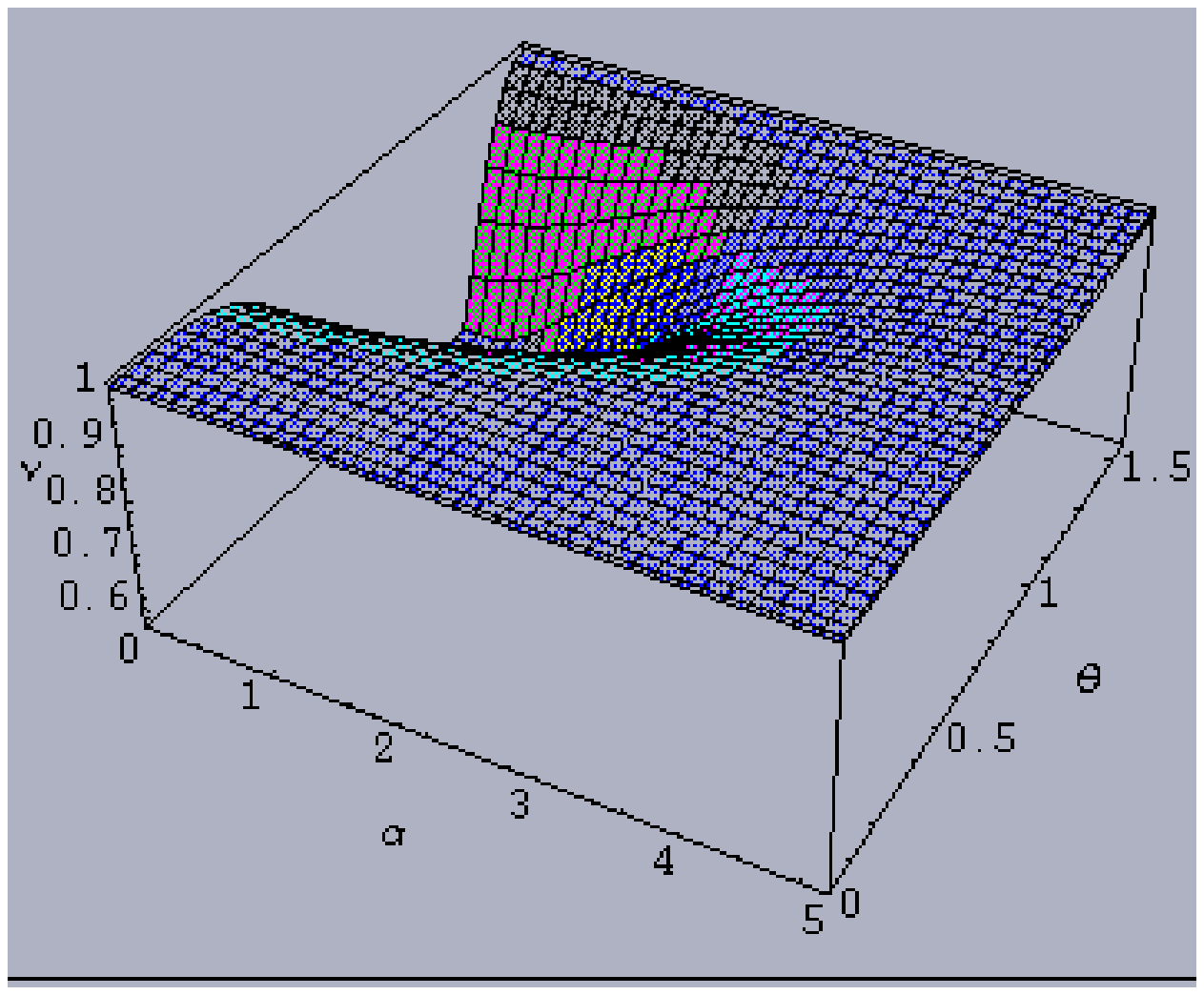}}
\caption{Sj\"oqvist {\it et al} visibility for Gibbsian spin-${1 \over 2}$ 
systems}
\label{kk2}
\end{figure}

\begin{figure}
\centerline{\psfig{figure=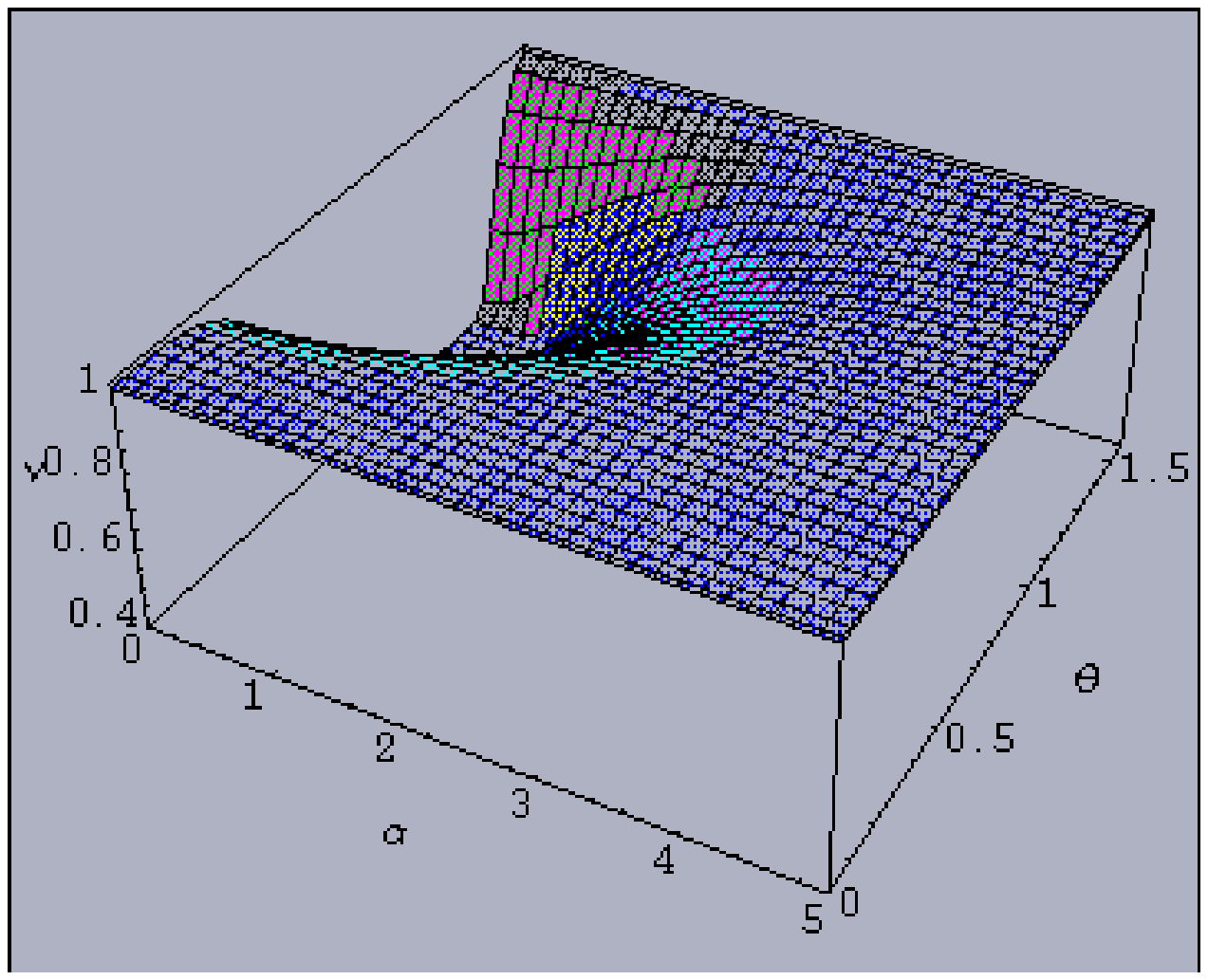}}
\caption{Sj\"oqvist {\it et al} visibility for Gibbsian spin-1 systems}
\label{kk3}
\end{figure}

\begin{figure}
\centerline{\psfig{figure=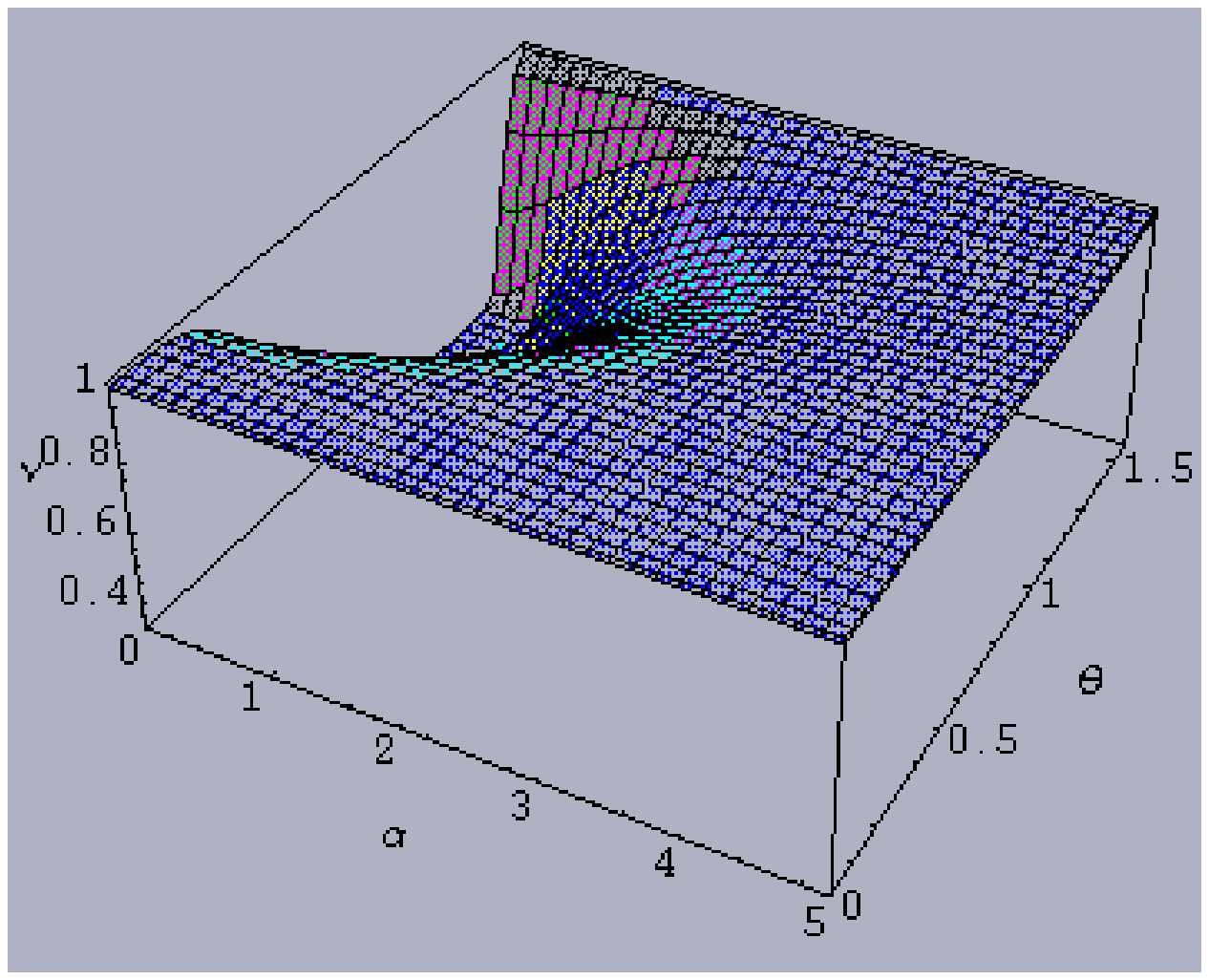}}
\caption{Sj\"oqvist {\it et al} visibility for Gibbsian spin-${3 \over 2}$ 
systems}
\label{kk4}
\end{figure}

\begin{figure}
\centerline{\psfig{figure=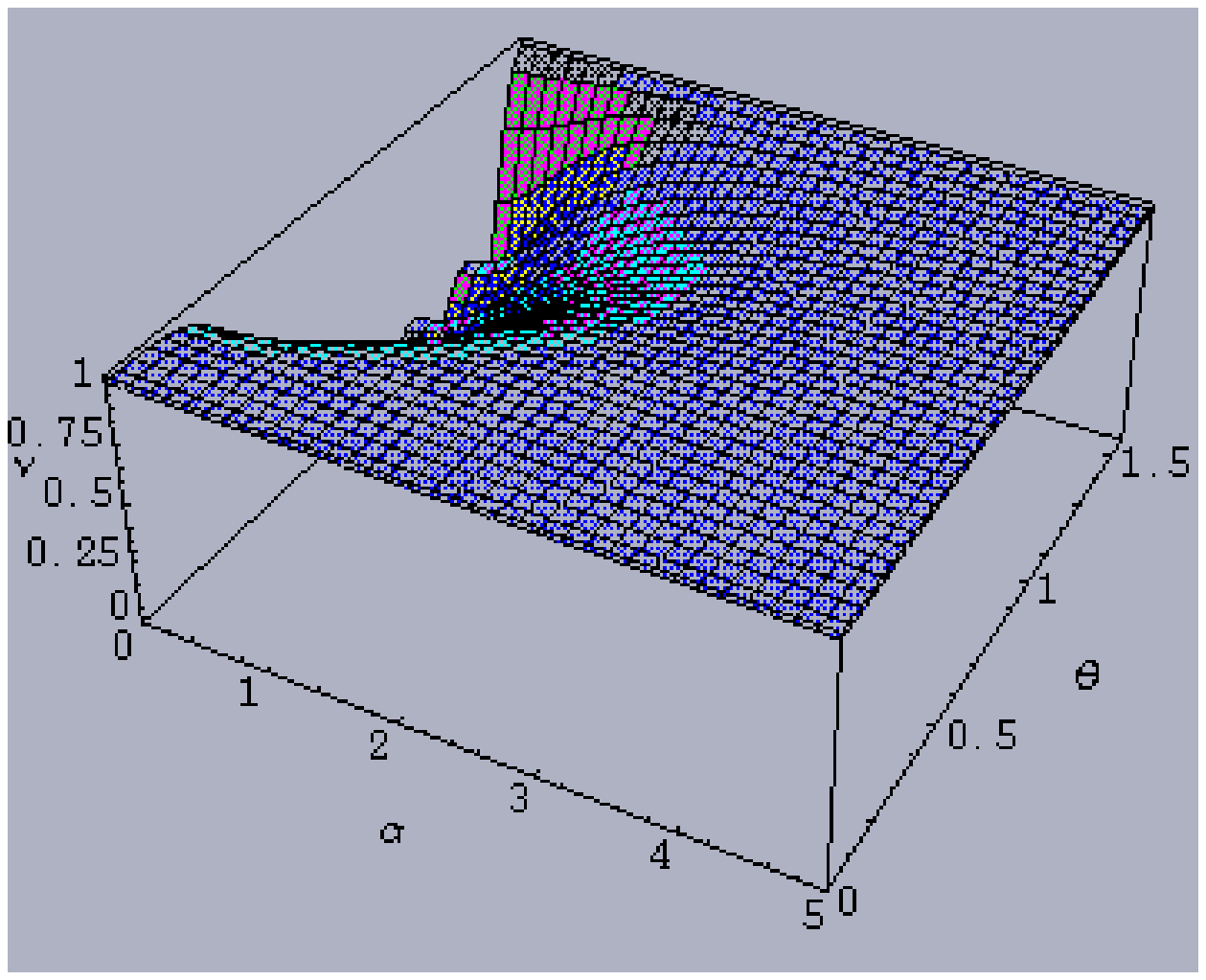}}
\caption{Sj\"oqvist {\it et al} visibility for Gibbsian spin-2 systems}
\label{kk5}
\end{figure}

\begin{figure}
\centerline{\psfig{figure=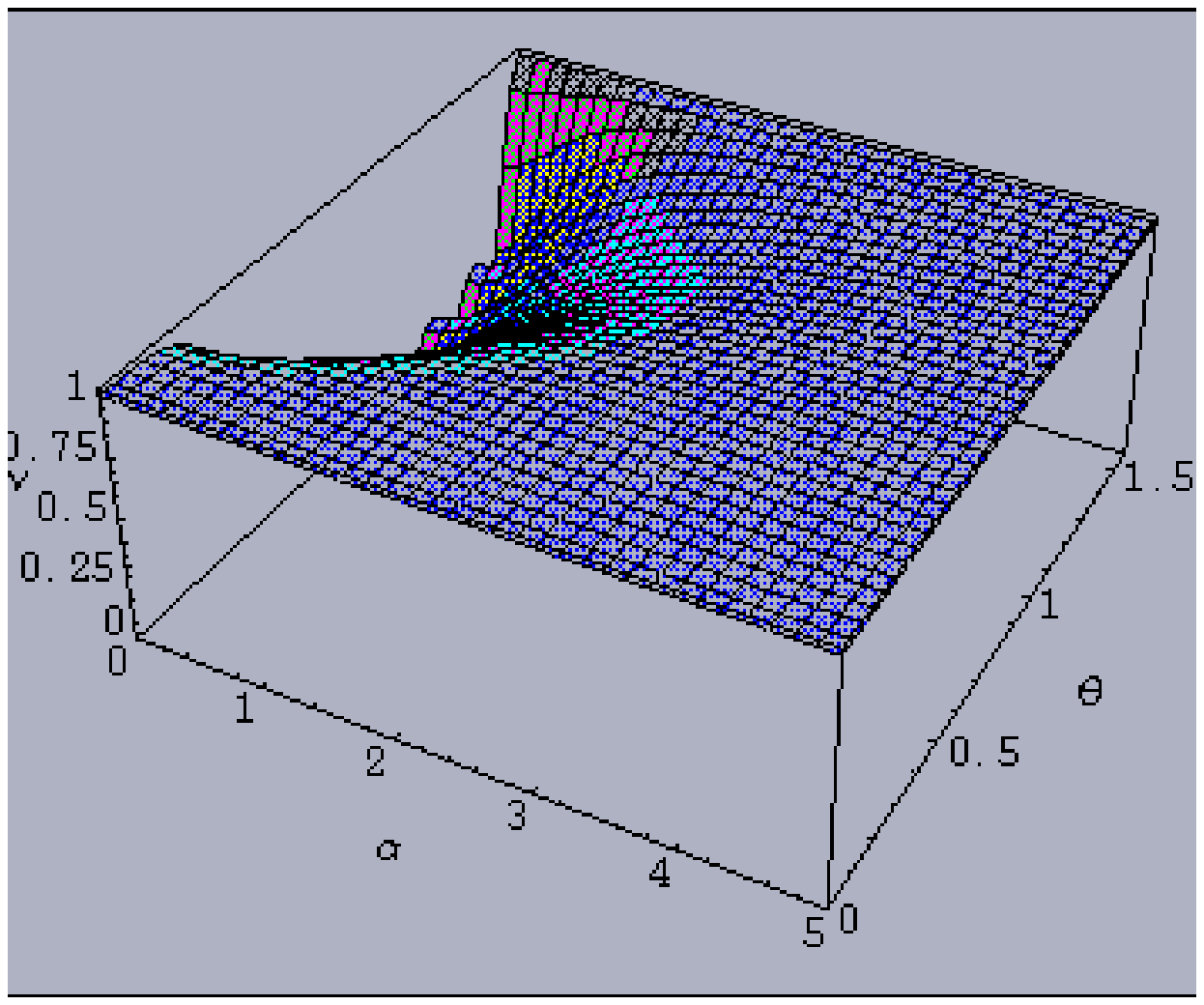}}
\caption{Sj\"oqvist {\it et al} 
visibility for Gibbsian spin-${5 \over 2}$ systems}
\label{kk6}
\end{figure}

\begin{figure}
\centerline{\psfig{figure=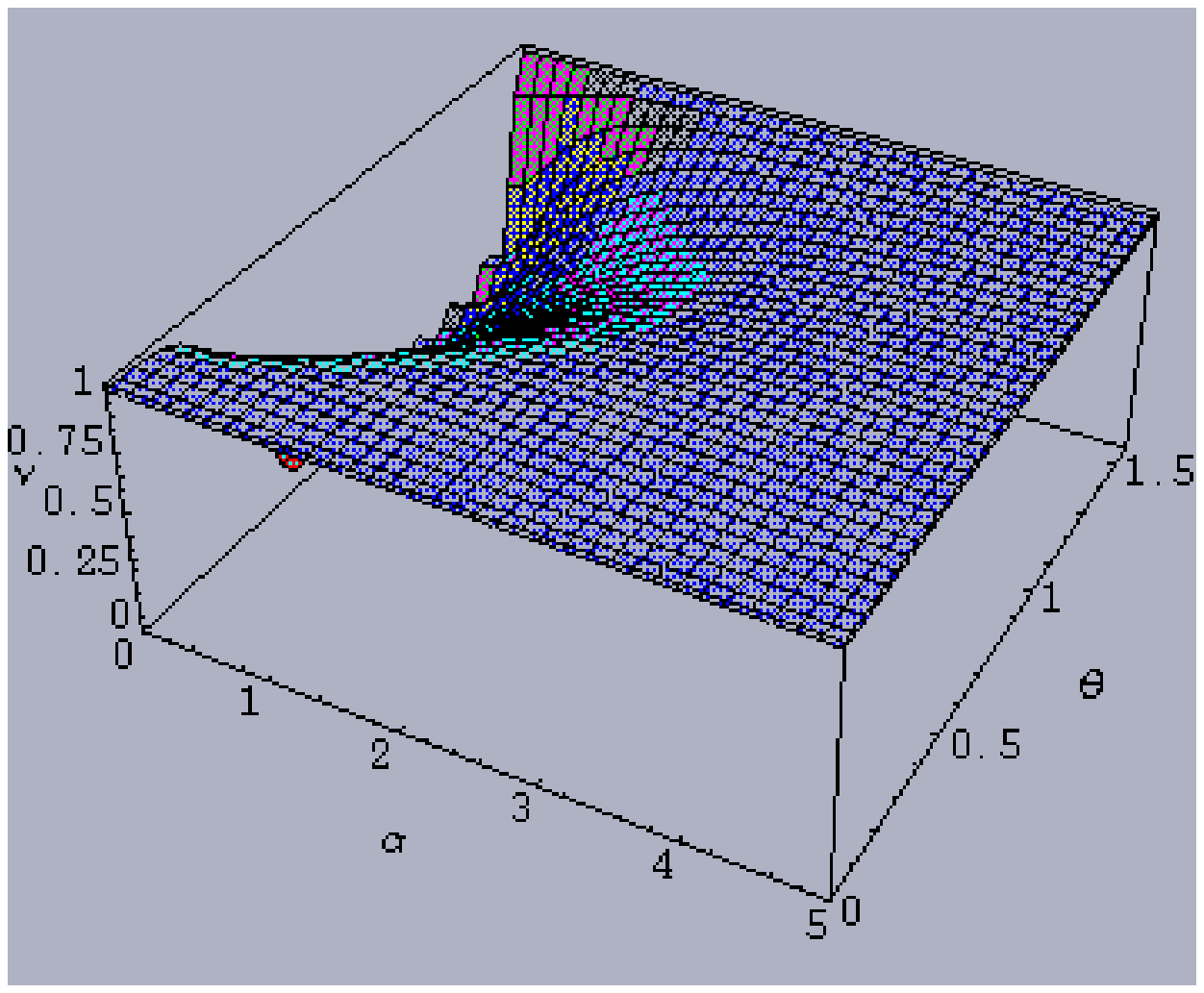}}
\caption{Sj\"oqvist {\it et al} visibility for Gibbsian spin-3 systems}
\label{kk7}
\end{figure}

\begin{figure}
\centerline{\psfig{figure=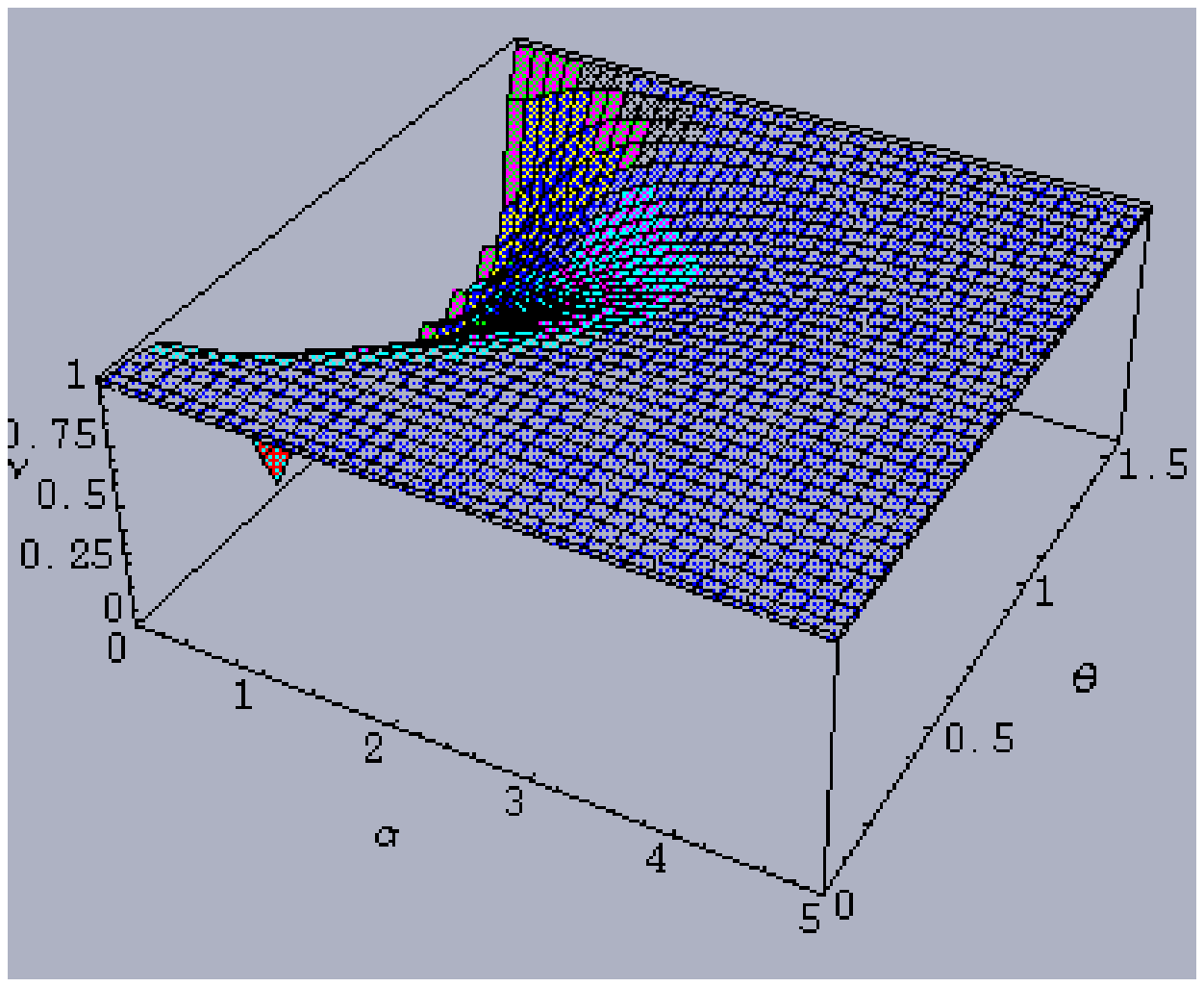}}
\caption{Sj\"oqvist {\it et al} visibility for Gibbsian spin-${7 \over 2}$ 
systems}
\label{kk8}
\end{figure}

\begin{figure}
\centerline{\psfig{figure=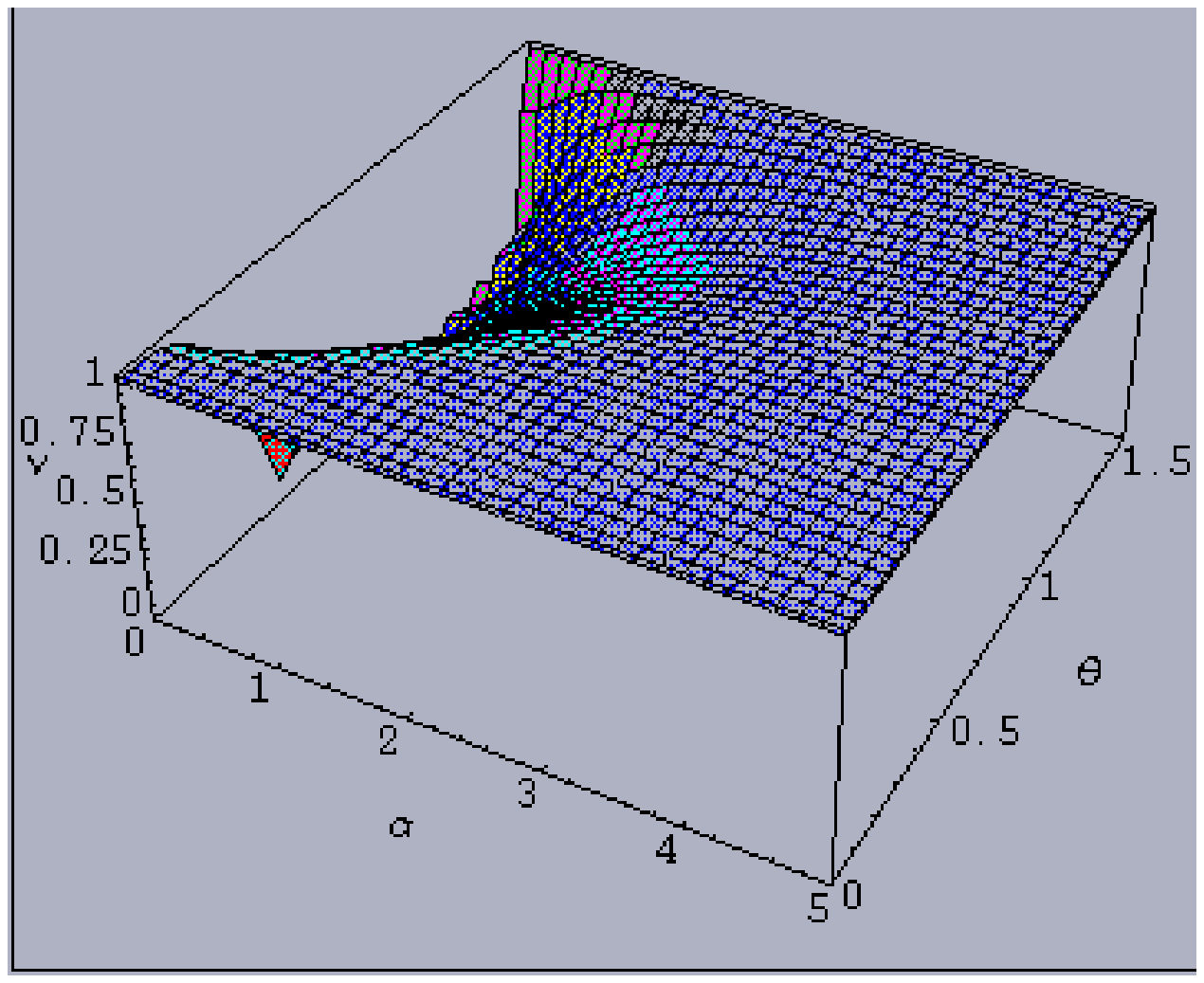}}
\caption{Sj\"oqvist {\it et al} visibility for Gibbsian spin-4 systems}
\label{kk9}
\end{figure}

\begin{figure}
\centerline{\psfig{figure=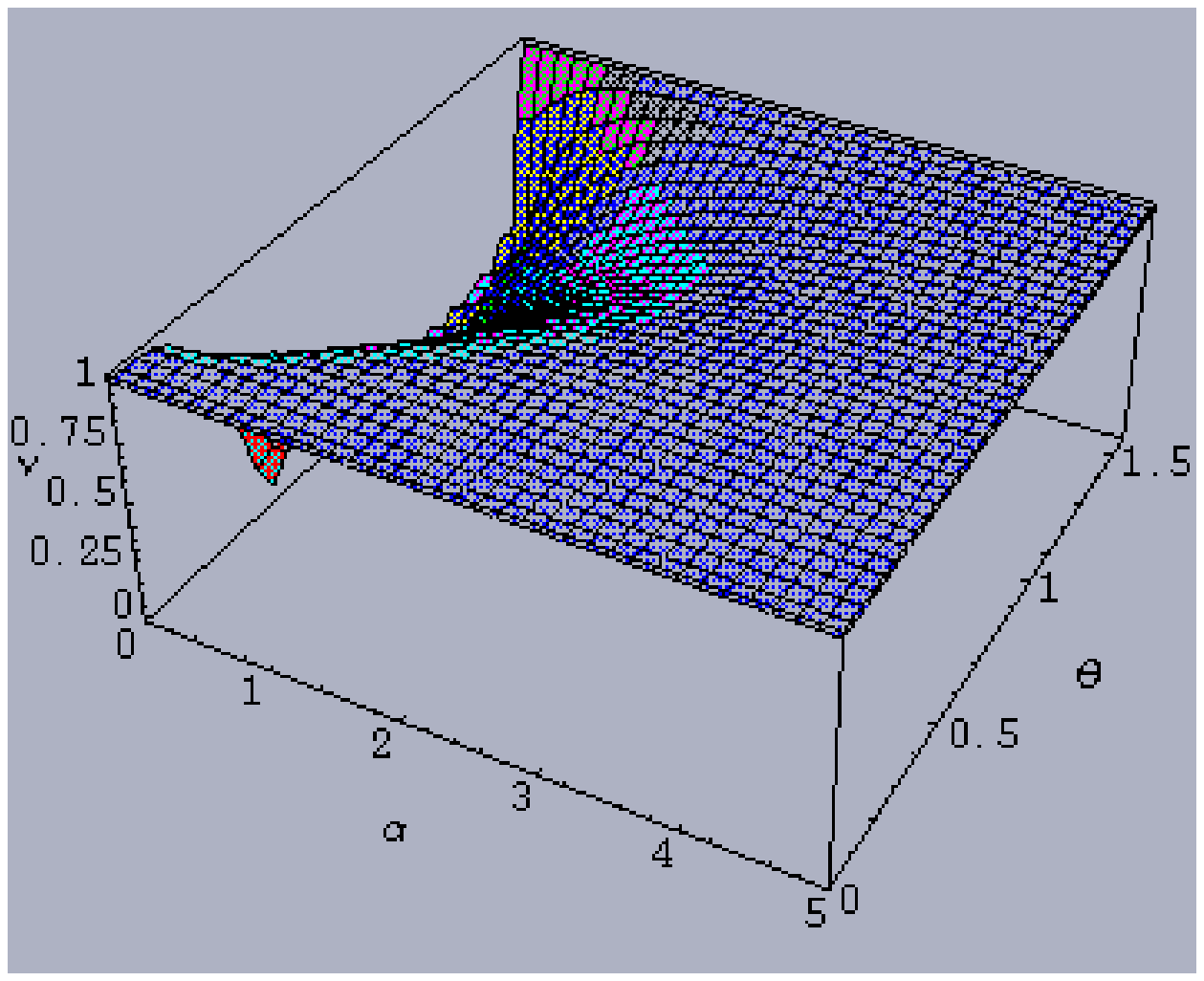}}
\caption{Sj\"oqvist {\it et al} visibility for Gibbsian spin-${9 \over 2}$ 
systems}
\label{kk10}
\end{figure}

\begin{figure}
\centerline{\psfig{figure=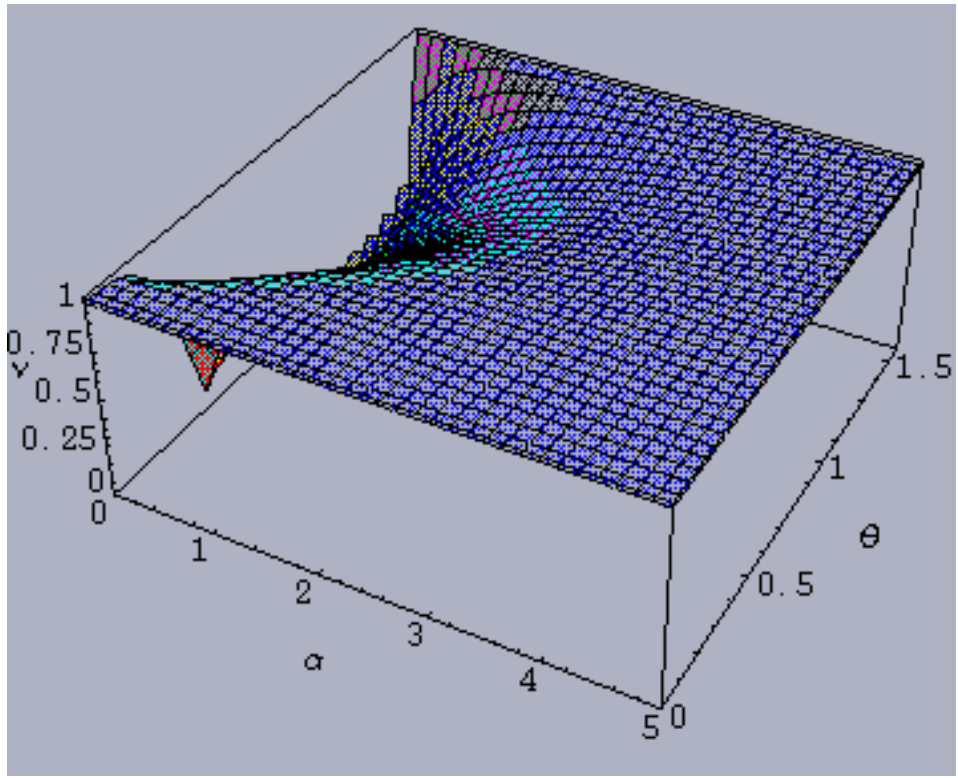}}
\caption{Sj\"oqvist {\it et al} visibility for Gibbsian spin-5 systems}
\label{kk11}
\end{figure}

In the approach of Sj\'oqvist {\it et al} \cite{sjo1} 
every pure state that diagonalizes
the initial density matrix is parallel transported separately.
By weighting the associated holonomy  of each such pure state 
by the corresponding eigenvalues,
we obtain the measure proposed in \cite{sjo1} ``in the experimental context
of quantum interferometry''. Since it is essentially trivial to obtain
the pure states that diagonalize the {\it initial} (diagonal) $n$-level 
Gibbsian
density matrix and the associated eigenvalues, and since Uhlmann
has presented the holonomy invariant (the trace of which is the Berry 
phase) 
\begin{equation} \label{bp}
W_{2 \pi} W^{*}_{0} = (-1)^{2 j} 
\rho_{0}^{1/2} e^{2 \pi i \cos{\theta} J_{z}} \rho_{0}^{1/2} = 
e^{- 2 \pi i m (1-\cos{\theta})} |m><m|,
\end{equation}
for the circular evolution of pure states, 
we can directly compute the Sj\"oqvist {\it et al} 
geometric phase ($\gamma$) and visibility ($\nu$) for the 
$n$-level Gibbsian density matrices. These are presented in Figs~\ref{k2} to
\ref{k11} and Figs.~\ref{kk2} to \ref{kk11}. 
The trace of the weighted (by the eigenvalues) sum of the 
holonomy invariants (\ref{bp}) for $n=2$ 
is
\begin{equation}
\cos{(\pi \cos{\theta})} + i \sin{(\pi \cos{\theta})} \tanh{\alpha \over 2},
\end{equation}
while for $n=3$, it is
\begin{equation}
{1 + 2 \cos{(2 \pi \cos{\theta})} \cosh{\alpha} + 2 i 
\sin{(2 \pi \cos{\theta})} 
\sinh{\alpha} \over 1  + 2 \cosh{\alpha}},
\end{equation}
for $n=4$,
\begin{equation}
{e^{- 3 i \pi \cos{\theta}} (1 + e^{\alpha + 2 i \pi \cos{\theta}})
(1 + e^{2 \alpha + 4 i \pi \cos{\theta}}) \over (1+ e^{\alpha}) 
(1 +e^{2 \alpha})}.
\end{equation}
and for $n=5$,
\begin{equation}
{1 + 2 \cos{(2 \pi \cos{\theta})} 
\cosh{\alpha} + 2 \cos{(4 \pi \cos{\theta})} \cosh{2 \alpha}
+ 2 i \sin{(2 \pi \cos{\theta})} \sinh{\alpha} + 
2 i \sin{(4 \pi \cos{\theta})} \sinh{2 \alpha} \over 1 + 2 \cosh{\alpha} + 2 
\cosh{2 \alpha}}.
\end{equation}
\section{Sj\"oqvist {\it et al} visibilities for Gibbsian $n$-level
systems ($n=2,\ldots,11$)} \label{r7}
\section{Comparisons across Gibbsian $n$-level systems of Sj\"oqvist
{\it et al} geometric phases and visibilities} \label{rr0}
\begin{figure}
\centerline{\psfig{figure=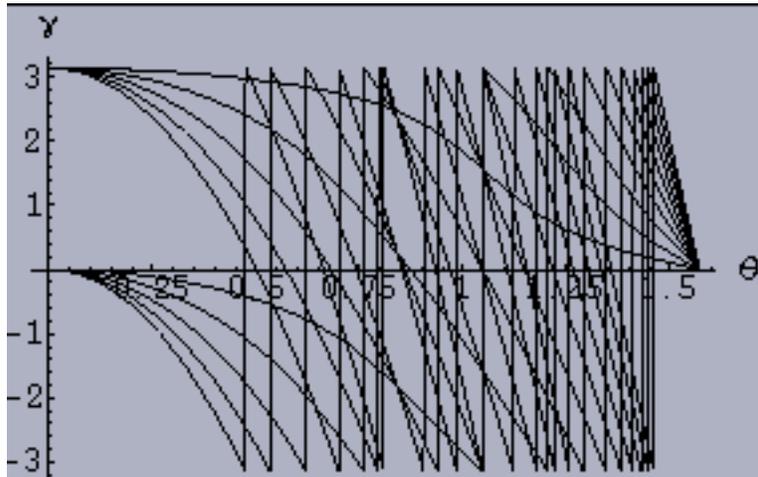}}
\caption{Sj\"oqvist {\it et al} geometric phases for $n$-level Gibbsian
systems ($n=2,\ldots,11)$ holding $\alpha=1$. At $\theta =.25$, the curve for
$n=2$ is dominant, followed in order by those for $n = 4, 6, 8, 10$, 
(all having 
{\it positive} values at $\theta = .25$) and $n = 3, 5, 7, 9, 11$ (all having 
{\it negative} values at $\theta =.25$), 
cf. Fig.~\ref{s1}}
\label{z1}
\end{figure}

\begin{figure}
\centerline{\psfig{figure=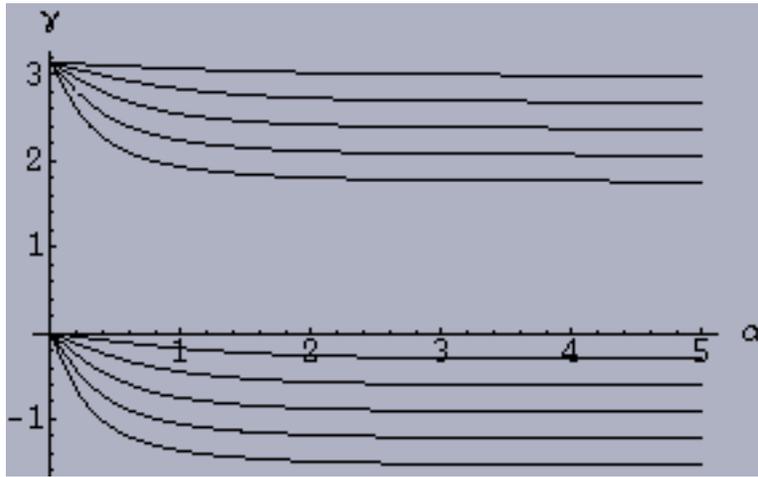}}
\caption{Sj\"oqvist {\it et al} geometric phases for $n$-level Gibbsian
systems $(n=2,\ldots,11)$ holding $\theta$ fixed at ${\pi \over 10}$.
In monotonically decreasing order are the curves for (the even) 
$n= 2, 4, 6, 8, 10$, followed by those for (the odd) $n= 
3, 5, 7, 9, 11$, cf. Fig.~\ref{s2}}
\label{z2}
\end{figure}

\begin{figure}
\centerline{\psfig{figure=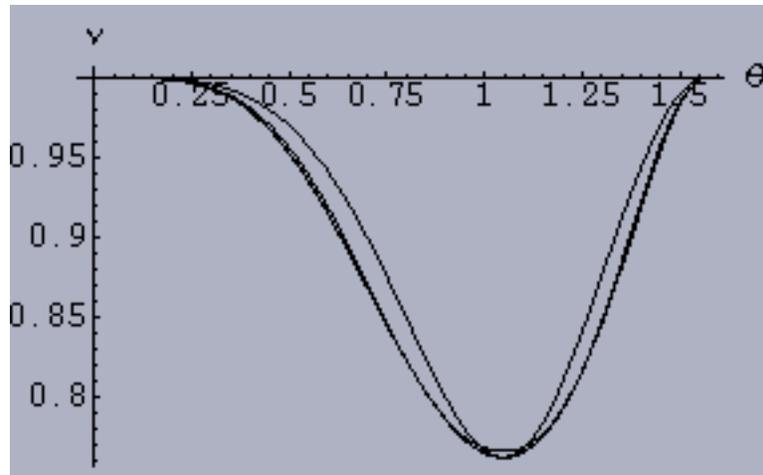}}
\caption{Sj\"oqvist {\it et al}  visibilities for $n$-level Gibbsian systems
($n=2,\ldots,11$) holding $\alpha$ fixed at 2. The curve for $n=2$
stands out from the other nine, cf. Fig.~\ref{s3}}
\label{z3}
\end{figure}

\begin{figure}
\centerline{\psfig{figure=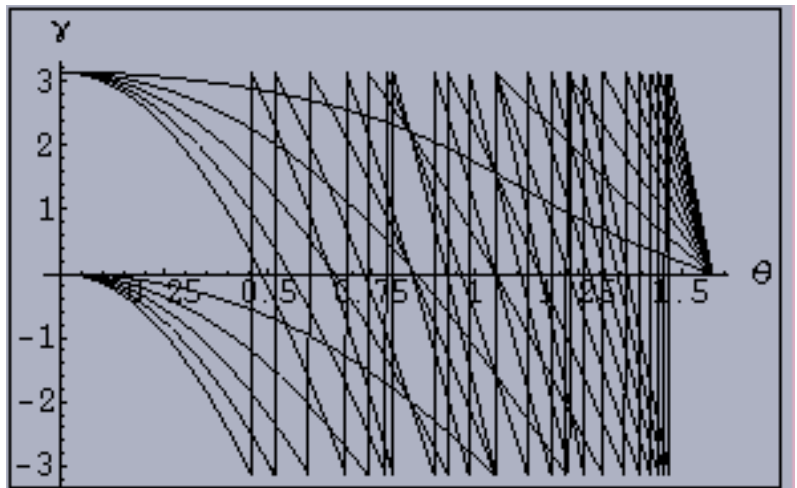}}
\caption{Sj\"oqvist {\it et al} geometric phases for $n$-level Gibbsian
systems $(n=2,\ldots,11)$ holding $\alpha =2$. In order of decreasing 
dominance at $\theta =.25$ are the curves for (the even)  $n = 2, 
4, 6, 8, 10$, followed by those for (the odd) $n=3, 5, 7, 9,  11$, 
cf. Fig.~\ref{s4}}
\label{z4}
\end{figure}

\begin{figure}
\centerline{\psfig{figure=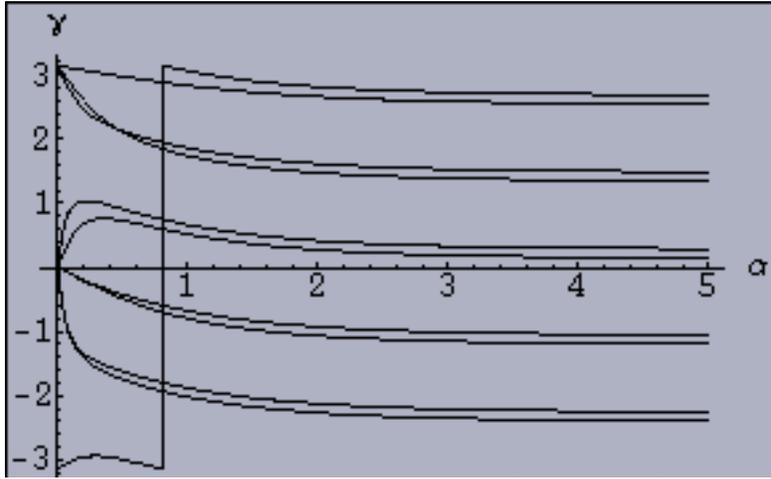}}
\caption{Sj\"oqvist {\it et al} geometric phases for $n$-level Gibbsian
systems ($n=2,\ldots,11$) holding $\theta = {\pi \over 5}$. At $\alpha=2$, 
in order of decreasing dominance are the curves for 
$n = 7,2,9,4,11,6,8,3,10,5$, cf. Fig.~\ref{s5}}
\label{z5}
\end{figure}

In Figs.~\ref{z1}-\ref{z5}, we display the direct counterparts of
Figs.~\ref{s1}-\ref{s5}, 
but now based on the methodology \cite{sjo1} of Sj\"oqvist 
{\it et al} rather than Uhlmann. (We omit the counterpart of
Fig.~\ref{s6} because it appears to consist of essentially noise.)
\section{Direct comparisons of Uhlmann and Sj\"oqvist {\it et al}
results} \label{r8}
\begin{figure}
\centerline{\psfig{figure=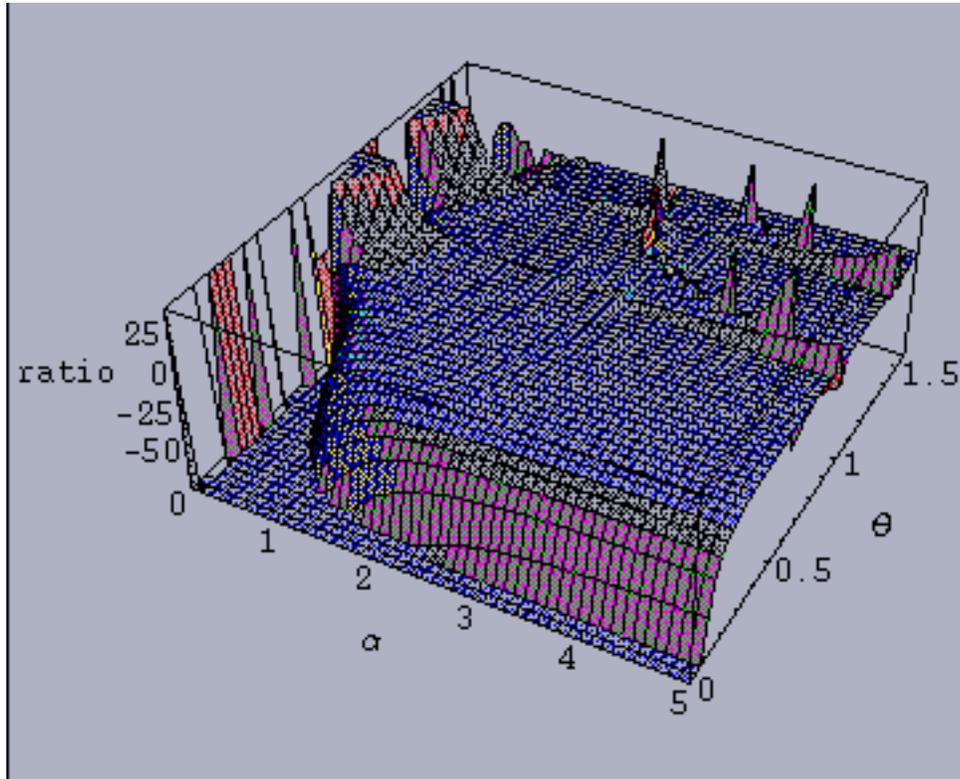}}
\caption{Ratio of the Sj\"oqvist {\it et al} geometric phase (Fig.~\ref{k6})
to the Uhlmann geometric phase (Fig.~\ref{g6}) for {\it six}-level
Gibbsian systems}
\label{o1}
\end{figure}

\begin{figure}
\centerline{\psfig{figure=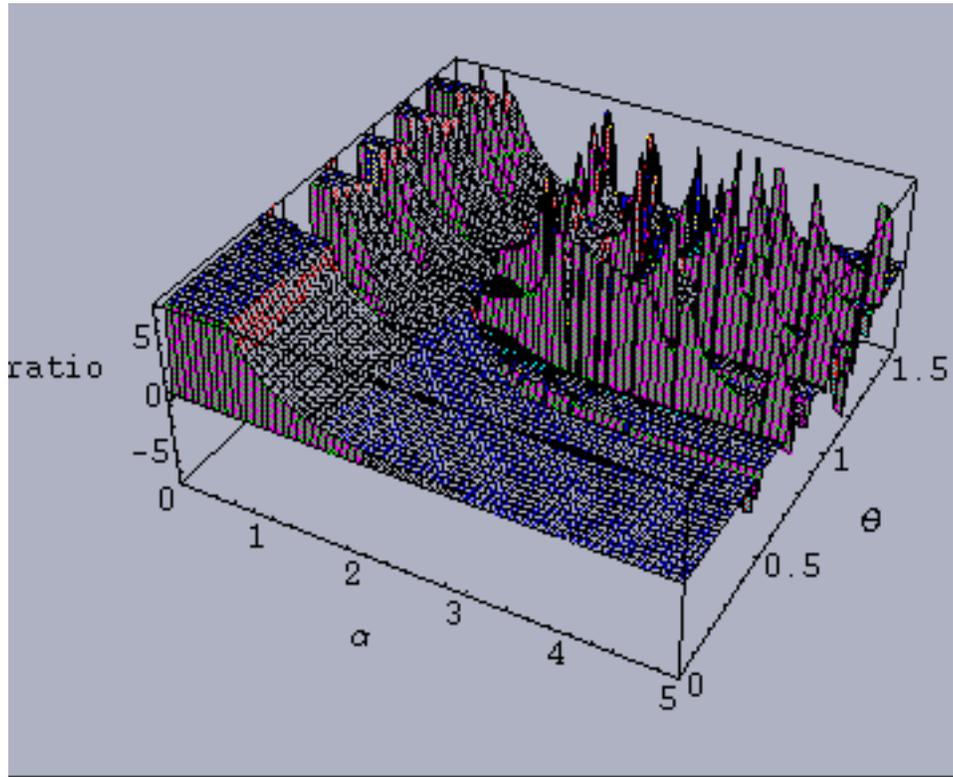}}
\caption{Ratio of the Sj\"oqvist {\it et al} geometric phase (Fig.~\ref{k11})
to the Uhlmann geometric phase (Fig.~\ref{g11}) for the {\it eleven}-level
Gibbsian systems}
\label{o2}
\end{figure}

\begin{figure}
\centerline{\psfig{figure=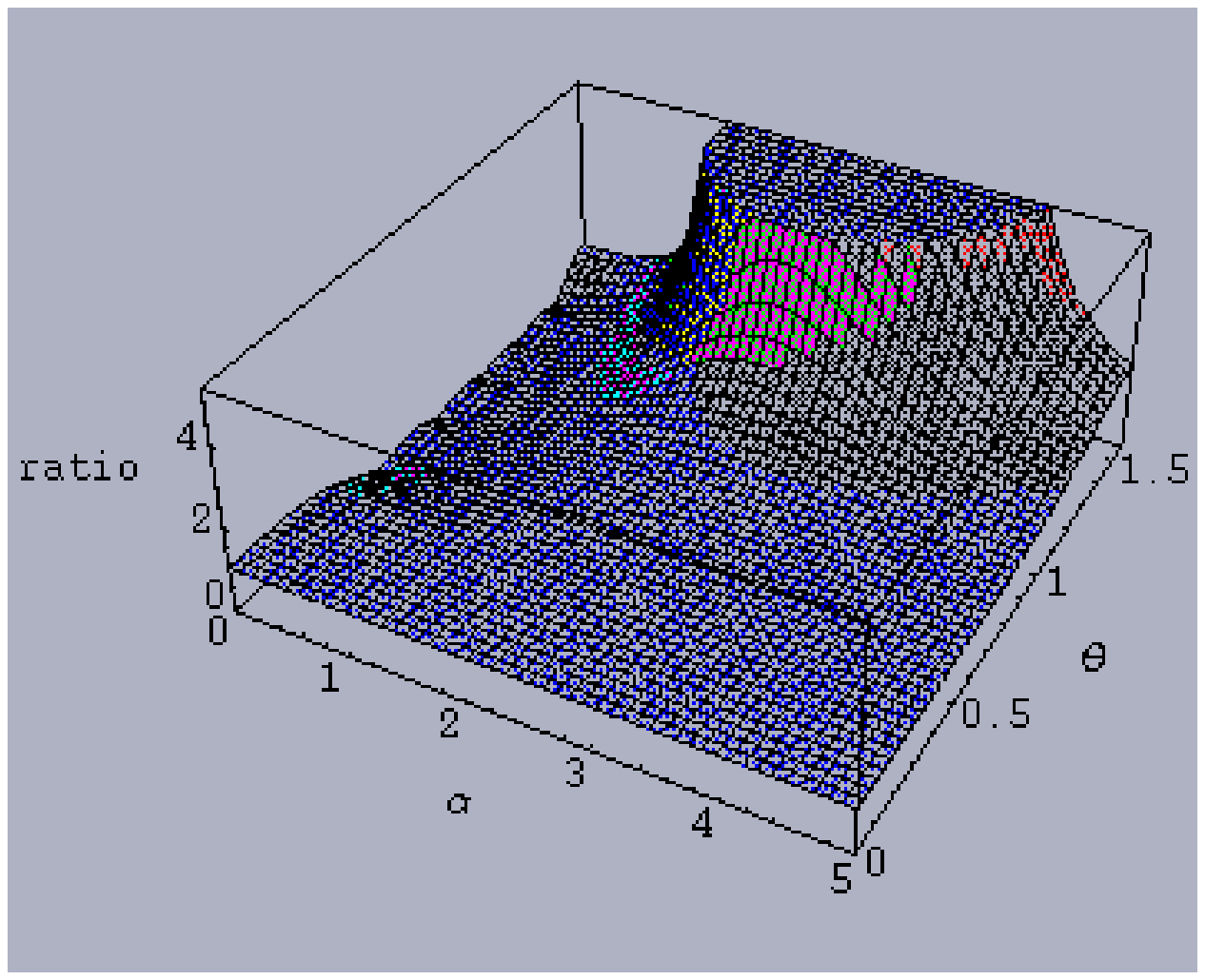}}
\caption{Ratio of the Sj\"oqvist {\it et al} visibility  (Fig.~\ref{kk6})
to the Uhlmann visibility (Fig.~\ref{v6}) for the {\it six}-level Gibbsian
systems}
\label{o3}
\end{figure}

\begin{figure}
\centerline{\psfig{figure=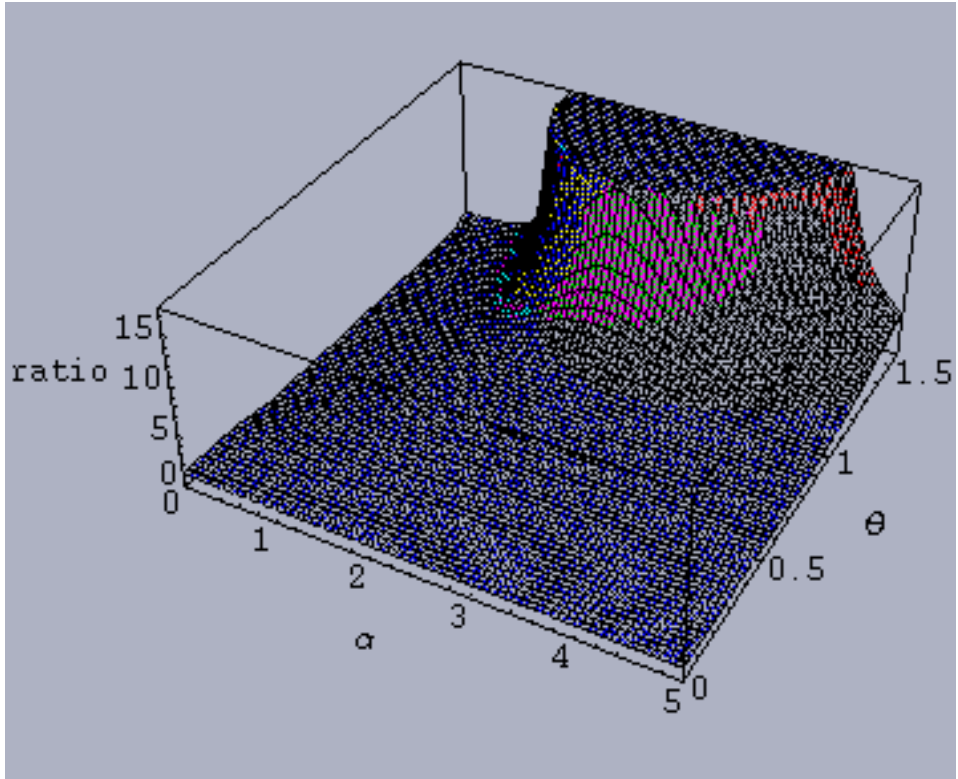}}
\caption{Ratio of the Sj\"oqvist {\it et al} visibility (Fig.~\ref{kk11})
to the Uhlmann visibility (Fig.~\ref{v11}) for the {\it eleven}-level
Gibbsian systems}
\label{o4}
\end{figure}

In Fig.~\ref{o1}, we plot for $n=6$ the ratio of the Sj\"oqvist {\it et al}
geometric phase (Fig.~\ref{k6}) to the Uhlmann geometric phase 
(Fig.~\ref{g6}), and in Fig.~\ref{o2}, the corresponding ratio for $n=11$.
In the next two figures (Figs.~\ref{o3} and \ref{o4}), we show the 
ratios of the associated visibililities.

\section{Summary} \label{r9}

We have reported an in-depth study here of two different methodologies
for determining geometric phases for {\it mixed} states \cite{sjo1,uhl1}.
The analysis has been framed in terms of rotations ($O(3)$-orbits) 
 of $n$-level
Gibbsian systems, as originally proposed by Uhlmann \cite{uhl2}.

There are, of course, many features in these results.
One of these is that the Sj\"oqvist {\it et al} geometric phases
(sec.~\ref{r6}) appear to be less sensitive to the inverse temperature 
parameter ($\alpha$) than do the Uhlmann geometric phases 
(sec.~\ref{r2}).

In particular, we have found (sec.~\ref{r4}) that 
as the number of levels of the Gibbsian density matrices increases 
(in the sample we have studied) from $n=2$ to 11,
the Uhlmann methodology often yields simple monotonic behavior.
On the other hand, this seems to be largely absent using
the Sj\"oqvist {\it et al}
methodology (sec.~\ref{rr0}), which  appears among other
things to be sensitive to the bosonic (odd $n$) or fermionic 
(even $n$) character
of the system under consideration,

It is certainly important to note that 
if one sets the parameter $a$ --- given in (\ref{aequation}) --- equal 
to zero in the approach of Uhlmann (sec.~\ref{r1}), the resulting
holonomy invariant (\ref{hi}) is simply {\it equal}
to $(-1)^{n+1}$ times the holonomy invariant of Sj\"oqvist {\it et al}
\cite[eq. (15)]{sjo1}.
This would appear to help to explain the added complexity of the plots
of the Uhlmann geometric phases (sec.~\ref{r2}) vis-\'a-vis those of the
Sj\"oqvist {\it et al} plots (sec.~\ref{r6}).

We have found compelling numerical evidence that for odd $n$
the central ($\lceil {n \over 2} \rceil$-th)  of the $n$ eigenvalues 
(ordered in terms of absolute value) of 
the Uhlmann holonomy invariant (\ref{hi}) is
always a real number (sec.~\ref{reigen}). 
 (By the fundamental theorem of algebra, for 
odd $n$, one of the 
$n$ eigenvalues must of course 
be real --- since complex roots come in conjugate 
pairs --- but clearly not necessarily the centrally
located eigenvalue.)
Also, the $(\lceil {n \over 2} \rceil,\lceil {n \over 2} \rceil)$-entry
of the Uhlmann holonomy invariant (\ref{hi}) is itself real, for 
odd $n$.
In terms of absolute values, the diagonal entries of 
the invariant (\ref{hi}) appear to be always
monotonically decreasing from the (upper left) 
(1,1)-entry to the (lower right) ($n,n$)-entry.
(We note that the angular momentum operator 
$J_{z}$ --- entering in our equations 
(\ref{fq1}), (\ref{aequation}) and
 (\ref{bp}) --- itself has declining [real] diagonal entries, that is, 
${n-1 /over 2},\ldots,-{n-1 \over 2}$.)

\acknowledgments

I would like to express appreciation to the Institute for Theoretical
Physics for computational support in this research.

\listoffigures

\end{document}